\documentclass[twocolumn]{aa}
\usepackage{rotating}   
\usepackage{longtable}   
\usepackage{graphicx, natbib, psfrag, latexsym, amssymb}
\bibstyle{aa}  
\def \bz{ }
\def \ba{ }       
\def \bc{ }       
\def \bx{ }
\def \etal   {{~et~al.~}}
\begin{document}
\title{The B3-VLA CSS sample. 
       VIII: New optical identifications from the Sloan Digital Sky Survey}
\subtitle{The ultraviolet-optical spectral energy distribution of the young 
radio sources}

\author{ C. Fanti     \inst{1}   \and 
	 R. Fanti     \inst{1}   \and
	 A. Zanichelli  \inst{1} \and
         D. Dallacasa   \inst{1,2} \and       
         C. Stanghellini \inst{1}      }

\offprints{A. Zanichelli\\
  \email{a.zanichelli@ira.inaf.it}}

\institute {Istituto di Radioastronomia -- INAF, Via Gobetti 101,
  I-40129 Bologna, Italy  
\and Dipartimento di Astronomia, Universit\`a di Bologna,
Via Ranzani 1, I-40127 Bologna, Italy}

\date{Received \today; Accepted ???}

      %
      %
      \titlerunning{UV - Optical SEDs of young radio sources}
      \authorrunning{C. Fanti \etal}

\abstract 
{Compact steep-spectrum radio sources and giga-hertz peaked spectrum radio 
sources (CSS/GPS) are generally considered to be mostly young 
radio sources. In recent years we studied at many wavelengths 
a sample of these objects selected from the B3-VLA catalog: the B3-VLA CSS 
sample. Only $\approx$  60 \% of the sources were optically identified. }
{We aim to increase the number of optical identifications and  
study the properties of the host galaxies of young radio sources.}
{We cross-correlated the CSS B3-VLA sample with the Sloan Digital Sky 
Survey (SDSS), DR7, {\bx and complemented the SDSS photometry with
available GALEX (DR 4/5 and 6) and near-IR data from UKIRT and 2MASS. 
}}
{We obtained new identifications and photometric redshifts for eight 
faint galaxies and  for one quasar and two  quasar candidates. 
Overall we have 27 galaxies with SDSS 
photometry in five bands, for which we derived the ultraviolet-optical spectral
energy distribution {\bx (UV-O-SED)}. {\bx We extended our investigation to 
additional CSS/GPS selected from the literature.
Most of the galaxies show an excess of ultra-violet (UV) radiation compared 
with the UV-O-SED of local radio-quiet ellipticals. 
We found a strong dependence of the  UV excess on redshift and analyzed it 
assuming that it is generated either from the nucleus (hidden quasar) or 
from a young stellar population (YSP)}. 

We also compare the UV-O-SEDs of our 
CSS/GPS sources with those of a selection of large size (LSO) powerful
radio sources from the literature.}
{If the major process of the  UV excess
is caused by  a YSP, our conclusion is that it is the
result of the merger process that also triggered the onset of the radio source
with some time delay. We do not see evidence 
for a major contribution from a YSP triggered by the radio sources itself.}


\maketitle

\section {Introduction}

Giga-hertz peaked spectrum (GPS) and compact steep-spectrum (CSS) radio 
sources (of subgalactic radio size) with a double-lobed structure, 
which are referred to
as {\it CSO}s and {\it MSO}s ({\it compact} and {\it medium-size symmetric 
objects}), have been suggested for several years
to be the young precursors of the large size powerful radio galaxies 
\citep{Fanti95, Readh96, O'Dea97}. Since then a considerable amount
of data have been produced to shed light on their properties and to develop
theoretical models about their radio evolution \citep[e.g.][and references 
therein]{Kaiser09}.

Much interest was given as well to the optical hosts of these radio 
sources. \cite{deVr98, deVr00} showed that 
the hosts are old giant ellipticals. Interactions of the radio source with 
the galaxy interstellar medium are clear from the study of the 
emission lines (see, e.g., \citet{Lab05, Holt2}). The presence of 
young stellar populations in these objects has been revealed several times 
\citep{Holt1,Lab08,deVr07,TADH}). 
Their origin is attributed to the shocks generated by the 
young radio source expanding into the interstellar medium \citep{Rees, 
Mellema, Bick00}, and/or as the result 
of a galaxy merger process which, at some later epoch, also
triggered the onset of the radio source \citep{Raimann,Holt1}.


{\bx In this paper we present UV-Optical data for a 
composite sample of CSOs/MSOs sources, which allows us to obtain the UV-O-SED 
of this class of radio sources and to test the presence of young stellar 
populations.}


The paper is organized as follows.

In Sects. 2 and 3 we describe the B3-VLA CSS sample  \citep{Fanti01} 
and its cross-correlation  with the SDSS. From this we obtained 
{\ba photometry in five bands for 35 radio sources, 12 of which are new 
identifications}. We discuss some properties of the sample. 

In Sect. 4 we derive the UV-O-SEDs 
of the B3-VLA CSS sources and of those CSOs/MSOs, taken from the literature, 
for which SDSS data exist. {\bx The SDSS data were supplemented with UV data 
from the GALEX surveys \citep{Martin} and
with near-IR data (J, H, K bands) from the 2MASS survey \citep{2MASS} and from
UKIRT literature data. 
We show that the hosts of these radio galaxies exhibit an excess of UV 
radiation ({\it UV-excess}) compared with the standard UV-O-SED of old 
elliptical galaxies.}

In Sect. 5 we discuss some properties of the {\it UV-excess} and its
possible origin.

In Sects. 6 and 7 we repeat the analysis for a sample of large size radio 
galaxies (LSO) and compare their UV properties with those of the small size 
sources.

Section 9 gives our conclusions.

{\bx Appendices contain supporting data tables, notes, and figures showing the
UV-O-SEDs for both compact and extended radio galaxies.}

\section{The B3-VLA CSS sample}
\label{B3CSSsample}

The B3-VLA CSS sample \citep[Paper I]{Fanti01} is a complete subset of the 
B3-VLA catalog \citep{vig1}. It counts 87  radio sources with angular size 
less than a few arcsecs, corresponding to a linear size 
$\la 20 $ kpc (for  H$_{\rm 0}$ = 100  km/s, q$_{\rm 0}$ = 0.5, 
values  used in Paper I),  
divided into two different flux density bins: 0.8--1.6 Jy and $\ge$ 1.6~Jy
at 408 MHz.
 
The sample was observed at several radio frequencies with the VLA \citep[4.9 
and 8.4 GHz, Paper I; 15 GHz][] {alex}. Several subsets were also 
observed with the Merlin and VLBI arrays at 1.6 GHz 
\citep{DDa, DDb} and  at 5.0 GHz \citep{orie}. Polarization studies have been 
presented in \citet{Fanti04} and in \citet{alex2}.
The high-resolution radio imaging showed that {\bc about 90 \% of the sources 
have a double structure and are therefore classified {\it CSO}s 
($\approx$ 30 \%) or {\it MSO}.}
The remaining {\ba sources are either core-jets or have a complex morphology.}


Initially \citep{vig1} systematic optical information was obtained only 
down to the limiting magnitude of the Palomar Observatory Sky Survey (POSS),
i.e.:

\noindent
{\it a)} all quasar candidates were observed spectroscopically and their 
redshifts measured  \citep{vig2};

\noindent
{\it b)} for galaxies visible on the POSS (m$\rm_R \la 20.5$), photometric 
redshifts (up to $\approx$ 0.5) were determined from the apparent 
red magnitude as described in  \citet{vig1};

\noindent
{\it c)} $\approx$ 75 \% of the sources (referred to as ``empty fields'' 
or ``E'' sources) had no optical identification.

Information for a few additional objects was also available from the literature
(see the notes in Sect.\ref{notes}).
Later on, deeper optical identifications in the R and K bands were carried out
at random for sources without an optical counterpart on the POSS \citep{McCa, 
PD,Th94,Max}. In this way a number of new identifications 
were obtained, mostly galaxies with redshifts either spectroscopic
(z$_{\rm sp}$) or photometric (z$_{\rm K}$) from the redshift-magnitude
relation in the K-band.

At the time of writing Paper I the percentage of the optically unidentified 
radio sources was still  $\approx$ 40 \% and it has not decreased much since 
then.

\section{Search for new optical identifications with the SDSS }
\label{srcId}

\subsection{The search criteria}
\label{srccrit}

Fifty-seven radio sources of the B3-VLA CSS sample are located in the sky 
area covered by SDSS, DR7 \citep{R7}. Optical counterparts have been searched 
for in the five wavelength bands $u$ ($\lambda=$3551 \AA), $g$ 
(4686 \AA), $r$ (6165 \AA), $i$ (7481 \AA), $z$~(8931~\AA) with
the procedures available on the SDSS site (CAS Database). 
The completeness magnitude limits of the SDSS are: $\approx$ 22.0 ($u$), 
$\approx$ 22.2 ($g$), $\approx$ 22.2 ($r$), $\approx$ 21.3 ($i$), $\approx$ 
20.5 ($z$), although objects fainter than these limits can also be found.

The primary search area around each radio source position was a circle 
of 1.2 arcsec in radius, adequate for the  accuracy of the optical and radio 
positions and for the uncertainties in the registration of the radio and 
optical reference frames.
The search was then repeated in a more extended area, 2.4 arcsec in radius,
which generally encompassed the total size of the radio source, in order to 
check for cases of displacement of the radio centroid from the optical position
and to evaluate the number of spurious coincidences. No additional object was 
found. 
{\bc We found 35 optical objects within the 1.2 arcsec search radius including
23 already known and hence confirmed objects. Furthermore the source 
\object{1016+443} is now definitely identified with another galaxy fainter 
than the one reported earlier. 
Finally the previous identification for \object{1350+432} is rejected 
because of a positional disagreement between the radio source and the optical 
object originally assumed as counterpart, and this source is now classified 
as ``E''.

A number of the 35 objects are rather faint, with magnitudes below the 
completeness limits in some photometric bands, four of them are below the 
limit in all bands. 
Because there may be some doubts about the reliability of these faint objects,
we made a blind search at about one hundred random positions, always with a 
search area of 1.2 arcsec radius, and found no object at all at any magnitude.
Therefore we are confident that we have no misidentifications with either 
real or spurious objects.}
{\bx Nevertheless some of the objects  have a signal-to-noise ratio $\la 3 
\sigma$ (magnitude  error $\approx$ 0.4), i.e. are undetected, in one or more 
bands. 
Specifically the detection rate is $\approx$ 50\% in the $u$ band, $\approx$ 
90 \% in the $g$ and $r$ bands and 100\% in the $i$ and $z$ bands. As a 
consequence $\approx$  49\% of sources are detected in all SDSS bands. Another 
$\approx$ 34\%  are detected in four bands, $\approx$ 14\% in three bands and 
only one object in two bands}.  

On the other hand, 10 galaxies that are known from the earlier optical 
identification projects were not found in the SDSS. This is not surprising,
because seven of them
have red magnitudes (from earlier works) much fainter than the SDSS limit 
and the remaining three, with unknown  red magnitudes, have high redshifts
(two spectroscopic $>$ 2.0 and one, photometric, z$ \approx 1.2$)  
and are likely fainter than the  SDSS magnitude limits.

\subsection{The new identifications}
\label{newId}

The SDSS classification of either quasar or galaxy is based on the  extent of
the optical object (star-like objects are quasar candidates).
Additional support for the classification is based on the location in the 
Hubble diagrams (Sect. \ref{Hdia}) and on the  UV-O-SEDs (Sect. \ref{SED}).

\begin{figure*}[t]
\centering
 {\includegraphics[width=10cm]{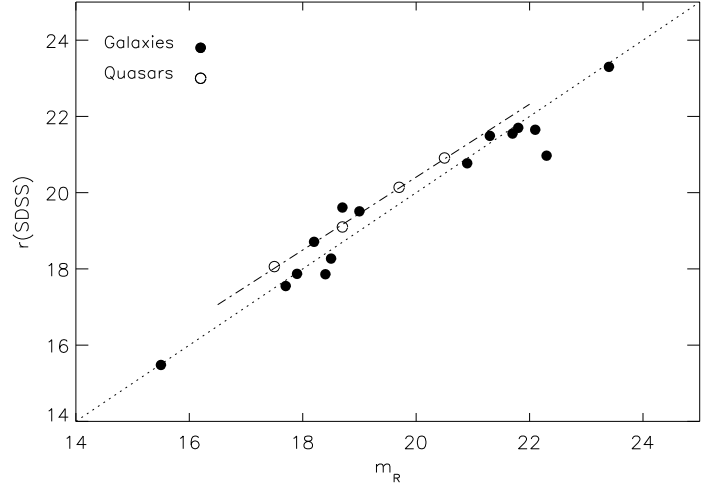}}
\caption{Comparison between the SDSS $r$ magnitudes and those taken from 
Paper I }
\label{r-comp}
\end{figure*}

\begin{figure*}
\resizebox{\hsize}{!}{\includegraphics{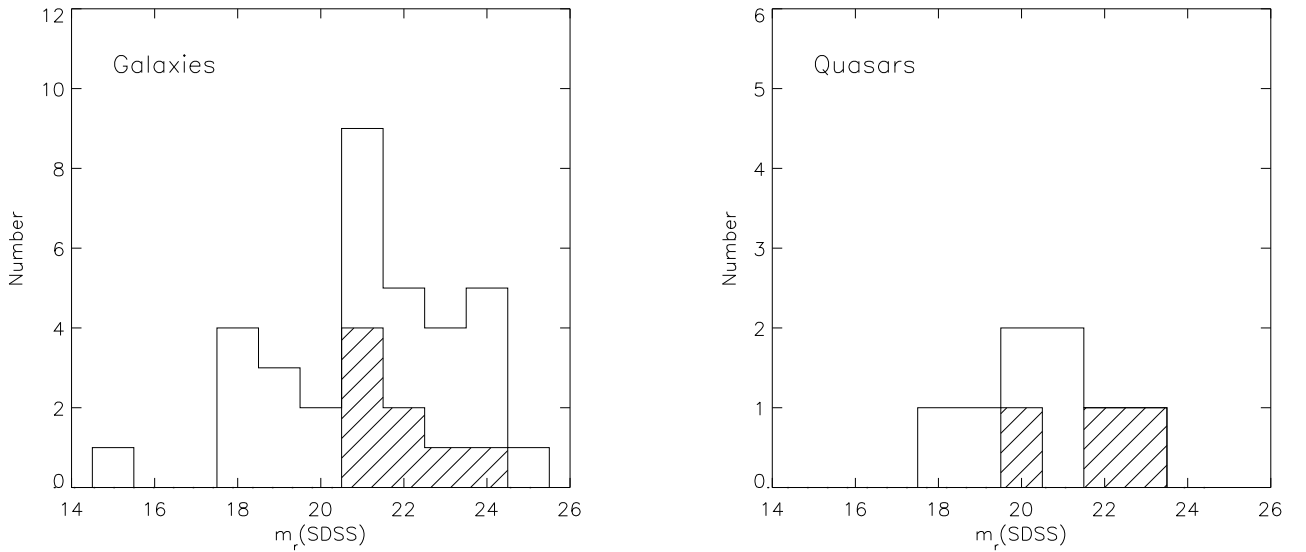}}
\caption{Distribution of the $r$ magnitudes for the optically identified 
galaxies (left) and quasars (right). Hatched bins indicate the new 
identifications (Q? included)}
\label{m-hist_G}
\end{figure*}

{\bc Eight new galaxies (including the change of identification for 
\object{1016+443}) and one new quasar with spectroscopic redshift 
were found. Finally there are
two other identifications with star-like objects, possible quasar candidates.
The quasar classification of one of them (\object{1055+404A}) is supported  
by its core-jet radio structure.
 
At present the fraction of optically identified B3--VLA CSS sources in this 
sub-sample matching the SDSS area has risen from  $\approx$60\% to  
$\approx$80\%.}
Table~\ref{B3tab1} gives the available optical data of the sample. 

The present identification status of the 57 SDSS radio sources  is then the 
following:

\noindent
a) Thirty-seven galaxies (G), all but three with redshift, either spectroscopic
(z$_{\rm sp}$) or photometric (z$_{\rm ph2}$, Sect. 3.3). Twenty-seven have  
SDSS photometry. 

\noindent
b) Six quasars (Q), five with z$_{\rm sp}$ and one with z photometric,
and two quasar candidates (no redshift);

\noindent
c) twelve still  unidentified sources (E), listed in Table~\ref{EF}. 
Actually, at the radio position of three of them very faint, 
unclassifiable objects have been detected by other authors 
(see notes in Sect. \ref{notes}). 

Figure \ref{r-comp} shows the comparison between the old  and the new $red$ 
magnitudes. {\bx  The old magnitudes (that we indicate with $m_{\rm{R}}$)  
are fairly heterogeneous, (see notes in Sect. \ref{B3tab1}) and  errors are not
quoted. In the comparison we did not apply galactic extinction corrections 
because of a lack of information on the filters used for several objects. 
However, because the  typical reddening is low
(E(B-V) $\approx$ 0.018), the extinction is expected to be similar in any red 
band used and therefore there should  be no effects in the magnitude 
comparison.
For galaxies the two sets of data have a relative dispersion of $\approx$ 0.5 
mag. with null offset. We suppose that the scattering is largely caused by the
(not quoted)  errors of the $m_{\rm{R}}$  magnitudes, because the errors on 
$r_{\rm{SDSS}}$ are generally smaller. 
For the four quasars though there is 
a significant offset of $\sim$ 0.4 mag. (SDSS being fainter) with small 
dispersion.  For them we found in the literature 
other measures, which were well consistent with  those from Paper I.
We found no explanation for this apparent offset.}

In Fig.~\ref{m-hist_G}
we show the r-magnitude distribution of the identified objects.

\subsection{ SDSS photometric redshifts}
\label{phrs}

{\ba The eight newly identified galaxies  are 
too faint to have been observed spectroscopically in the Sloan Survey.
We derived their} photometric redshifts (z$_{\rm ph}$, z$_{\rm ph2}^{\rm cc2}$
and z$_{\rm ph2}^{\rm d1}$) from the multi-band SDSS photometry,  using the 
SDSS  available routines. The z$_{\rm ph}$ is based on galaxy templates, 
while the  z$_{\rm ph2}$ is based on neural nets acting in a five-color 
space (cc2) or in a magnitude space (d1), and also employs concentration 
indices (see description in ``Help - algorithms -Photoz'', in the SDSS site).
The z$_{\rm ph2}$ is recommended by the SDSS group for faint objects, as in 
our case.

For six out of the eight galaxies the SDSS routines gave consistent values of 
z$_{\rm ph2}^{\rm cc2}$ and z$_{\rm ph2}^{\rm d1}$. The z$_{\rm ph}$ are 
instead underestimated compared with the z$_{\rm ph2}$.
For the two remaining new galaxies (\object{0814+441} and 
\object{1441+409}) the fitting routines failed in giving z$_{\rm ph2}$, while 
the z$_{\rm ph}$  have very low ($\approx 0.15$), implausible values for 
objects with $r \ga 23.0$.

In order to evaluate the reliability  of these six photometric redshifts we 
determined z$_{\rm ph}$ and z$_{\rm ph2}$ for the 18 galaxies with known 
z$_{\rm sp}$ as well. Only for one source 
(\object{1159+395}) no  z$_{\rm ph2}$  fits were obtained, and  z$_{\rm ph} = 
0.62 
\pm 0.2$ strongly disagrees with z$_{\rm sp} = 2.37$.
For the other 17 sources the z$_{\rm ph2}^{\rm cc2}$ and 
z$_{\rm ph2}^{\rm d1}$ again generally agree well, 
while the z$_{\rm ph}$ are underestimated with respect to  
z$_{\rm ph2}$ for z$\ga 0.5$.
 
\begin{figure}[t]
\resizebox{\hsize}{!}{\includegraphics{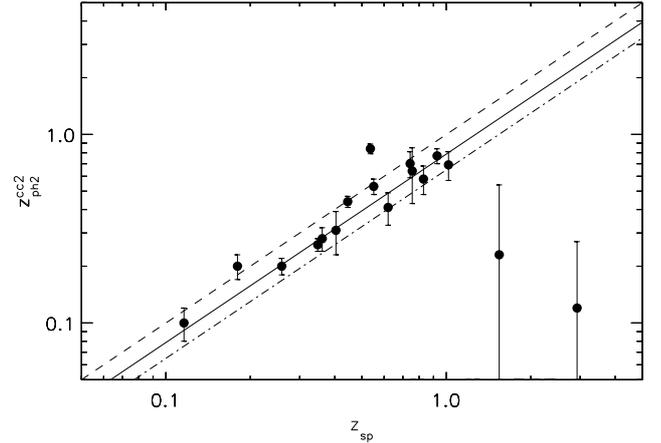}}
\caption{Comparison between photometric z$_{\rm ph2}^{\rm cc2}$ and 
spectroscopic 
redshifts. The two very discrepant large error objects are   \object{0744+464} 
and {\object{1314+453A}}. The central line is for z$_{\rm ph2}=$ 
z$_{\rm sp}/1.23$, (see text)  the two others represent the data dispersion}
\label{z_ph-z_sp}
\end{figure}

In Fig.~\ref{z_ph-z_sp} we show the comparison between z$_{\rm sp}$ 
and z$_{\rm ph2}^{\rm cc2}$. 
Clearly there is a good correlation up to  z$ \approx$ 1, with a systematic 
difference of $\approx$ 23 \%  (z$_{\rm ph2}$ being lower than z$_{\rm sp}$) 
and a scatter of $\approx$ 20 \%.  Additionally  z$_{\rm ph2}^{\rm {cc2}}$
is totally wrong for the two high-redshift 
(z$_{\rm sp} \ga$ 1.5) objects (\object{0744+464} and \object{1314+453A}). 
The case of \object{1159+395}, mentioned just above, is a similar one.
{\bz A similar result is obtained, with a $\approx$ 29 \% offset, using  
z$_{\rm ph2}^{\rm d1}$.}
 {\bz Hence for the newly identified galaxies we give the average of the two 
z$_{\rm ph2}$, each one corrected for its own  offset.}  

The reason for this systematic discrepancy between z$_{\rm ph2}$ and 
z$_{\rm sp}$ and for the wrong photometric values at high redshifts will be 
discussed in Sect. \ref{SEDG} {\bc and is illustrated in Fig~\ref{all_z}}.

\subsection{Hubble diagrams }
\label{Hdia}

\begin{figure*}[t]
{\includegraphics[width=17.5cm]{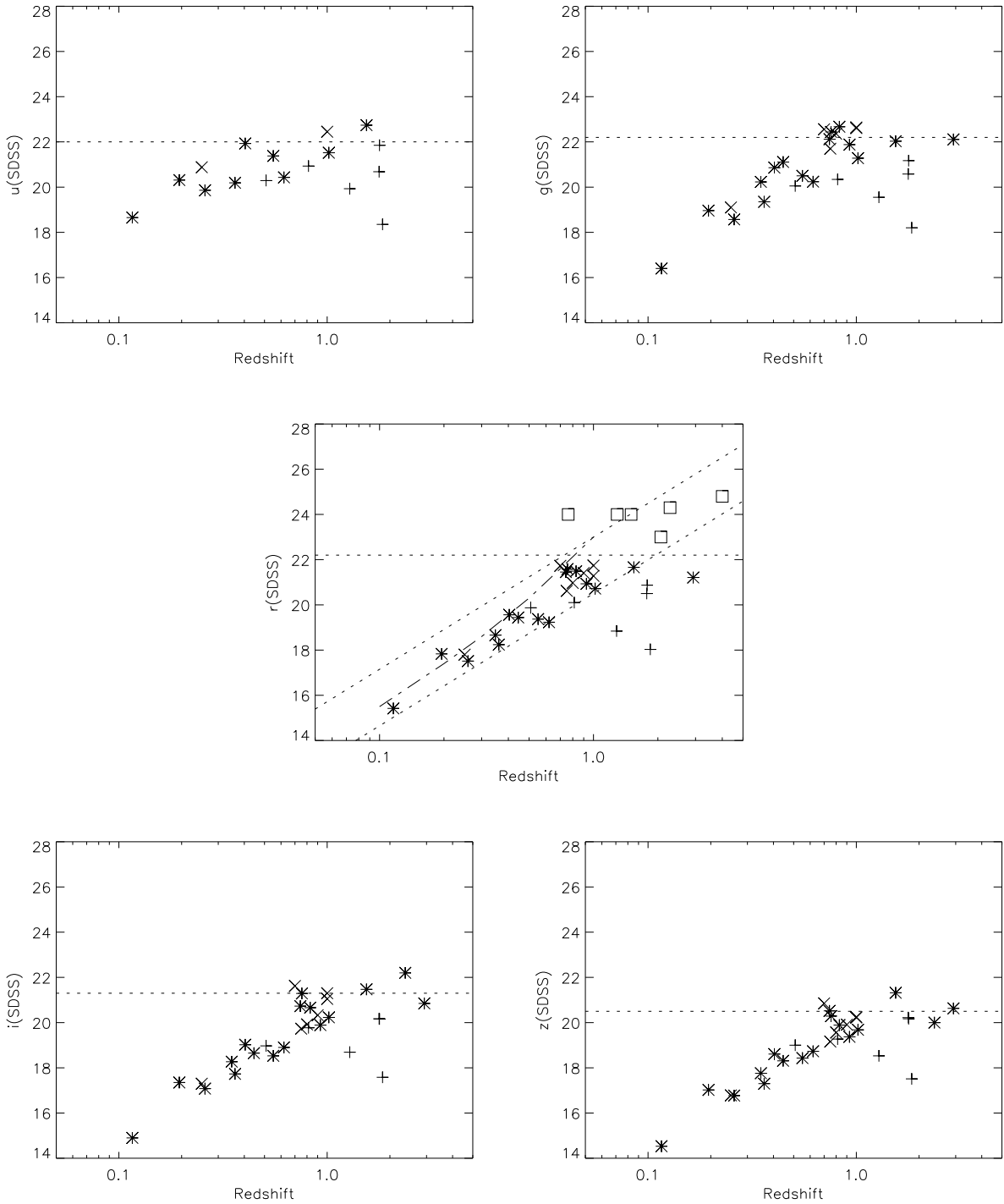}}
\caption{Hubble diagrams in the five SDSS bands for {\bx quasars (+), galaxies
with $z_{sp}$ ($\ast$) and with z$_{\rm ph2}$ (x) only. The squares are the 
faint galaxies undetected in the SDSS.}
The horizontal lines represent the declared magnitude limits of SDSS. 
In the central panel the two parallel lines  are from \citet{Lab07} and the 
curve is the best-fit line for GPSs from \citet{Sn02}. 
} 
\label{H-dia}
\end{figure*}

We built the Hubble diagrams  in the five SDSS bands for both radio 
galaxies and quasars (Fig.~\ref{H-dia}). 
{\bx  Only the magnitudes corresponding to significant ($\ge 3 \sigma$) 
detection were used. 
We applied  corrections for galactic extinction according to \cite{Schl98}.}
In the diagrams we also plotted the {\ba newly identified galaxies with 
photometric redshifts} (z$_{\rm ph2}$ corrected for the systematic difference 
with respect to z$_{\rm sp}$ discussed in Sect. \ref{phrs}). Because we do not
have specific information on emission lines, we cannot make corrections for 
them. Nevertheless we made a statistical estimate of their effect, considering
the relations  between radio power and emission line luminosities 
\citep[see][and references therein]{Lab08b}, typical of [\ion{O}{ii}]3727, 
[\ion{O}{iii}]4959/5007 and H$_\alpha$. 
The expected effect is generally a contribution of some percent, with a maximum
effect for lines well centered in a given band of up to $\approx$ 20 \% in 
flux for the highest radio power sources. Accordingly no major problems are 
expected to occur.

In the $r$ band Hubble diagram, central panel in Fig.~\ref{H-dia}, we also plot
the faint galaxies (squares) that are undetected in the SDSS, using m$_{R}$, 
i.e. the old red magnitudes (Col. 3 of Table \ref{B3tab1}).
We also plot two lines from a recent plot of radio galaxies with z$ \le 1$  
\citep{Lab07}, which includes more than 90\%  of those plotted objects. 
We also show the best-fit line for GPS galaxies by \cite{Sn02}.

In our plot the large majority of galaxies are distributed within the two 
limiting lines by \citet{Lab07} and are slightly 
brighter than  a typical GPS.  We note two galaxies out of those
boundaries. One is the highest spectroscopic redshift 
source (\object{0744+464}), which appears very luminous and well 
out of the extrapolation of the radio galaxies band up to its redshift,
{\bc but it is too weak for a morphological classification. It 
shows strong broad lines  \citep{McCa}, which may significantly contribute to 
the SDSS magnitudes, and is classified  as  {\it broad line radio galaxy} 
(and not as a {\it quasar}) on the basis of details of the Ly$_\alpha$, 
\ion{He}{ii} and \ion{C}{iv} lines}. The other one, \object{1143+456} 
(z = 0.762), which is {\bx not detected in the SDSS}, is at least two
magnitudes fainter than  other objects with a similar redshift.

The objects with corrected z$_{\rm ph2}$ (Sect. \ref{phrs}) fit the 
distribution well (crosses in the figure). 
Had we not introduced the correction factor, they would have been
systematically fainter than the galaxies with z$_{\rm sp}$ in the same range.

Four out of the six quasars with redshift are located below 
the galaxies region (are brighter), as expected, and another 
(\object{1242+410}) is at the 
borderline of the two classes of objects. The last one (\object{0800+472}) 
is  mixed, instead, with  the galaxies (see notes in Sect.\ref{notes}). 

The $g$, $i$, and $z$ band Hubble diagrams reproduce the main 
features of the $r$ diagram. The dispersion of the points is somewhat 
reduced in the $i$ and $z$ bands. In these diagrams six of the galaxies 
with corrected z$_{\rm ph2}$ fit very well. The remaining one 
(\object{1016+443}) is about 1.5 magnitudes too faint in the $i$ and $z$ 
diagrams and, according to its magnitudes, could have a redshift up to 1.2.

We used the $r$, $i$, $z$ Hubble diagrams best fit lines to make an 
estimate of the redshifts for the two galaxies with multi-band photometry for 
which the SDSS routines failed to give a z$_{\rm ph2}$. For \object{0814+441} 
we estimate z$\approx$ 1.2. For \object{1441+409}, whose magnitudes are out of
the ranges covered by our objects, the  extrapolation would give z$ \approx $ 
2.0.

{\bx The $u$ band Hubble diagram is less populated because only 
$\approx$ 50 \%  of the identified objects are detected in this band. The 
magnitudes show a smaller dependence on redshift compared with the other 
Hubble diagrams.}

\subsection {Completeness of the identifications}
\label{IdCompl}
 
\begin{figure}[t]
\resizebox{\hsize}{!}{\includegraphics{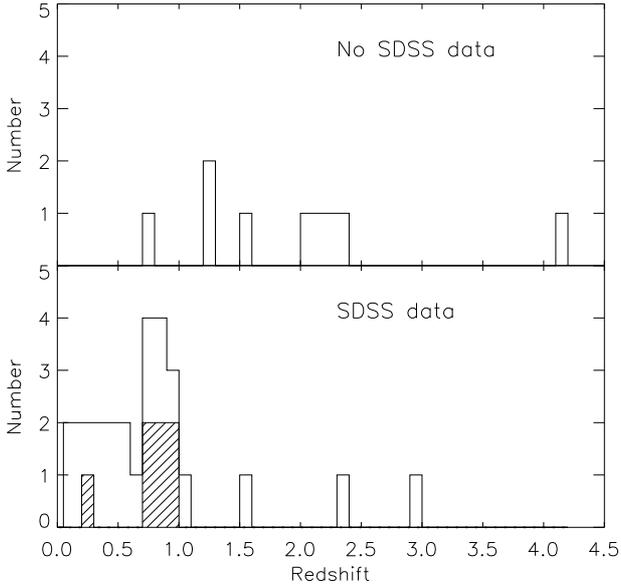}}
\caption{Distribution of redshifts for galaxies found in (below) and missed by
(top) the SDSS. 
The hatched bins represent the (corrected) photometric redshifts. }
 \label{z-distr}
\end{figure}

The Hubble diagrams show that the identified galaxies with redshift (either 
spectroscopic or photometric) appear to 
end at z$\approx 1$, and this effect seems essentially due to 
the SDSS magnitude limits. This led us  to assume that the identifications are
reasonably complete up to that redshift.
In order to reinforce this conclusion, we show in Fig.~\ref{z-distr} 
the redshift distribution for the detected (bottom) and 
undetected galaxies (top) by the SDSS. Only one of the SDSS non detections
(\object{1143+456}, already mentioned in Sect. \ref {Hdia}), is at z$<1$.

It is likely that the remaining E sources have the same redshift distribution
of the set of galaxies above z$_{\rm sp}\ge 1.2$.

\section{UV-optical spectral energy distributions of CSOs/MSOs}
\label{SED}

{\bx We supplemented the B3-VLA CSS sample with a set of  $CSO$s/$MSO$s with 
SDSS photometry, both quasars and galaxies (11 and 12 respectively), taken 
from the 3C and PW catalogs \citep{Fanti95}, from \cite{Lab07} and 9C catalog 
\citep{Insk06b}.
All these sources are reported in Table~\ref{SDSS_sample_2}. The samples are 
separated by horizontal lines and are, in the order from top to bottom: B3-VLA,
3C, PW, \citet{Lab07} and 9C.}

We transformed the SDSS magnitudes into flux densities 
following the prescriptions given in the SDSS site.  
The SDSS photometry is given in terms of {\it asinh magnitudes}  
\citep{Lupton}, defined as

$m = - (2.5/\ln 10)~ \Big [{\rm asinh}~ \left((f/f_{0})/2 b\right) + \ln(b)\Big] $,

\noindent
where $f$ is the flux density of the object and $f_{0}$ = 3631 Jy is the
zero point of the AB scale, to which the SDSS photometry is calibrated.
The {\it asinh magnitudes} are more  appropriate for 
faint objects and become identical to the traditional magnitudes at high 
signal-to-noise ratios. 
The parameter $b$, determining the flux level over which the {\it asinh 
magnitudes} are similar to the traditional magnitudes is tabulated, e.g.,
in \cite{R7}. 
The flux densities in the various SDSS bands were derived from the 
magnitudes by inverting the above formula.

{\bx In order to extend the spectral coverage at wavelengths shorter
and longer than those of the SDSS we
made a systematic search in the GALEX survey catalogs (DR 4/5 and 6;  
FUV band $\lambda_{\rm{eff}} \approx$ 1539 \AA, 
NUV band $\lambda_{\rm {eff}} \approx$ 2316 \AA) and in the
near-IR (J, H, K bands) 2MASS (point sources survey for quasars, extended 
sources survey for galaxies) and UKIRT literature data.
The search was made within a circle of 4 arcsec in radius around each radio 
position, which is adequate for the positional accuracy of the UV and near-IR 
objects. The positions of the objects found were checked with those of the 
SDSS and were found to be consistent with them.

All magnitudes were converted into flux densities with 
the appropriate conversion formul\ae. 

Briefly we comment that

a)  the coverage of the GALEX surveys is incomplete and is dependent on the sky
region. For our objects the best coverage, $\approx$ 80 \%, is obtained with 
the {\it All Sky Survey (AIS)}, which is the less deep of the GALEX surveys we
searched for. 
Deepest surveys, as {\it MIS, NGS, GI} have a lower sky coverage.
The majority of the detections are from those deep surveys.
All observed quasars were detected in one or both bands, while the detection 
fraction for galaxies is $\approx$ 42\%, and the incompleteness is more
pronounced for redshifts $\ga$ 0.6. 
Instead all but one of the objects  undetected in the SDSS are also undetected 
by GALEX.
We derived luminosity upper limits for the undetected objects, but they 
were too high to be useful.
  
b) the 2MASS extended objects survey gives {\it total magnitudes}. The UKIRT 
magnitudes are at fixed apertures, and we always took those with the largest 
ones, but we suspect that in a few cases they may not have been 
wide enough. When both types of magnitudes exist for the same galaxy we chose
the 2MASS.
In total we have near-IR  data for $\approx$ 30 \% of our objects. 
Dividing these into quasars and galaxies, the percentages become $\approx$ 60
\% and 25 \%. None of the SDSS undetected objects was found.  

The GALEX and near-IR data that we found are reported in 
Table~\ref{GalexIR_Tab}.}
 
{\bx After correction for galactic extinction  \citep{Schl98}}, luminosities at
the ``source-frame'' (s.f.) wavelength,  
$\lambda_{\rm s.f.} = \lambda_{\rm obs}/$(1+z), were computed 
according to the Concordance Cosmology parameters (H$_0$ = 73 km s$^{-1}$ 
Mpc$^{-1}$, $\Omega_m =$0.27, $\Omega_{\Lambda} = 0.73$), after correction 
for the (1+z) factor of the k-correction,  and are expressed
in units of $10^7$~L$_{\sun}$~\AA$^{-1}$.

\begin{figure}[t]
\resizebox{\hsize}{!}{\includegraphics{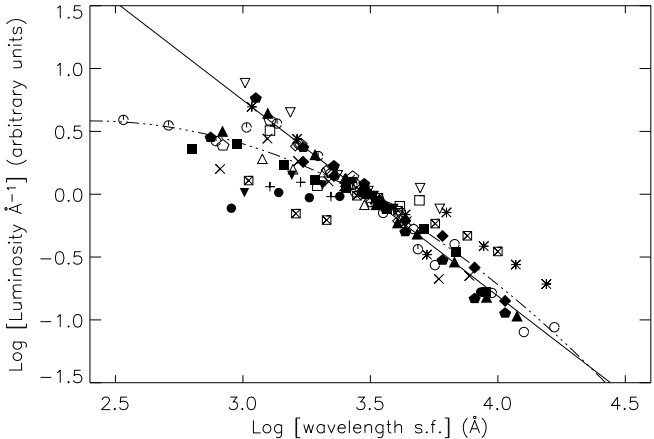}}
\caption{Composite plot of the UV-O-SED of quasars in the source-frame 
{\ba arbitrarily normalized to each other at $\lambda\approx$ 3000 \AA}. 
The curved line is the best fit of the data in the range 1500 - 5000~\AA.
The straight line is from \citet{VanB01} (see text).}
\label{QS-Sed}
\end{figure}

\subsection{Quasars }
\label{SEDQ}

{\bx Overall we have six quasars from the B3-VLA sample and 11 from the 
literature samples.

The UV-O-SEDs (not shown)  are generally rising at shorter wavelengths and can
be well fitted in a minority of cases by a single power law, 
L($\lambda) \propto \lambda^{-\alpha_\lambda}$, or, more commonly, by
two power laws matching at $\approx$ 3000 - 3500 \AA,  the one at the
shorter wavelengths being the flatter.\footnote{{\bx The UV-O-SEDs may be 
affected by emission lines, because we do not have enough information to 
subtract them.}} 
It is also worth noting that the four quasars with data points at $\lambda 
\la$ 1000~\AA~ (B3 0701+392, B3 0805+406, 3C186 and 1442+101) show a sharp 
drop in  luminosity below that wavelength.}
The UV-O-SEDs of the two  quasar candidates without  redshift 
(\object{1055+404A} and \object{1340+439}) have an indeterminate shape.

Table~\ref{QssTab} gives the parameters of the individual UV-O-SEDs. 
 
{\bx The $\alpha_{\lambda}$ coefficients above $\approx$ 3000 \AA\ have a mean 
value $1.44 \pm 0.14$, in agreement with  the composite quasar spectrum  by 
\citet{VanB01} ($\alpha_{\lambda}$ = 1.56 in the range $\approx$ 1500 - 6000 
\AA), obtained from a homogeneous data set of over 2200 spectra from the SDSS,
and derived from two emission-line free widely spaced spectral regions. On the
contrary, at $\lambda$s $\la$ 3000 \AA, $\approx$ 40 \% of our quasars have a 
significantly flatter spectrum.
These findings are emphasized by the composite plot of the 17 UV-O-SEDs 
(Fig.~\ref{QS-Sed}).
The power law fit of the composite quasar spectrum by \citet{VanB01} is 
with a few exceptions a good representation of our data for  $\lambda \ga$ 
2600 \AA.  At shorter wavelengths there are many more significant 
discrepancies because of the individual UV-O-SEDs curvature.}

Besides effects caused by contamination from emission lines, we mention
among the possible explanations for the above discrepancies: 

{\bx 
i) at $\lambda \la$ 2600~\AA~:  internal dust reddening has been suggested by
\citet{Baker} from an analysis of the optical spectra of CSS quasars from the 
Molonglo Quasar Sample. They  quote a spectral index $\alpha_{\nu}$ = 1.5 
(f($\nu$) $\propto \nu^{-\alpha_{\nu}}$), steeper than in other quasar classes,
which is well consistent with the $\alpha_{\lambda}$ values we obtained in the 
short wavelengths range. Their suggestion of internal reddening is also
supported by the relatively prominent Balmer decrement ($A_V \approx 4$) they 
find in these objects. Nevertheless, they also mention some contradictions with
such a high extinction. 
We tried to check the consistency of the curved shape
of our composite spectrum with the hypothesis of dust reddening, using the 
extinction models by \cite{Cardelli}.
We find an acceptable consistency for 0.05 $\la$ A$_V \la 0.15$ and 1.0 $\la$ 
R$_V \la$ 2.0.
To our knowledge, however, these R$_V$ values seem too low. An alternative
possibility is that the curvature of the spectrum is intrinsic.

At $\lambda$s $\la$ 1000 \AA ~the Lyman $\alpha$ forest is responsible for the
fast drop, as is well shown, e.g., by the SDSS spectrum of 1442+101 available 
from the SDSS site}

{\bx ii) at $\lambda \ga$ 4000~\AA~:  the luminosity contribution of the host 
galaxy, in case of an under-luminous nuclear emission. B3 0800+472, 
B3 1242+410 (see Sect.~\ref{Hdia} and notes in Sect. \ref{notes}) and 1153+32 
are  possible cases.}
 
The median logarithmic luminosity at 3000 \AA~ is $3.3 \times 10^8~L_{\sun}$
\AA$^{-1}$, or $\approx 8 \times 10^{45}$ erg~s$^{-1}$ integrated in the 
optical-UV band, with a dispersion of a factor $\approx$ 6.

\subsection{Galaxies }
\label{SEDG}

{\bx Overall we have 25 galaxies from the B3-VLA sample with redshift 
(z$_{sp}$ or z$_{ph2}$) and another 12 from the literature samples.

The individual UV-O-SEDs in the source-frame are shown in 
Figs.~\ref{plot_sp}.1, \ref{plot_ph}.2  (B3-VLA galaxies with spectroscopic 
and photometric redshift respectively) and Fig. \ref{plot_ph}.3 (literature 
samples). They also include data from  GALEX NUV and FUV bands (14 objects) 
and from 2MASS and UKIRT (9 objects).}
To facilitate the comparison, we overimposed to each UV-O-SED an 
elliptical galaxy model of 13 Gyr by \cite{Bruz03}, hereafter B\&C, normalized
at the longest SDSS wavelengths.
This model describes the UV-O-SED of the present epoch elliptical 
galaxies\label{here}.\footnote{Indeed, because our galaxies cover a range of 
redshifts, each one ought to be compared with the model of age appropriate 
to its epoch. However,
the differences in the explored redshift interval are minor ones and 
unimportant 
in this context. They will be examined in more detail later on.}

\begin{figure}[t]
\resizebox{\hsize}{!}{\includegraphics{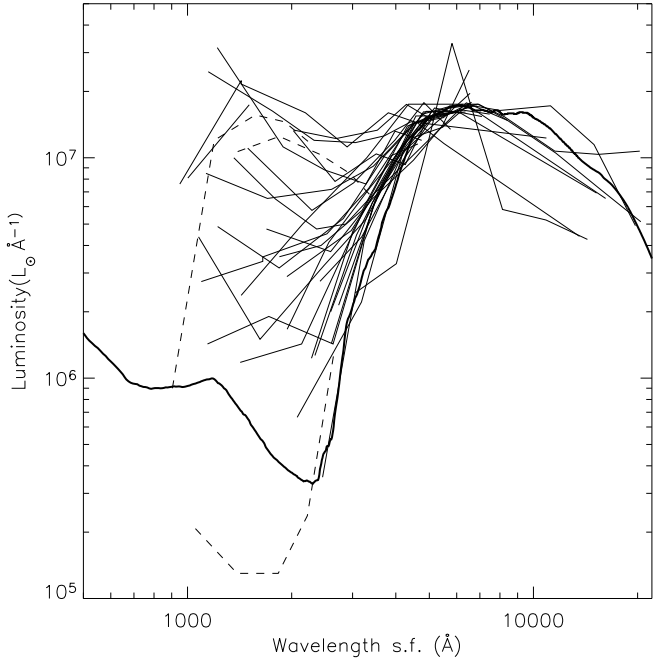}}
\caption{Composite plot of the SEDs in the source-frame for the radio 
galaxies with spectroscopic redshift.   The thick line is the B\&C model of an
elliptical galaxy of 13 Gyr and 10$^{12}$M$_\odot$ (see footnote on page 
\pageref{here}). The individual UV-O-SEDs are normalized to this model. 
The broken lines indicate the sources \object{0744+464}, \object{1159+395}, 
\object{1314+453A}, and \object{3C237}, which do not have data-points at 
$\lambda > 4000$~\AA~ and whose UV-O-SED normalization is therefore quite 
arbitrary.}
\label{RG-Sed}
\end{figure}

Depending on the galaxy redshift, the five SDSS bands cover different 
rest-frame wavelength ranges; three galaxies at z$ \ga$~1.5 (\object{B3 
0744+464}, \object{B3 1159+395}, and \object{B3 1314+453A} have their entire 
UV-O-SED at $\lambda \le 4000$ \AA, and  in these cases the normalization is 
rather arbitrary. 

The majority of the sources shows a decrease in luminosity at $\lambda \la$ 
4000 \AA, as in the elliptical galaxy model. {\bx But this decrease is 
definitely less pronounced than in the model, and in some cases it is followed
by a turn-up at shorter wavelengths 
(see e.g.: \object{B3 0809+404}, \object{B3 1049+384}, 
\object{B3 1241+411}, \object{3C237}, \object{3C241}). The UV-O-SED of 
\object{1314+453A}  is rising monotonically and can be fitted by a power law 
as for quasars, with $\alpha_\lambda=1.2$. Nevertheless, because the optical 
object is {\ba clearly extended (see notes in Sect. \ref{notes}), we keep 
the ``galaxy'' classification.}

In order to better emphasize the above  findings,} we show 
in Fig.~\ref{RG-Sed}  a rest-frame composite plot of the UV-O-SEDs of
galaxies with z$_{\rm sp}$, each one scaled in luminosity, at  wavelengths 
$\lambda \ge 4500$ \AA, to the 13 Gyr age, $10^{12}$ M$_{\sun}$  stellar mass,
elliptical galaxy model by B\&C. 

Clearly in the large majority of cases there is an excess in luminosity at 
$\lambda \la 4000$~\AA, (referred to as {\it UV-excess}), {\bx which becomes 
more prominent when the GALEX data are added}.
Therefore, in addition to the old stellar component, another source of light 
(whose luminosity we indicate by L$_{\rm UV}$) is required, which is 
responsible for most of the radiation at wavelengths shorter than 4000 \AA. It
is likely that L$_{\rm UV}$ does also contribute to some extent to the 
UV-O-SED at $\lambda \ge 4000$ \AA.

In order to properly visualize the {\it UV-excess}, we remark that  one should 
have normalized the UV-O-SEDS, in Fig.~\ref{RG-Sed} as well as in 
Figs.~\ref{RG_Seds}.1, \ref{RG_Seds}.2 and \ref{RG_Seds}.3
at  wavelengths much longer than 4500 \AA, where the 
contribution of the $new$ component would be small or negligible. Our 
normalization instead necessarily leads to some underestimate of the 
{\it UV-excess}. 
Near-IR data (e.g. J, H, K band), which are more representative of the 
UV-O-SED of an elliptical galaxy, would be very useful. {\bx However, as 
already mentioned, only nine galaxies of this combined sample
have some near-IR data. For them our normalization 
produces an underestimate of
the {\it UV-excess} from a few \% up to $\ga$ 60\% compared with  
that based on the near-IR data.}

Anyhow, in spite of the normalization uncertainties, the {\it UV-excess} 
appears quite clear:  almost all the UV-O-SEDs lie above the B\&C model
at $\lambda \la 4000$~\AA.

{\bx  It is worth commenting on which information GALEX adds to the SDSS. Eight
objects with  GALEX data show a rise of the UV-O-SED in the NUV or FUV  band, 
which indicates a second peak at $\lambda \la 1500$ \AA, with a luminosity 
comparable to that at $\lambda$s $\ga 4000$ \AA.
The other six instead show a decreasing or constant luminosity in the GALEX 
bands, which is still compatible with a peak in the far-UV, but of much lower 
luminosity.
}

\subsubsection{The {\it UV-excess}}
\label{UV-exc}

\begin{figure*}[t]
\centering
\resizebox{\hsize}{!}{\includegraphics{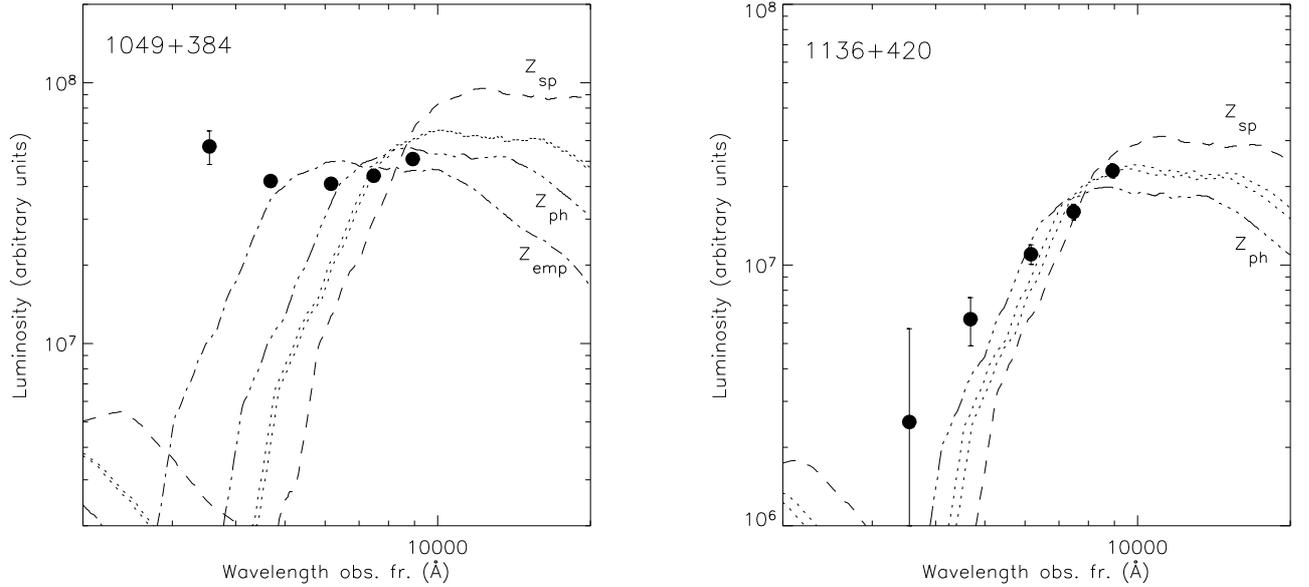}}
\caption{ Fit  of the B\&C old stellar population galaxy model to the sources
\object{1049+384} (left) and \object{1136+420} (right), for different 
redshifts. 
Two of the lines are marked with z$_{\rm sp}$ and z$_{\rm ph}$; the  two 
intermediate lines represent the two z$_{\rm ph2}$. 
In \object{1049+384} the dash-dotted line (marked with z$_{\rm emp}$, for 
empirical) gives the best $\chi^2$ of all and corresponds to z$=0.05$. 
In \object{1136+420} a blind fit would provide z$\sim 0.45$ i.e. identical 
to z$_{\rm ph}$ and it is not plotted.}
\label{all_z}
\end{figure*}

{\bx We quantify  the {\it UV-excess} by means of the source-frame luminosity,
L$_{0.30}$, at 3000 \AA~ (0.30 $\mu$m), where the excess over the elliptical 
galaxy model is mostly evident,  compared with  the luminosity, L$_{0.45}$, at
4500 \AA\ (0.45 $\mu$m), and their ratio, R$^{0.30}_{0.45}$}.
{\ba For a standard elliptical galaxy 13~Gyr old at the present 
time (z$=0$) that has experienced passive evolution {\ba (Sect. \ref{Or_UV}}),
R$^{0.30}_{0.45}$ is $\approx$ 0.16  and increases slowly
up to $\approx$ 0.23 at z = 1.5 (Eq. \ref{k}, page \pageref{page_k})}.

The choice of these two wavelengths is a necessary compromise because of the 
source-frame wavelength range available to us. {\bx L$_{0.30}$ is known for all
the sources at z $\la$ 1.0, including those with photometric redshifts only, 
while luminosities at $\lambda$s $<$ 3000~\AA ~ are available only for a 
fraction of the objects owing to the GALEX detection incompleteness. For 
instance L$_{0.20}$, at 2000~\AA, is known for $\approx$ 70 \% of 
the sample, with higher incompleteness at higher redshifts.
On the other side,} without near-IR data, for a few sources with z$ > 1.0$ 
we do not have  L$_{0.45}$ and R$^{0.30}_{0.45}$, because their SDSS 
wavelengths are all $\le$~4500~\AA ~ in the source-frame.

{\bx L$_{0.30}$, L$_{0.45}$ and their ratios are reported in 
Table~\ref{SDSS_sample_2}.}

The {\it UV-excess} allows us to understand the
systematic differences between photometric and spectroscopic redshifts.

{\bc In Fig.~\ref{all_z} we show two sample galaxies in which the SDSS data 
points are compared {\ba with a B\&C elliptical galaxy model} shifted to the 
observed wavelength by using z$_{\rm sp}$, z$_{\rm ph2}^{\rm cc2}$, 
z$_{\rm ph2}^{\rm d1}$ and z$_{\rm ph}$. 
{\ba The galaxy {\bz stellar} mass is the best-fit free parameter}.
Each curve is the one which gives the best $\chi^2$ for the corresponding 
redshift.  The fit agreement improves as we move from z$_{\rm sp}$ to z$_{\rm 
ph}$. In addition we show in the plot of \object{1049+384} the curve 
(dash-dotted) obtained by blindly fitting the data with the B\&C  model. It 
provides the absolutely best $\chi^2$ and would correspond to a redshift of
z$=0.05$.} We note that the photometric redshifts, independent of the method 
used to derive them, give a better description of the UV-O-SEDs in terms of a 
standard  elliptical galaxy and therefore, in order to account for at least a 
part of the {\it UV-excess}, the fits require a redshift lower than the true 
one.

\subsubsection{ Dependencies of the {\it UV-excess} on source parameters}
\label{UV-dep}

{\bx
Before examining the behavior of L$_{0.30}$, L$_{0.45}$ and  
R$^{0.30}_{0.45}$, against  (1+z), P$_{1.4}$ and LS, the parameters to which
the {\it UV-excess} could be more likely related, we mention some 
{\it caveats} about our composite sample.

i) Correlations of any quantity with redshift and radio power may suffer for 
the {\it degeneracy problem} caused by the observationally induced correlation
between redshift and radio power. 
This is particularly severe for flux-limited samples.}   

ii) More generally, our composite sample (B3-VLA plus literature objects) 
does not uniformly cover the 3D-space {\it radio power - size - redshift}, 
so that, in addition to the {\it degeneracy problem}, some 
apparent effects could be artifacts caused by  the inadequate coverage. 

iii) The statistics in our sample is limited. 

\begin{figure}[t]
\resizebox{\hsize}{!}{\includegraphics{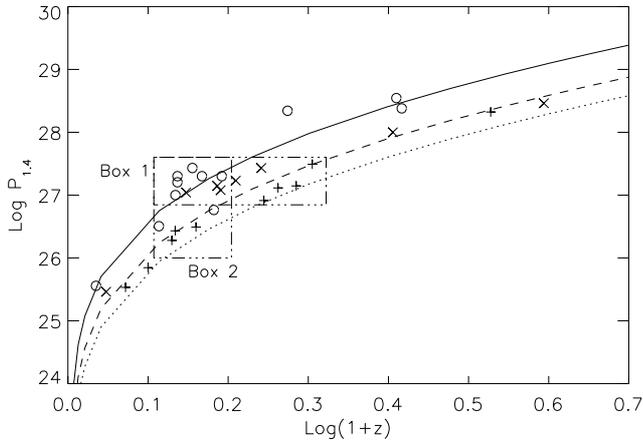}}
\caption{ Distribution of Log(radio power) {\it vs}  Log(1+z) for the 
galaxies  of the composite sample. Empty circles represent 3CR\&PW sources; 
crosses and plus signs are the two B3-VLA sub-samples. The lines correspond to
flux densities $S_{1.4}$ = 2.1 Jy, 0.66 Jy and 0.33 Jy, representing the flux 
density limits of the three samples extrapolated at 1.4 GHz.}
\label{P_z}
\end{figure}

{\bx Concerning the first caveat, we point out that our composite sample does 
not have  one single flux density limit,  but it is built with three sets of 
sources (two sub-sets of the B3-VLA catalog, Sect. \ref{B3CSSsample}, and 
3CR+PW and Labiano lists) that have different flux density limits ($\approx$ 
0.33. 0.66 and 2.1 Jy respectively when extrapolated  at 1.4 GHz). 
Their combination allows us to remove the degeneracy in some ranges of radio 
power and redshift. 
Indeed we can see in Fig.~\ref{P_z} that in the P$_{1.4}-$(1+z) plane there is
an area ({\it Box~1}) where 15 objects with approximately the same values of 
$P_{1.4}$ ($7\times 10^{26} \le$ P$_{1.4}$(W~Hz$^{-1}$)$~\le 4\times 10^{27}$)
cover the redshift range from $\approx 0.3$ to $\approx 1$.}

\begin{figure*}[t]
{\includegraphics[width=17.5cm]{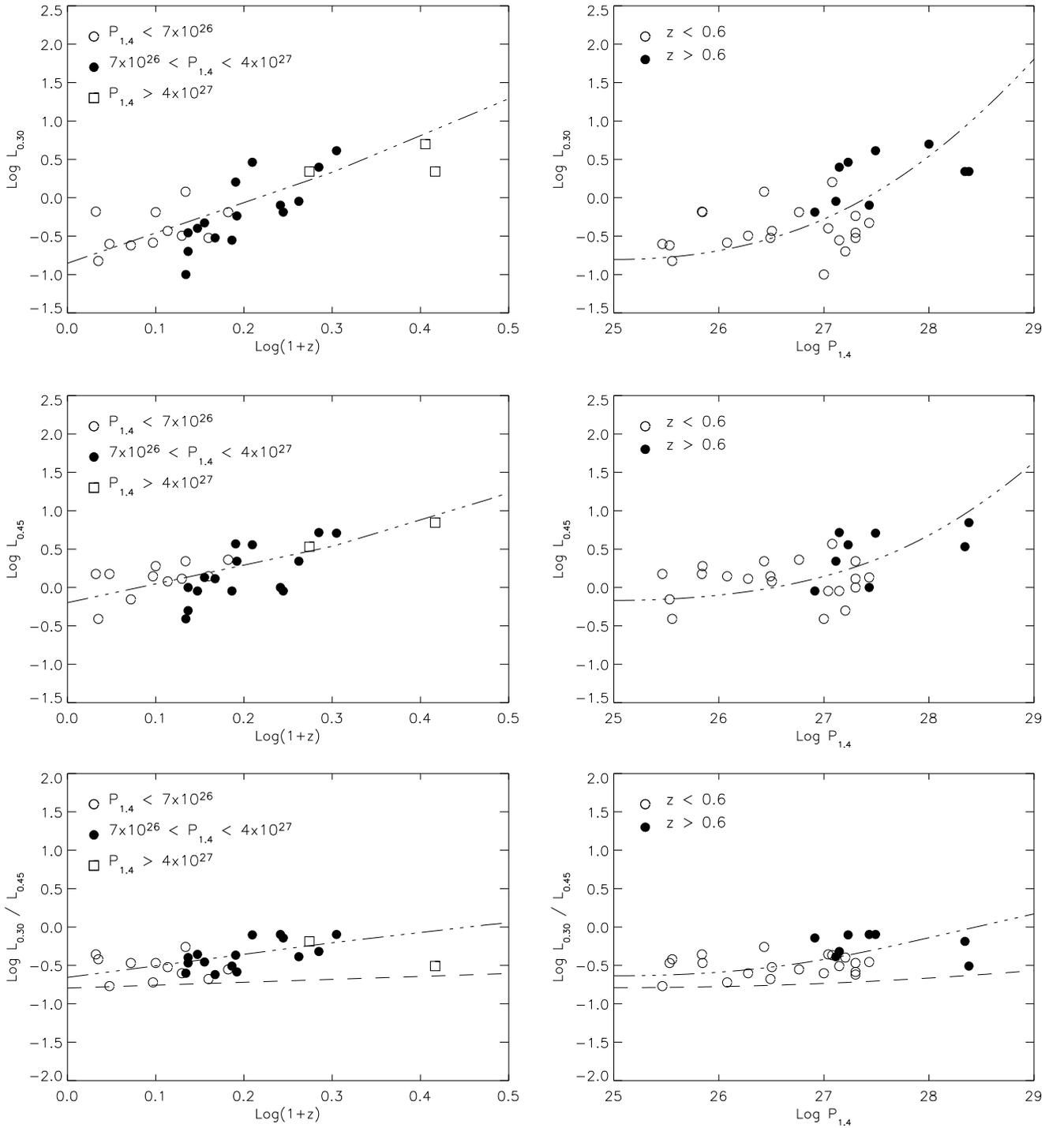}}
\caption{Correlations between Log(L$_{0.30}$), Log(L$_{0.45}$) 
and Log(R$^{0.30}_{0.45}$) 
with Log(1+z) (left) and Log(P$_{1.4}$) (right) for CSOs/MSOs. The curved 
lines in the left-hand plots represent the fits to the data obtained using Eq.
\ref{L_lam} (page \pageref{eq2}). 
The curves in the right-hand plots show how the relations with (1+z) reflect
on P$_{1.4}$  because of the P$_{1.4}$--(1+z) dependence (Fig. \ref{P_z}), for
a flux density limit S$_{1.4}$=1.5 Jy.
The almost horizontal dotted lines in the two bottom panels represent the 
expected value of  Log(R$^{0.30}_{0.45}$) for an old elliptical galaxy.}

\label{UVexc}
\end{figure*}

A second area ({\it Box~2},~~  $0.11 \le \log(1+z) \le 0.2$ and 
P$_{1.4} \ge 10^{26}$ W~Hz$^{-1}$) contains 14 objects with roughly the same 
 range of (1+z), which cover a radio power range of a factor $\approx$ 30. 
In these two boxes we may break the degeneracy between the two parameters.
   
{\it Revenon a nos moutons} \footnote{``La Farce de Maitre Pathelin''
(XVe siecle), author unknown.} 

\begin{figure*}
{\includegraphics[width=17.5cm]{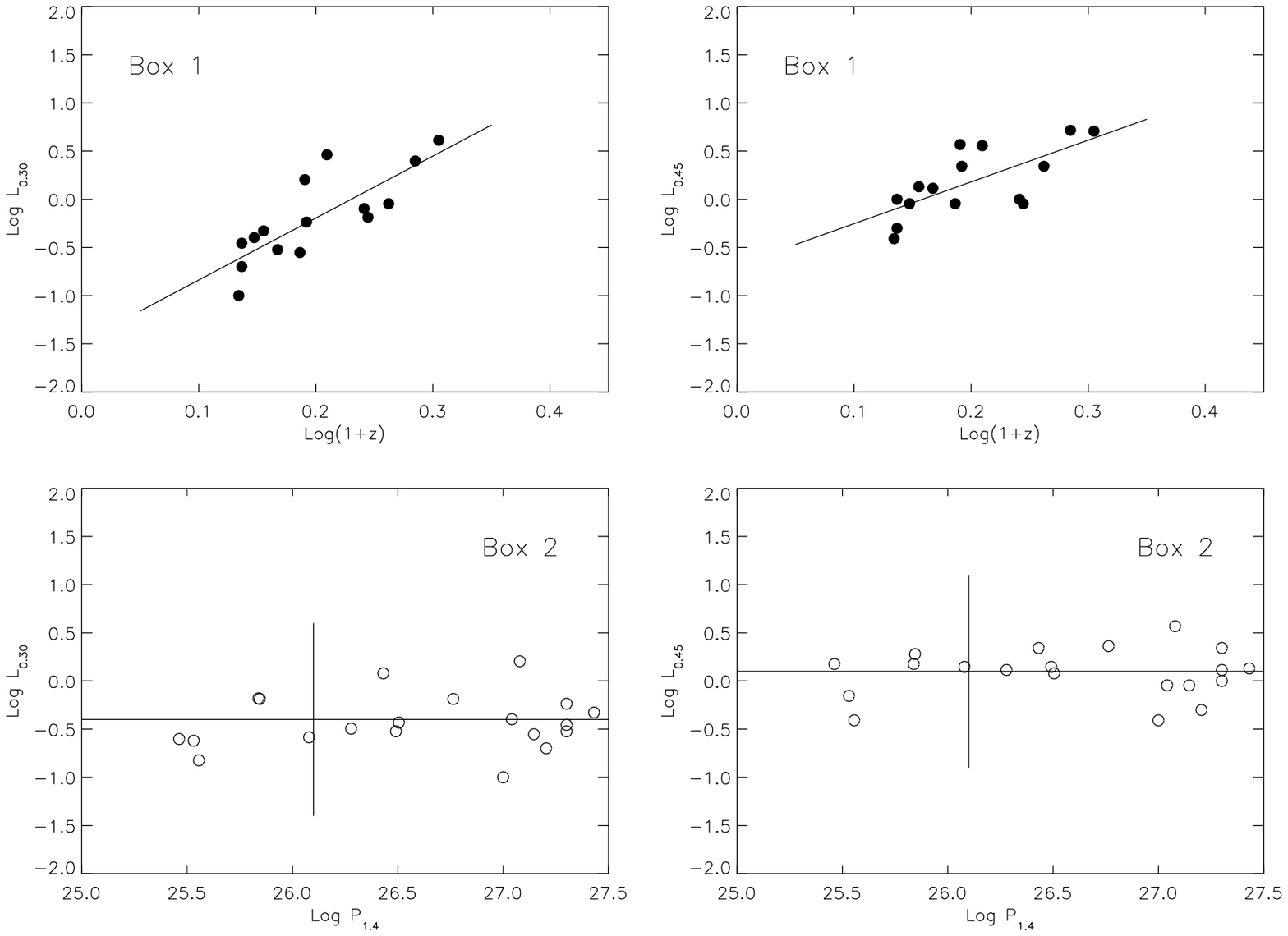}}
\caption{Top panels: enlargement of left hand panels of Fig.~\ref{UVexc} for 
the sources in {\it Box~1}. 
Bottom panels: enlargement of right hand panels of Fig. \ref{UVexc} 
for the sources in {\it Box~2} (to the right of the vertical line), and for 
the sources with z $\le$ 0.3 (to the left of the vertical line). 
The lines through the plotted data are  the best-fits. }
\label{UVexc_red}
\end{figure*}

\begin{figure*}[t]
\centering
{\includegraphics[width=17cm]{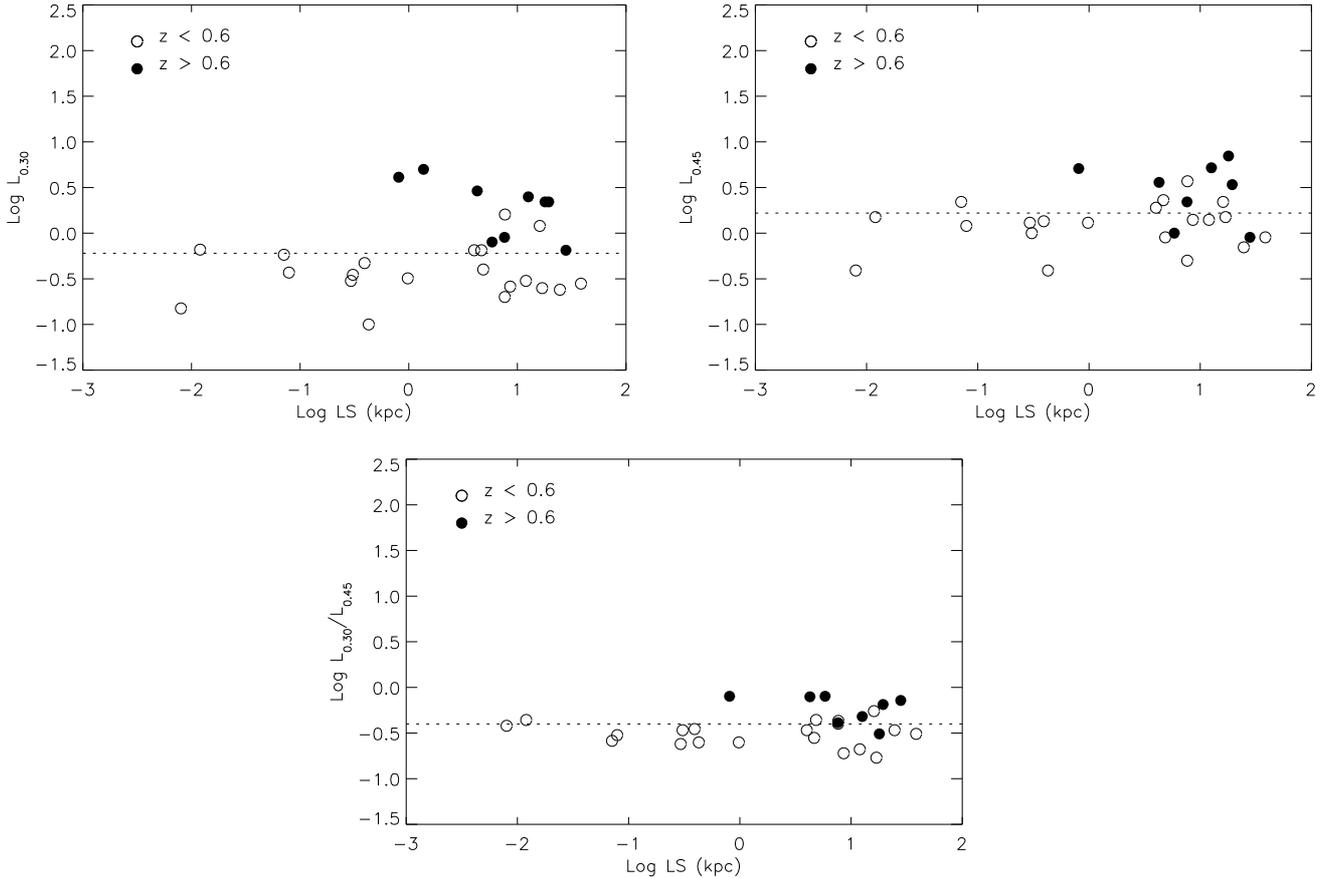}}
\caption{Same as in Fig.~\ref{UVexc}, but {\it vs} Log(LS). The horizontal 
lines  approximately separate  objects with z$<0.6$ and  z$>0.6$. }
\label{UVexc_LS}
\end{figure*}

We  plot in Figs. \ref{UVexc}, \ref{UVexc_red}, and \ref{UVexc_LS} L$_{0.30}$,
L$_{0.45}$, R$_{0.45}^{0.30}$ {\it vs} (1+z), P$_{1.4}$ and LS.

{\bx
We recall that some data-points are missing, at high (1+z), in the plots of 
L$_{0.45}$ and of R$^{0.30}_{0.45}$, because the rest-frame UV-O-SEDs 
at z $> 1.5$ are shifted to such short wavelengths that
without IR observations it is impossible to determine L$_{0.45}$.
In the plots {\it vs} (1+z) (Fig.~\ref{UVexc}, left) the data-points
are grouped into three classes of radio power  (in W~Hz$^{-1}$ at 1.4 GHz): 
P$_{1.4} \le 7\times 10^{26}$ (11 sources), 
$7\times 10^{26} < $P$_{1.4} < 4\times 10^{27}$ (15 sources)  and P$_{1.4} > 
4\times 10^{27}$ (6 sources) 
and are plotted with different symbols; in the plots against P$_{1.4}$ and LS
(Fig.~\ref{UVexc}, right and Fig.~\ref{UVexc_LS}) they are instead grouped 
into two classes of redshift, z$<0.6$ and z$>0.6$  
and are again plotted with different symbols.}

In spite of the data point dispersion some trends are visible in 
Fig.~\ref{UVexc}: 

{\it a)} both L$_{0.30}$ and L$_{0.45}$  
significantly increase with redshift, the L$_{0.30}$ {\it vs} (1+z) dependence 
being the steeper. As a consequence also R$^{0.30}_{0.45}$ increases 
with (1+z), but less steeply. {\bc The scatter of the data around the median 
is somewhat large, $\approx$ 0.3 in Log, for L$_{0.30}$ 
and L$_{0.45}$, but definitely smaller for R$^{0.30}_{0.45}$, 
indicating that the scatter  of L$_{0.30}$ and  L$_{0.45}$ is intrinsic 
and not caused only by errors.}

{\it b}) Correlations are found also with P$_{1.4}$.

 {\it c})  Sub-kpc sources have smaller L$_{0.30}$, L$_{0.45}$ 
and R$^{0.30}_{0.45}$, compared with the larger ones (Fig.~\ref{UVexc_LS}).

The findings at points {\it a)} and {\it  b)} are not independent because of
the P$_{1.4}-$(1+z) degeneracy. 
However, if we consider in Fig.~\ref{UVexc} the objects of    
{\it Box~1}  ($7\times 10^{26}<$P$_{1.4}$(W~Hz$^{-1})<4\times 10^{27}$, filled
symbols, amplified in Fig.\ref{UVexc_red}, top), we see that the dependencies 
of L$_{0.30}$, L$_{0.45}$ and  also R$^{0.30}_{0.45}$ (not shown) on (1+z), 
for $ 0.3 \le z \le 1$, are clearly visible. 
They cannot be induced  by a P$_{1.4}$ dependence, as P$_{1.4}$ does not vary 
appreciably in the explored redshift interval. {\bx The  galaxies with 
photometric redshift only, not plotted, also very nicely fit the redshift 
dependence}.

Conversely, looking at the objects of {\it Box~2} ($0.3 \le z \le 0.6,
P_{1.4} \ge 10^{26} W Hz^{-1})$, 
we do not find (Fig.\ref{UVexc_red}, bottom) any significant dependence
of either L$_{0.30}$ and L$_{0.45}$ on P$_{1.4}$. 
{\ba The same constancy of  L$_{0.30}$ 
and L$_{0.45}$ with P$_{1.4}$ extends also to sources with z $\le$ 0.3 and
P$_{1.4} \le 10^{26} W Hz^{-1}$, as shown in Fig.~\ref{UVexc_red} (left of
the vertical line).}

So the only clear dependence seen in the data is  with (1+z). This 
dependence appears to change across the explored redshift interval. 
For objects in the intermediate 
radio power  and redshift range ({\it Box~1}) it is stronger than  the 
average dependence over the whole redshift range 
\Big(L$_{0.30} \propto$ (1+z)$^{5.8\pm 1.2}$ and L$_{0.45} \propto$ 
(1+z)$^{3.6\pm 1.1}$ against L$_{0.30} \propto$ (1+z)$^{4.5\pm 0.4}$ and 
L$_{0.45} \propto $(1+z)$^{2.4\pm 0.5}$\Big).
At low redshifts (z$ \la 0.3$) there might be no dependence at all 
on (1+z) for  both luminosities, which is consistent with \citet{Lab08}.

In order to give a simple physical interpretation of these behaviors, we  
assume that  L$_{0.30}(z)$ and L$_{0.45}(z)$ are the combination of a
term, L$_{\rm UV}(\lambda, \rm {1+z})$, responsible for the {\it UV-excess} 
(whose nature we ignore at this point, but we shall discuss it in Sect. 
\ref{Or_UV}) and a second one from the underlying old elliptical galaxy.   
We assume the simplest possible model, i.e. a power law  for 
L$_{\rm UV}$($\lambda,1+z$) and an elliptical galaxy model of 13 Gyr  age 
at the present epoch (B\&C) undergoing passive evolution 
(Sect.~\ref{Or_UV}). 
 A galaxy {\bz stellar} mass of  $4 \times 10^{11}$ M$_\odot$ is taken 
(Sect. \ref{YSP}) and its contribution to the luminosity, 
$k_{\lambda}$, at the two considered wavelengths, 
can be approximated from the B\&C models\label{page_k}

\begin{equation}
k_{0.30} \approx 0.08 \times \rm{(1+z)}^{1.58} ~~~~ 
k_{0.45} \approx 0.5  \times \rm{(1+z)}^{1.2}  
\label{k}
\end{equation}

\noindent
($k_\lambda$ in units of $10^7 {\rm L}_\odot {\rm \AA}^{-1}$).
 Therefore

\begin{equation}
L_{\lambda}= L_{\rm UV}[\lambda,\rm{(1+z)}] + k_{\lambda} 
           = B_{\lambda}\times \left(\frac{\rm{1+z}}{1.6}\right)^{a_{\lambda}} 
              + k_{\lambda},
\label{L_lam}
\end{equation}
\label{eq2}
\noindent
where the  power law L$_{\rm UV}$ has been normalized at z = 0.6, which is the 
median value of the redshift for the sample. L$_{\lambda}$ and  B$_{\lambda}$ 
are in units of $10^7$ L$_{\sun}$ \AA$^{-1}$.

We fitted this law to the L$_{0.30}(z)$ and L$_{0.45}(z)$ data,
with a$_{\lambda}$, B$_{\lambda}$ as free parameters. We limited the fit at  
z$\le 1.5$, 
in order to have the same objects at both wavelengths. We obtained

\begin{eqnarray*}
a_{0.30} &=& 4.9\pm 1.1 \hskip 1.4 cm a_{0.45}  = 4.1\pm 1.0 \nonumber \\
B_{0.30} &=& 0.60 \pm 0.05 \hskip 1cm B_{0.45} = 0.89 \pm 0.09. \nonumber \\
\end{eqnarray*}

The fits of this model to the data are shown in Fig.~\ref{UVexc} (left panels).

The values of the parameters $a_{0.30}$ and $a_{0.45}$ are not significantly  
different from each other, and their average value  would allow a fair fit 
for both L$_{0.30}$ and L$_{0.45}$.
However, the same value for $a_{0.30}$ and $a_{0.45}$ implies an almost null 
dependence of R$^{0.30}_{0.45}$ on (1+z), contrary to the observations. 
Therefore the difference between the $a_{\lambda}$ coefficients must be kept. 

Finally,  the  apparent effect of smaller L$_{0.30}$, L$_{0.45}$ 
and R$^{0.30}_{0.45}$ for sub-kpc sources, compared with the 
larger ones (point c), visible in Fig.~\ref{UVexc_LS}, arises because in our 
composite sample there are basically no sub-kpc objects at z$ \ge 0.6$. 
If the analysis is restricted to objects with z$ \le 0.6$, namely comparing 
sub-kpc and larger size objects in the same redshift range, the effect 
disappears. 

{\bx In spite of the data incompleteness, we also checked the subset of 
luminosities at 2000~\AA, L$_{0.20}$, and found results fully consistent with 
those for L$_{0.30}$.}

Our conclusion is therefore that all trends we observe are largely 
accounted for by a dependence on redshift of  the L$_{\rm UV}$ component.
Nevertheless, because of the limited statistics and the non uniform
coverage of the {\it redshift - radio power - size} space we cannot exclude 
second order {\ba relations} with the other parameters.

\section{The origin of the {\it UV-excess} in CSOs/MSOs galaxies}
\label{Or_UV}

There are different physical processes that can be responsible 
for the {\it UV-excess} \citep{Best98,TADH,Lab08}:

\noindent
 1) AGN radiation either from an unobscured nucleus or  scattered by 
the medium surrounding an obscured nucleus;
 
\noindent
2)  radiation from  young stellar populations (YSP);

\noindent
3) nebular continuum.

\begin {figure*}
\centering
{\includegraphics[width=17cm]{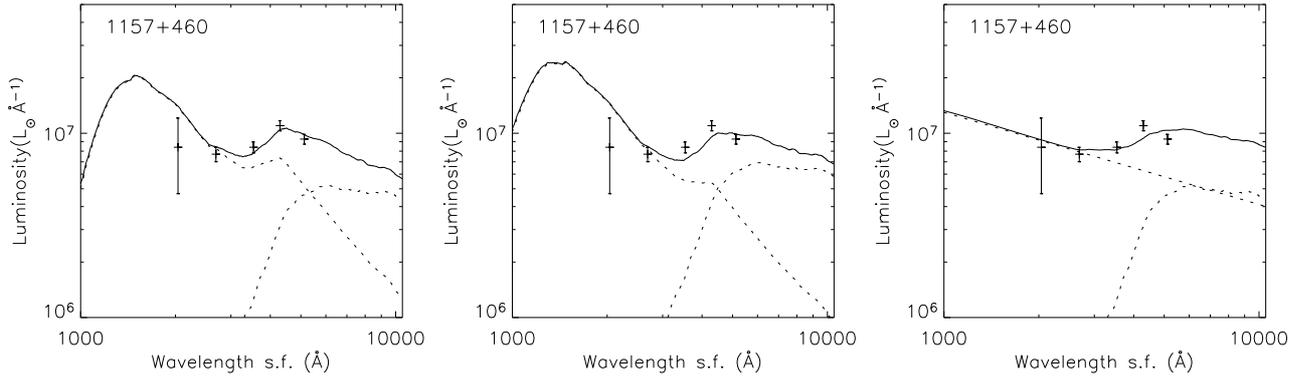}}
\caption{Example of different models yielding similar fits: {\it left hand}
and {\it center} panels fits with two YSP models of ages 0.1
and 0.06 Gyr respectively;  {\it right hand} panel  fit with the power 
law model. The dotted
lines represent an OSP model (at the longer wavelengths) and either a power law
(right-hand panels) or two different YSP models (left hand and middle panels).
The line through the data points is the sum of the two sub-components. } 
\label{plotYSPfit_0}
\end{figure*}

As discussed by \cite{TADH}, the nebular continuum can be derived from the 
$H_{\beta}$ flux. However, because we do not have this spectroscopic 
information we cannot make any assessment about this process.

Hence we considered only AGN light contribution either direct or scattered, 
and a contribution from young stellar population(s). 
Accordingly each UV-O-SED was fitted {\bz twice} with a combination of 
an old stellar 
population (OSP) elliptical galaxy model plus  a power law  
(Sect.\ref{AGN-contr.}) or a YSP model (Sect.\ref{YSP}). {\bz Of course
we cannot exclude partial contributions in the UV region from both 
processes in the same object}.

{\ba For the UV-O-SEDs of the OSP and YSP we used the stellar synthesis models
by B\&C, computed for an instantaneous star-burst, with the initial mass 
function (IMF) by \cite{Chabrier} and with solar metallicity}.
For the OSP we assume an age of 13 Gyr at  z~=~0 
(corresponding to a galaxy formation redshift $\ge 10$) 
and progressively  shorter ages, according to the galaxy redshift,
down to $\sim$ 4.3 Gyr at z=1.5. Galaxies are assumed to experience 
passive evolution as they age.

In this range of redshifts/ages the luminosity of an elliptical galaxy was  
higher in the past (with a small dependence of this trend on $\lambda$), 
approximately  $\propto $(1+z) for $\lambda$s $\ga$ 4500 \AA\ 
(to be compared with Eq. \ref{k} for shorter $\lambda$s). Therefore the 
UV-O-SEDs maintain approximately the same shape, apart from a scaling factor. 
The other scaling factor between model and data represents the 
OSP {\bz stellar} mass, 
which is a free parameter of the fit. The other free parameters are either 
the power law parameters or the YSP {\bz stellar} masses and ages.

{\bx For each source we used all the available data in the fit.
GALEX data are very important to constrain both the {\it AGN model} and 
the {\it YSP model}.
However, because these data exist for only 
a redshift-biased minority ($\approx$ 42 \%) of objects, we also made fits 
without GALEX in order  to derive statistical 
results for a redshift-unbiased set of objects. 
We then examine the systematic 
differences between the two sets of fits with and without GALEX data.}

The best fits are obtained by minimizing the $\chi^2$ 
considering only the formal photometric errors. 
The $\chi^2$ vary from reasonably good to quite bad. 
Excluding the trivial explanation of a totally wrong model, there 
are several reasons that may justify a bad $\chi^2$: 

\noindent
i) The formal errors on 
magnitudes might be underestimated (for instance assuming a 
minimum error of 0.05 mag. most of the bad $\chi^2$ improve significantly); 

\noindent
ii) Effects caused by emission lines, which we could not remove. In some 
cases a data point at $\lambda_{\rm s.f. } \approx 
6500$ \AA~ is too high for any possible model probably because of a 
strong contribution from H$_{\alpha}$. {\ba In such cases we ignored that data
point in the fit}.

\noindent
Moreover, for the two stellar population models:   

\noindent
iii) We used a grid of YSP models at discrete values of age {\ba (Sect. 
\ref{YSP})}; 

\noindent
iv) because in many cases we have only five data points we use only a 
two-population model, but we 
know that more than one YSP may be present \citep[e.g.][]{Holt3}; 

\noindent
v) There may be 
uncertainties in the shape details of the population models themselves,
{\bz related to special parameter choices such as IMF, metallicity, reddening,
etc.} 

For all these reasons we paid more attention to a general visual agreement 
of the model shape with the data than to the actual values of $\chi^2$.
We also anticipate that in a number of cases the fit quality of the YSP 
models is more or less similar to that of the power law models, and it is
often difficult to distinguish between these two possibilities. 
(see, e.g., Fig.~\ref{plotYSPfit_0}). 

{\bx The results of the fits without GALEX are reported in 
Table~\ref{SDSS_sample_2}. 
End-of-table notes refer to model changes required by GALEX data. For 
\object{1159+395} no fit was possible, as all  data are at $\lambda_{rest}$
shorter than 3000 \AA~ (Fig.~\ref{plot_sp}.1). For 
\object{1014+392} the SDSS data are contradictory with the near-IR and   GALEX
data (Fig.~\ref{plot_sp}.1)
and we derived a model using only the last two sets.}

\begin{figure*}[t] 
\centering
{\includegraphics[width=17cm]{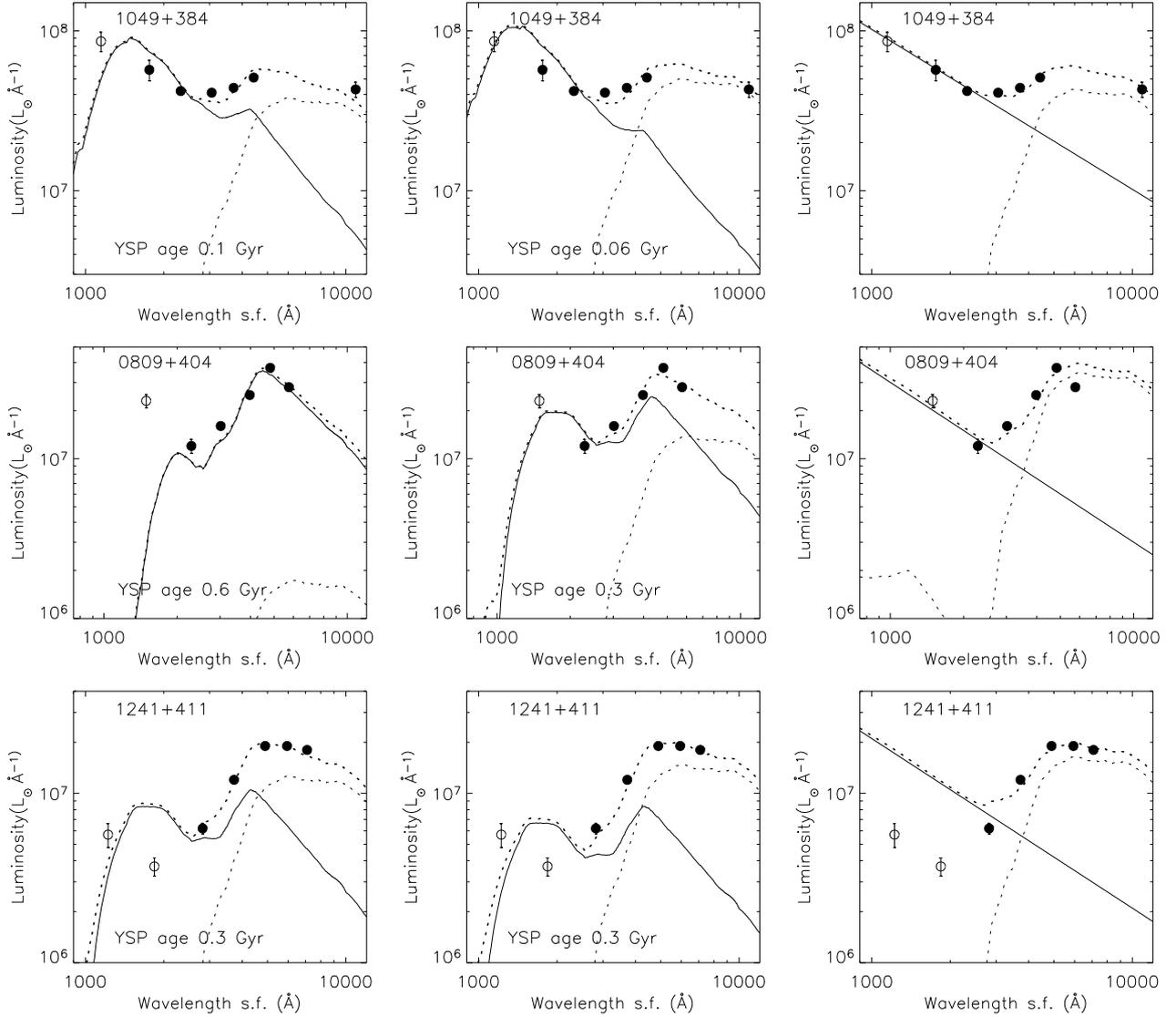}}
\caption{ Examples of model-fitting. 
The thin 
lines represent an OSP model (at the longer wavelengths) and either a power 
law (right-hand panels) or two different YSP models (left hand and middle 
panels) which give similar fits to the data. The line through the data points 
is the sum of the two sub-components. GALEX data, not used in the fits, 
are overplotted as {\it empty circles}.
{\it Top row}: The GALEX point does not discriminate among the three models 
based on SDSS data.
{\it Center row}: The GALEX point is in better agreement with the second, SDSS
based, YSP model, which can be slightly improved by minor changes.
{\it Bottom row}: The left hand and right hand panels show the best fits
with the SDSS data only. The center panel shows a slight modification of the 
0.3 Gyr  YSP model which, although unsatisfactory at the shortest wavelengths,
represent the best model when the GALEX data are added. }
\label{plotYSPfit_1}
\end{figure*}

\subsection{AGN contribution ?}
\label{AGN-contr.}

Because we do not have spatial resolution \citep[like, e.g., in][] {Lab08}, we
cannot state whether the {\it UV-excess} comes from either an extended volume 
or from a nuclear point source.
We cannot exclude nuclear scattered light, because not only we have no 
information on spatial distribution and orientation, but we also  
have no optical polarization information. 

We remark also that with the possible exceptions of 
\object{0744+464} and \object{1314+435A}, the 
possibility that some galaxy is a misclassified quasar can be excluded, 
because quasars are a factor $\approx$ 50 brighter in the UV than our galaxies.
{\bc On the other hand, we cannot exclude the presence of 
low/moderate luminosity AGNs in some galaxies.}

For the AGN model,  we used the power law L$_{\rm UV} =$ L$^*_{0.3}$(1+z)$ 
\times \lambda^{-\alpha_\lambda}$ with $\alpha_\lambda$ consistent with the 
average quasar UV-O-SED (Sect. \ref{SEDQ}).
We find that often this agrees fairly well with the data, although  there 
are cases in which the agreement is  definitively bad.

{\bx In three cases out of 14 the  GALEX data allow us to reject the power law
model. In the remaining cases GALEX does not provide significant differences: 
bad fits remain bad and good ones remain good}.

The results of the fit are given in  Cols. 14 - 16. of 
Table~\ref{SDSS_sample_2}.

At 3000 \AA~ the power law
contribution dominates over that of the underlying galaxy. The opposite is true
at $\lambda$s $\ga 4500$ \AA. 
{\ba L$^*_{0.3}$ turns out to be increasing with
redshift as  (1+z)$^{3.5\pm0.9}$. The OSP {\bz stellar} masses are in the 
range 
of $(2 -20)\times 10^{11} M_\odot$  and are $\propto$~(1+z)$^{1.5\pm 0.6}$.} 
{\bz This strong dependence  (increase) of the OSP {\bz stellar} mass with 
redshift is hard to explain in the light of models of galaxy evolution, so 
we are led to conclude that a model with AGN contribution only is not 
acceptable.

\subsection{Young stellar population(s) ?}
\label{YSP}

The presence of young stellar populations (YSP) in young radio
galaxies has been pointed out by several authors. 
\citep[see, e.g.,][] {Holt1, Lab08}.
Two major processes are invoked for their origin: jet induced star 
formation and galaxy merger.
In the first  case one would expect  
the YSP to have an age approximately equal to that of the radio source,
typically in the range $\sim 10^3$ to $\sim 3\times 10^5$ yrs for our sources
\citep[see, e.g.,][]{Murgia03}.
In the second one, instead, the YSP could be much older because a long time 
delay ($0.1 - 1$ Gyrs) is 
expected between the star burst triggered by the merger and the onset
of the radio source.

For the YSPs we considered a wide discrete range of
ages (1, 0.6, 0.3, 0.1, 0.06, 0.03, 0.01 Gyr and 6, 3, 1, 0.6 Myr).} 
Note that for a fixed  stellar mass, the younger the YSP is, the shorter is 
the wavelength at which L$_{\rm UV}$ dominates the UV-O-SED, {\bx hence the 
importance of short wavelengths to reveal very young stellar populations.}
 
We present the results in Cols. 8--13 of Table~\ref{SDSS_sample_2}.

We stress that for each UV-O-SED there are often  two or three more or less 
equivalent models (Fig.~\ref{plotYSPfit_0}), differing from each other 
in age and {\bz stellar} mass of the YSP and
somewhat  less in  the UV-OSP {\bz stellar} mass. Since the wavelengths are 
sampled in  a limited range, the UV-O-SEDs of  YSPs 
differing in age by a factor $\approx 3$ are similar in shape, 
therefore, by properly choosing the {\bz stellar} mass, they 
can reproduce  more or less the same light contribution in the SDSS bands.  
The fitting model is more  constrained when a broader range of 
wavelengths {\bx (e.g., GALEX and near-IR)}  is available.
In addition, for some  objects  at high redshift (z $>$ 0.7) and without 
near-IR data, the OSP {\bz stellar} mass is little constrained. In these cases 
we report the maximum {\bz stellar} mass allowed  by the data. 
For the high-redshift sources \object{0744+464} and 
\object{1314+453A}
the UV-O-SEDs could   be fitted by an YSP alone.  
In general the YSP contribution is dominant at $\la$3000 \AA, but at high 
redshifts it becomes important  also at 4500 \AA and may overwhelm the OSP 
contribution.

{\bx It is interesting to compare the stellar populations parameters derived
with and without the GALEX for the objects with GALEX data.
We find that for $\approx$ 40 \% of objects the model is unchanged or 
marginally changed, and that for the remaining ones 
a YSP model a factor $\approx$ 2  older, and hence with masses
$\approx$ a factor two larger, is preferred. 
The fits tend to be worse in the cases
where the luminosity increases again in the GALEX NUV band and shows, or
suggests, a peak at $\lambda \la$ 1500~\AA. Actually this observed peak appears
narrower than or somewhat displaced from the peak present in the YSP models.
 
In Fig. \ref{plotYSPfit_1} we show examples of objects with GALEX data to 
demonstrate their effect in the fits.
}

{\bx The YSP luminosities at 3000 \AA~ and at 4500 \AA ~are 
$\propto$(1+z)$^{4.4\pm1.0}$ and  $\propto$(1+z)$^{3.7\pm0.9}$ respectively,
which is consistent with the L$_{\rm UV}$ dependences on (1+z) at the two 
wavelengths that we found in Sect. \ref{UV-dep}.}

{\bx 
The YSP ages derived without GALEX data are generally in the 
range 0.06 - 0.8 Gyr, with a median value $\approx$ 0.2 Gyr.
The average YSP  stellar mass is $\approx 10^{10}$ M$_{\sun}$, with a 
dispersion of a factor $\approx$ 4. 
We find that the YSP mass is correlated with (1+z) and with L$_{0.3}$,
while the YSP age is independent of (1+z).}

{\bx If instead we also use GALEX data when available, the YSP ages of  most 
objects with z $\la$ 0.5 are shifted to the range 0.2-1.2 Gyr, and the median 
value is $\approx$ 0.4 Gyr.
While the majority of the higher redshift objects are not detected by GALEX, 
for three out of four with these data, the models have marginal non-systematic
changes, while the fourth one (3C237) requires the YSP age to be 
$\approx$ 4 times shorter.
The correlation between YSP mass and (1+z) may disappear, but 
because of the large incompleteness at higher redshift 
we leave this as an open question.} 

{\bx The OSP stellar masses are fairly independent of GALEX data. After 
correction for passive evolution, they  are in the range of $(1 - 10)~ 
10^{11}$ M$_\odot$,  with an average value of $4 \times 10^{11}$ M$_{\sun}$,
independent of (1+z), radio power and radio size.}


The above findings are consistent with an OSP undergoing a passive evolution
(as in B\&C) plus a YSP of a small {\bz stellar} mass (a few \% of the OSP
stellar mass), which makes a major contribution to the UV luminosity, and a 
minor one at longer wavelengths, except for the high redshift, where the 
contribution can be substantial.

With our data we have no evidence for a YSP with an age comparable to that of 
the radio source, namely $< 10^{6}$ yr. If present, it would be a minor 
contributor to the {\it UV-excess}. Therefore we prefer the scenario in which 
the YSP responsible for the {\it UV-excess} is caused by a merger event,   
which at a later stage triggered  the onset of the radio source.

\section{A comparison sample of  large size radio galaxies}
\label{EXT}

It has been well known for a long time, even earlier than for CSOs/MSOs, that 
there is an {\it UV-excess} in the hosts of powerful large  size radio 
galaxies (several hundred kpc; LSOs).
This UV emission is generally aligned with the  
radio structure and the phenomenon is referred to as the {\it alignment effect}
\citep[see, e.g.,][]{Ch87, Mc87}. A set of 3CR and 6C sources at 
redshift $\approx$ 1 has been investigated with great detail in the nineties
\citep {Best97, Best98, Best99}
to understand this effect \citep[see also][and references therein]{Insk06}. 
More recently  the {\it UV-excess} has been found in local samples of 
radio galaxies as well \citep {Aretx, Raimann}.

{\bx In order to compare the UV-O-SEDs of the objects of our composite sample 
of young radio sources with the hosts of radio sources with large linear size
(LS $\ga$ 30 kpc) and with a redshift distribution similar to ours,  
we selected the 3CR FRII $LSO$s that fall in the area covered by the SDSS.
This sample, containing 37 radio galaxies  (the set of 3C sources at
$z \approx$ 1  studied by \cite{Best97} is included), was supplemented by the 
six  6C sources with z$\approx$1, from \citet{Best99}, with SDSS data. This 
second small sub-sample, with radio power lower by a factor $\approx$ 7 
compared with 3CR sources at the same redshift, allows us to manage the 
degeneration between radio power and redshift, which cannot be avoided by
dealing with the 3CR sample alone.

Overall we have a sample of 43 large-size radio galaxies with SDSS data
(see Table \ref{Tab_LSO}). For these, we have almost complete data in the 
near-IR (mostly UKIRT data). In addition, $\approx$ 75 \% have been observed 
by GALEX (mostly $AIS$), two thirds of which were detected in NUV and 
eventually FUV band with slightly larger incompleteness at highest redshifts.

We produced the UV-O-SEDs following the same procedure as for the CSOs/MSOs. 
They are presented in Fig. \ref{3C_ext_SEDs}. Visually we see the {\it UV 
excess} in the majority of cases. 
The most impressive result is the bright peak around 1000-1500 \AA~ in the 
high-redshift galaxies (see, e.g. 3C265, 6C1011+36). 
This peak was also noted in a fraction of CSOs/MSOs, but here it is sometimes 
very bright compared with the $\lambda >$ 4000 \AA~ peak/plateau. Hints for 
these peaks can be seen in the SDSS $u$ band alone as well, but they become 
very clear when adding GALEX data. 
}

For these sources we repeated the analysis described in Sect. \ref{UV-dep} and
Sect. \ref{Or_UV}.

\subsection{Correlations}
\label{Ext-corr}

\begin{figure*}[t]
\centering
{\includegraphics[width=17cm]{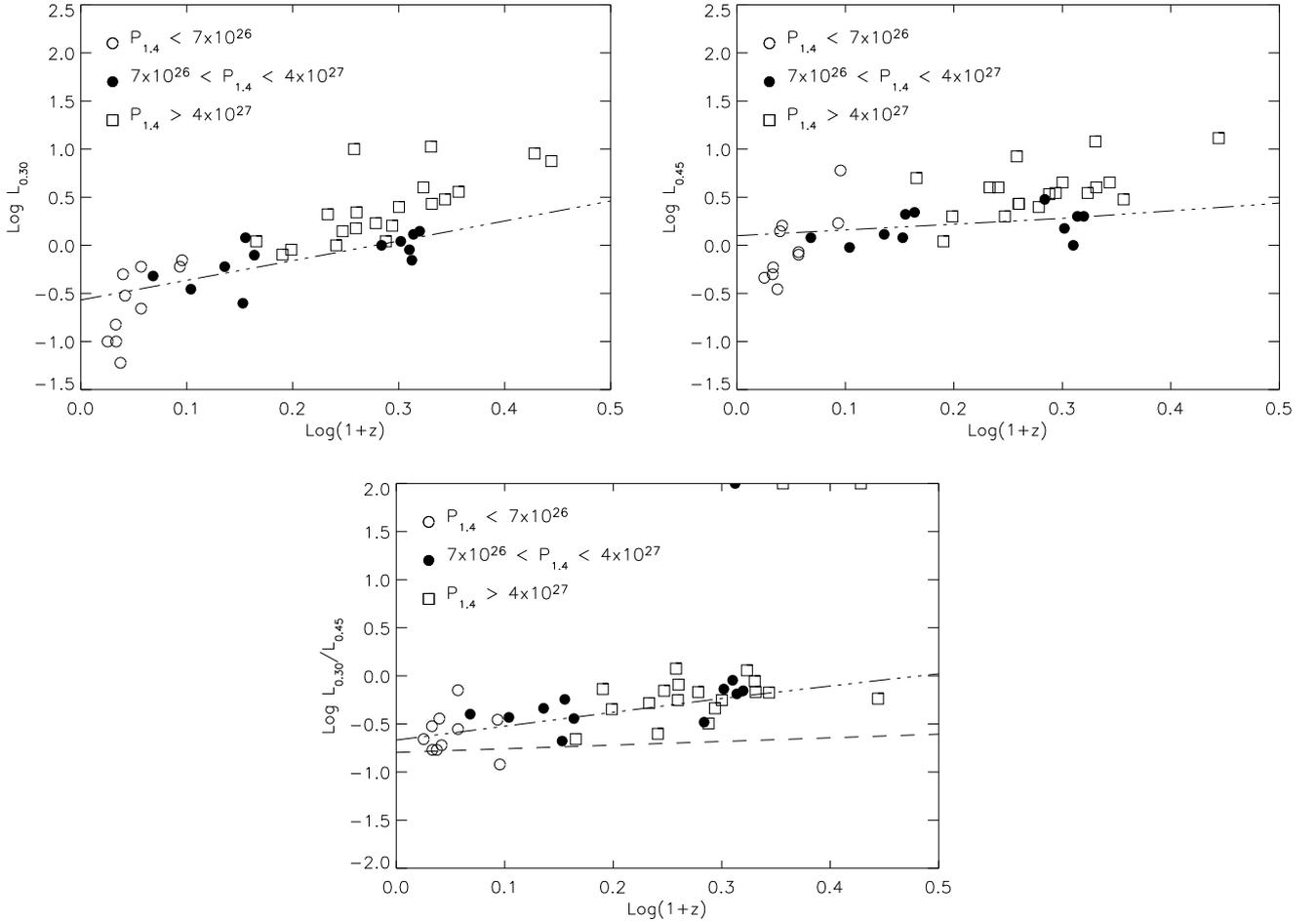}}
\caption{Plots of Log(L$_{0.30}$), Log(L$_{0.45}$) and  Log(R$_{0.45}^{0.30}$)
{\it vs} Log(1+z) for LSOs.The dotted lines represent the fit to the data 
obtained using Eq.\ref{L_lam-P}. The almost horizontal line in the bottom 
panel represents the expected value of  Log~R$^{0.30}_{0.45}$ for an old 
elliptical galaxy. 
}
\label{UVexc_Ext}
\end{figure*}

The same {\it caveats} as discussed in Sect. \ref{UV-dep} hold also for 
the analysis of the LSO sample. 
{\bx But the use of the small set of 6C sources in addition to the 3CR sample 
allows us, as for the $MSO$s/$CSO$s,  to select two subsamples:

\noindent
i)  {\it Subsample 1}:~  
thirteen sources in a narrow redshift range (0.9 $\la z \la$ 1.2) and in a 
broad radio 
power range \Big($ 7 \times 10^{26} \la $P$_{1.4}$(W~Hz$^{-1}~ \la 3 
\times 10^{28}$  \Big);

\noindent
ii)  {\it Subsample 2}:~  twelve sources in a narrow radio power range 
\Big($7 \times 10^{26} \le$ P$_{1.4}$(W~Hz$^{-1}$)$ \le 4 \times 10^{27}$\Big) 
and in a broad redshift range (0.2 $\la z \la$ 1.2).

These two subsamples allow us to examine the relations with radio 
power at $\approx$ constant redshift and vice versa.

In Fig.~\ref{UVexc_Ext} we present plots of L$_{0.30}$, L$_{0.45}$ and 
R$_{0.45}^{0.30}$ {\it vs} (1+z) only to save space. 

We repeated the analysis described  for the small sources in Sect. \ref{YSP}.  
From the plots in Fig.~\ref{UVexc_Ext} the following trends are apparent:
 
i) for the sources of {\it subsample 2} (filled symbols in Fig. 
\ref{UVexc_Ext}) there is a  clear dependence of $L_{0.30}$ and 
$R^{0.30}_{0.45}$ on (1+z), and a much smaller one, if any, for $L_{0.45}$.  
These dependences extend reasonably  well also to the lower power sources with
z $\la$ 0.20. Therefore, for the sources with 
$P_{1.4} \le 4\times 10^{27}~ W Hz^{-1}$,
$L_{0.30}$,  $L_{0.45}$ and $R^{0.3}_{0.45}$ can be described with a dependence
on (1+z) only, independent of radio power. 
The curves plotted in Fig.~\ref{UVexc_Ext} are derived by fitting the same 
model used for the CSOs/MSOs (Eq.\ref{L_lam}), with the appropriate 
parameters;}

ii) the more powerful radio galaxies, 
P$_{1.4}$(W~Hz$^{-1}$)$ \ge 4 \times 10^{27}$ (squares), describe a separate 
sequence compared with the less powerful objects at the same redshift, on 
average  a factor $\approx$ 2 brighter in both L$_{0.30}$ and L$_{0.45}$.
{\bx If we isolate among them those of {\it subsample 1} (0.9 $\le z \le$ 1.2),
we can derive for these high-power sources the dependence
of $L_{0.30}$ and  $L_{0.45}$ on $P_{1.4}$.}

\begin {figure*}
\centering
{\includegraphics[width=17cm]{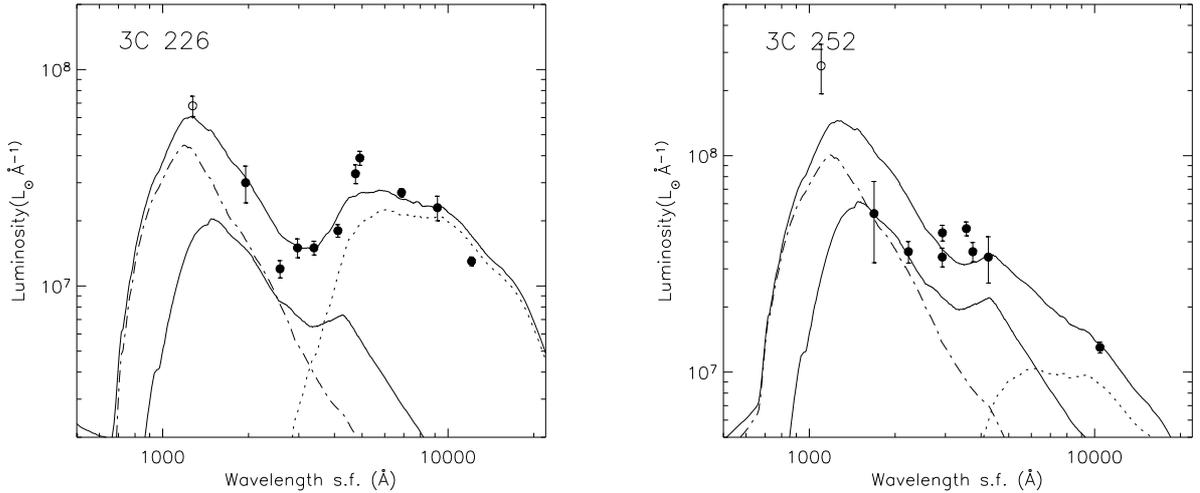}}
\caption{Examples of fits with three stellar population  models: one OSP, one 
YSP of intermediate age (0.01 -  1 Gyrs) and a very young population with the 
age of the radio source ($< 10^{-2}$ Gyrs, for an assumed expansion speed of 
0.1 c). 
} 
\label{VeryYoung}
\end{figure*}

We attempted to describe quantitatively the above findings with a 
modification of Eq. \ref{L_lam}. In order to take into account the radio 
power dependence, we added a new  term, dependent on radio power 
only. This new term introduces 
the possibility of an additional and independent mechanism  for the  
production of the {\it UV-excess}:

\begin{equation}
L_{\lambda}= \left[B'_{\lambda}\times \left( \frac{\rm{1+z}}{1.6}\right) ^{a'_{\lambda}} 
+ k_{\lambda}\right] 
+ v_{\lambda}\times\left( \frac{{\rm P}_{1.4}}{10^{27}}\right) ^{w_{\lambda}}
\label{L_lam-P}
\end{equation}
\noindent
($k_\lambda$ from Eq. \ref{k}, Sect.\ref{UV-dep}).
Units are $10^7$ \AA$^{-1}$ for L$_{\lambda}$ and W~Hz$^{-1}$ for P$_{1.4}$.

For this model we find
\begin {eqnarray*}
  a'_{0.30}&=&2.2\pm 0.7 \hskip 1.1cm  a'_{0.45}= - 0.1 \pm 1.4 \nonumber \\ 
  B'_{0.30}&=&0.52\pm 0.03 \hskip 0.6cm B'_{0.45} = 0.76 \pm0.12 \nonumber \\ 
  v_{0.30}&=&0.024\pm 0.01 \hskip 0.7cm v_{0.45}  =  0.054 \pm 0.02 \nonumber \\ 
  w_{0.30}&=&1.6 \pm 0.15 \hskip 1cm w_{0.45} \approx 1.48 \pm 0.14. \nonumber \\  
\end {eqnarray*}
\vskip -0.4cm
After the dependences on (1+z) and P$_{1.4}$ are taken off the data,   
no significant residual dependence is found on LS.

{\bx As for CSOs/MSOs, in spite of the incompleteness, we also examined the  
luminosities, L$_{0.20}$, at 2000 \AA~ and found results well consistent 
with those for L$_{0.30}$.}

\subsection{YSP models}
\label{Ext-YSP}

We applied the two stellar population galaxy models, as in Sect. \ref{YSP}, and
derived OSP and YSP {\bz stellar} masses and ages \footnote{{\bx For the 3CR 
sources of \cite{Best97} we used in the fits their HST data as well 
(5400$\le \lambda_{obs.fr.} \le 8600$}).}.

{\bx The ages of the YSP derived without GALEX data span a range of values 
from 0.04 to 0.8 Gyr, with a mean value of $\approx$ 0.16 Gyr, similar to what
was found for CSOs/MSOs, but the addition of the GALEX data tends to indicate 
younger YSP models, especially for the high redshift-galaxies. 
This effect is often caused by the rapid increase of luminosity in the NUV (and
FUV, when detected) band, namely by the {\it bright peak} mentioned earlier.
On average the effect is of rejuvenating the YSPs by a factor $\ga$ 3 
and consequently of decreasing their masses by about the same factor. 
Nevertheless, in general the models fail in properly reproducing the fast 
rising of the peak with decreasing wavelength.  

We attempted to see if the {\it bright peak} can be caused by an additional 
contribution of UV radiation and modeled the UV-O-SEDs of a few sources with 
the best-quality data with three stellar populations: one OSP, one YSP of 
intermediate age (0.01 -  1 Gyr) and a very young population  with the age of 
the radio source ($\la 10^{-2}$ Gyr), derived from the source linear size and 
an assumed expansion speed of 0.1 c. We display two examples in Fig. 
\ref{VeryYoung}.
The results suggest that a second very young YSP may be possible, although 
these fits are insignificantly better than those with a single YSP only.}

The OSP {\bz stellar} masses are fairly independent of the GALEX data and are 
similar to those of the CSOs/MSOs sample.
The   average mass is $4 \pm 0.5 \times 10^{11}$ M$_\odot$. 
There are no significant relations of the OSP {\bz stellar} mass with redshift,
radio power or radio linear size
\footnote{{\bx The low-redshift radio galaxies \object{3C236}, \object{3C285},
and \object{3C321} have been studied spectroscopically by \cite{Holt3}, with 
the aim of determining the properties of the YSPs. These authors also use in 
their fits in addition to the UV-O-SEDs stellar absorption lines, power law 
contributions, and various reddening recipes. The comparison of their results 
with ours shows that our age estimates differ within a factor $\approx$ 3 from 
theirs. The luminosities of their and our YSPs are also different by the 
amount expected from the differences in age. We attempted to fit our SDSS data
using their ages, but the corresponding fits are worse than ours.
We think that the differences in the results are not surprising because the
data and parameters that we are using are different from those 
in \citet{Holt3}.}}.

\subsection{AGN contribution?}
\label{AGN-LSO}

As for CSOs/MSOs, we attempted to fit the UV-O-SEDs of these large sources with
an old elliptical galaxy template  plus a power law spectrum (AGN) as well.
Differently from what was found for CSOs/MSOs, in these sources {\bx the great
majority  of which has near-IR and GALEX data}, there are a number of cases 
where these fits are worse than those of the YSP model, and none with better 
fits.
We stress the difference with our sample of small size sources, for which we 
have near-IR and GALEX data only in a minority of cases,  and often we could 
not distinguish between AGN and YSP models. With data in a broader wavelength 
range, it is easier to discriminate between the two models.

Anyhow the ranges of values for L$^*_{0.30}$, $\alpha_\lambda$ and 
M$_{\rm OSP}$ are the same as those we derived for CSOs/MSOs. Also 
L$^*_{0.30}$ and M$_{\rm OSP}$ show the same dependence with (1+z) of 
CSOs/MSOs. As for CSOs/MSOs, we consider the dependence of $M_{\rm OSP}$ on 
(1+z) unrealistic.

\section{Young versus old radio galaxies}
\label{disc}

\begin{figure}[t]
\resizebox{\hsize}{!}{\includegraphics{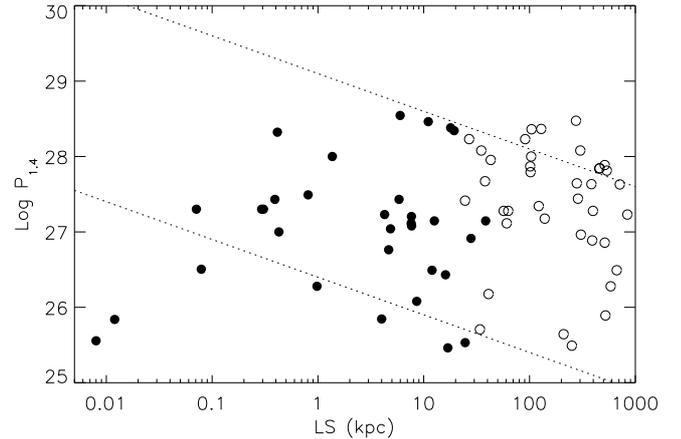}}
\caption{Distribution of the CSOs/MSOs (black points) and LSOs (empty circles)
in the {\it radio power - linear size} plane. The two lines represent an 
evolutionary model where P$_{1.4} \propto$~LS$^{-0.5}$ (see text).
}
\label{P-LS}
\end{figure}

In the previous sections we analyzed the UV-O-SEDs and the {\it UV-excess} of 
small (CSOs/MSOs) and large size (LSOs) radio galaxies. 
{\bx The two samples, very different in source radio size, have a similar 
redshift distribution and span the same range in radio luminosity, although 
with a difference of a factor $\approx$ 2.5 in their median values (see Fig. 
\ref{P-LS}).

Within the {\it evolutionary scenario}, in which the CSOs/MSOs are the young 
and the LSOs the aged phases of the extragalactic radio source population, any
comparison between properties of the two classes should be made between samples
selected in radio power and size according to the evolution scenario. 
Typically radio sources are expected to dim while growing in size. As an 
example, in Fig. \ref{P-LS} we overplotted two evolutionary lines of a model 
in which P$_{1.4} \propto$ LS$^{-0.5}$ \citep[see][]{Baldwin82, Fanti95}. 
According to this model CSOs/MSOs (black points) within those lines would 
evolve into the LSOs (open circles) enclosed within the same lines. 
Other models with  different dependences between P$_{1.4}$ and LS were 
also considered.
The previous analysis on CSOs/MSOs (Sect. \ref{UV-dep} and \ref{Or_UV}) and on
LSOs (Sect. \ref{EXT}) was made considering the two samples separately and 
without taking any evolutionary model into account. Therefore we repeated
the analysis, selecting from each set of sources the objects within the 
evolutionary  band. We found negligible changes and, therefore we use below 
the results reported in Sects. \ref{UV-dep}, \ref{Or_UV}, 
and \ref{EXT}.

}  

The main points in  comparing  the results obtained for the two classes of 
sources are: {\bx

i) the luminosities L$_{0.30}$ and L$_{0.45}$  and their ratio have  
similar ranges of values in  small ($\la  30$ kpc) and in  large 
($ \gg 30$) kpc radio galaxies. 

ii) L$_{0.30}$ and L$_{0.45}$  show a dependence on (1+z), both  in CSOs/MSOs
and LSOs. However, these dependences are different at $\ga$ 2 $\sigma$ level, 
the LSOs showing a weaker dependence.

iii) In LSOs   a dependence of L$_{0.30}$ and L$_{0.45}$ on P$_{1.4}$, not 
found for CSOs/MSOs, shows up for $P_{1.4} \ga 4 \times 10^{27}$W~Hz$^{-1}$. 
In the fitting procedure we described the dependence on $P_{1.4}$ with an
additional term (see the parameters of Eq.~\ref{L_lam} and Eq.~\ref{L_lam-P}).
Note, however, that among the small sources we have only a few objects in that
radio power range and we cannot exclude that we missed a P$_{1.4}$ dependence.

iv)  At $\lambda < 2000$~\AA~ the high-redshift (high-radio power) LSOs very 
often show pronounced peaks in their UV-O-SEDs, which are also seen, although 
not as bright, in the CSOs/MSOs. 

v)  The YSP ages associated with high-redshift high-radio power LSOs appear 
shorter by a factor $\ga$  3 with respect to the CSOs/MSOs.
     
vi)  The OSP  have similar {\bz stellar} masses in the two sets of radio 
sources.
}

Under the assumption that the {\it UV-excess} is predominantly caused by the 
presence of a YSP population, we discuss the similarities and the differences 
in the framework of a {\it composite scenario} that combines the two 
different processes which may generate YSP populations. 

The first process is the {\it galaxy merger}, which generates a burst 
(or several bursts) of star formation (the YSP$_{\rm merger}$).
With a time delay $\la$ 1 Gyr some residuals of the {\bz stellar} mass reach 
the super-massive  black hole of the more massive galaxy or the two black holes
of the merging galaxies coalesce, thus triggering the formation of the 
radio source.
The second process is the {\it generation of shocks} in the outer medium, 
caused 
by the expansion of the radio source,   
which, at the end, produces another burst of star formation (YSP$_{\rm r.s.}$).
This new population is expected to be more or less distributed along the 
source axis, explaining the {\it alignment effect}, and  would 
have an age comparable with that of the radio source, {\bz namely $\la 0.01$
Gyr for an expansion speed $\approx$ 0.1 c \citep{A.L.}}.
Looking at the UV-O-SED of the host galaxy of a radio source one would observe
at the same time the YSP$_{\rm merger}$, with an age of $\la 1$ Gyr, the 
YSP$_{\rm r.s.}$ with an age of $\la 10^7$ yrs  and the old elliptical galaxy, 
with an age of {\bz several Gyr}.
Owing to its longer time scale the  YSP$_{\rm merger}$ should appear similar 
in small (CSOs/MSOs) and in large size (LSOs) radio sources.
Furthermore the  (YSP$_{\rm r.s.}$) would be expected to show properties 
related to radio source parameters, like radio power (amount of shocks 
generated) and linear size (time required to build up the very young YSP$_{\rm
r.s.}$ and its later decay).

The dependence of a fraction of L$_{\rm UV}$ (see Eq.\ref{L_lam-P}) on 
radio power in the powerful LSOs could  be a marker of the YSP$_{\rm r.s.}$.
This fraction would be responsible for the {\it alignment effect}
seen  in these objects \citep{Best98}.
On the contrary, the lack of dependence of L$_{\rm UV}$ on radio power in 
small radio sources would indicate that the YSP$_{\rm r.s.}$ is still in 
the stage of being formed and the YSP$_{\rm merger}$ dominates.
{\bx The UV-O-SEDs that we obtain using GALEX data
for the most powerful LSOs
leave open the  possibility for an additional YSP with an age comparable
with that of the radio source (see Fig. \ref{VeryYoung}). 

The major problem with the {\it composite scenario} is that in the LSOs the 
redshift-dependent L$_{\rm UV}$  component has a dependence 
on  (1+z) significantly different  from that of CSOs/MSOs.  
If, as we assume, this component is caused by the YSP$_{\rm merger}$, which has
a time scale much longer than the radio source life-time, it should display the
same properties in both classes of sources and, in particular, it should 
exhibit the same (1+z) dependence.
We attempted to force the same redshift dependence  on both small and 
large sources, but we only achieved a very poor agreement.

We might wonder whether a YSP$_{\rm r.s.}$ may be produced not only by the 
very high-radio power sources (P $> 10^{27}$ W Hz$^{-1}$), but also by the 
less powerful ones, without a dependence on radio power, weak or null 
dependence on (1+z) and be dominant in L$_{UV}$, in a way  to wash out 
the redshift behavior of the L$_{UV}$ of the YSP$_{\rm merger}$.
But then we would expect the LSOs to be systematically 
brighter than  the CSOs/MSOs, which seems not to be the case.

Alternatively we could imagine that the LSO life-times are longer than 
assumed and comparable with those of the $YSP_{\rm merger}$, so that when the  
$YSP_{\rm r.s.}$ is fully developed, the other is already fading out.

If instead the origin of the {\it UV-excess} is the merger process only, 
we should
conclude that more powerful radio sources are produced when major mergers 
that involve an amount of fuel (stellar mass) above a certain threshold 
occur. 
Furthermore, in order to explain the {\it alignment effect}, we should assume  
that mergers occur anisotropically and that the radio jet, when
born, is oriented along the direction of falling material \citep[see ][as 
cited by \cite{Insk06} for a model of this kind]{West}.
However, the different redshift dependences of the UV-excess
remain an open problem.

From the above our data do not allow us to reach an unquestionable conclusion 
although there are some hints in favor of the {\it composite scenario}.

Perhaps our approach and our data are not sufficient to deal with this subject,
and a more refined analysis, additional data (e.g. spectroscopic, as in 
\citet{Holt3}), and a more complex scenario is required that also includes 
other processes such as AGN contribution, emission lines, and nebular continuum
effects.
}

\section {Conclusions}
\label {concl}

The main results from this work can be summarized as follows.

1) The Sloan Digital Sky Survey (SDSS) data allowed us to obtain new optical 
   identifications for radio sources in a sub-set of 57 objects  from the 
   B3-VLA CSS sample laying in the area covered by the SDSS.
   Eight of the new identifications  are with galaxies 
   and three with quasars or quasar candidates. In this sample the fraction of 
   identified sources has increased from $\approx$ 60 \% to 
   $\approx$ 80 \%. 

2) In total, including also earlier identifications, we have photometry
   in the five SDSS bands for 27 galaxies and 8 quasars or quasar candidates.

3) {\bz Photometric redshifts (z$_{\rm ph2}$) for 17 of the  identified 
   galaxies with z$_{\rm sp}$
   were obtained using the SDSS routines. A comparison between the two types 
   of redshifts showed that the z$_{\rm ph2}$ are  underestimated }
   on average by  $\approx$ 26 \%. A correction of this amount was 
   applied to the z$_{\rm ph2}$ of the newly identified galaxies.
   With these corrected redshifts the new identification are well located 
   in the Hubble diagrams, mixed with  the galaxies with spectroscopic
   redshifts.

4) Including also  galaxies with photometric redshifts only, the 
   optical identifications are considered 
   complete up to z$ \approx$ 1.0.

5) After extending the B3-VLA CSS sample with 
    CSOs/MSOs from the literature, we 
    derived the source-frame optical 
   spectral energy distribution (UV-O-SED) for galaxies and quasars with the 
   SDSS 5-band photometry.
   {\bx We added to the SDSS data those in the far UV from the GALEX surveys 
   and
   in the near-IR from the 2MASS survey, and from literature the UKIRT 
   observations.
   Although they are incomplete, these data give very valuable information.}

   {\bx In quasars the UV-O-SEDs are described to a good approximation 
   by a single power law, or, more commonly, by
   two power laws matching at $\approx$ 3000 - 3500 \AA,  the one at the
   shorter wavelengths being the flatter.}
   The UV-O-SEDs of galaxies show in the large majority of cases a  
   luminosity excess  in the UV ($\lambda < 4000$ \AA), when compared with the 
   UV-O-SED  of an old elliptical galaxy.   

6) We analyzed this excess in terms of the source-frame luminosities 
   at 3000~\AA~ and 4500~\AA~ and of their ratio as a function of redshift, 
   radio 
   luminosity, and radio linear size. We found that the main dependence is on
   the redshift, while there are no significant dependences with the other 
   parameters. 
   \noindent
   We described the {\it UV-excess} as caused by a 
   source of radiation 
   overimposed on the UV-O-SED of an old elliptical galaxy undergoing passive 
   evolution. The UV component becomes more and more important as the redshift 
   increases: $\propto $(1+z)$^a$ with $a \approx$ 5.0 and 4.0 at 3000 \AA~ 
   and 
   4500~\AA~ respectively. {\bx For the fraction of objects with GALEX 
   observations 
   we also examined the behavior of L$_{0.20}$, which agrees with the 
   results of  L$_{0.30}$.}

7) We attempted to describe this UV component with two models: an AGN
   component with a quasar-like power law spectrum, and a Young Stellar 
   Population (YSP) component, described by the models developed by 
   B\&C. For CSOs/MSOs both models are equally plausible for 
   the majority of sources, and we cannot make a clear choice between them. 

   {\bz Yet there is a different 
   implication between the two: the AGN  model requires an evolution 
   for the OSP {\bz stellar} mass with time, with {\bz stellar} masses 
   larger in the past than at present 
   epoch, while in the YSP model the OSP {\bz stellar} mass is roughly 
   constant.}
   
8) We performed an analysis of the UV-O-SEDs for a composite sample of 
   large size ($\gg$ 30 kpc) sources with radio power and redshifts similar 
   to CSO/MSO's, taken from the literature, using  the SDSS photometry 
   {\bx plus UV GALEX data and near-IR data
   often available from the literature}, in order to explore the relations with
   our smaller size radio sources, which are considered to be  the young 
   precursors of those with  larger size.
   The two sets of sources show a number of common properties as well as 
   some clear differences, so that it is difficult to draw firm conclusions. 
   The popular scenario where the {\it UV-excess} is caused by a 
   first generation  YSP  
   from a galaxy merger, which later triggered the radio source, and a second
   generation YSP produced by shock of the expanding radio source, has some 
   hints
   of support and some contradictions as well.

\begin{acknowledgements}

We thank an anonymous referee for pointing us to the use of the UV  
GALEX data and for several useful comments, which helped to improve the 
presentation of the paper.

This search is based on the  use of the SDSS Archive, funding for the creation
and distribution of which  was provided by the Alfred P. Sloan Foundation, the
Participating Institutions, the National Science Foundation, the U.S. 
Department of Energy, the National Aeronautics and Space Administration, the 
Japanese Monbukagakusho, the Max Planck Society, and the Higher Education 
Funding Council for England. The SDSS was managed by the Astrophysical 
Research Consortium for the Participating Institutions. 

The authors also made use of the data from the NASA Galaxy Evolution Explorer 
database, from the Two Micron All Sky Survey and from the NASA/IPAC 
Extragalactic Database (NED).
GALEX is operated for NASA by the California Institute of Technology under 
NASA contract NAS5-98034.
The Two Micron All Sky Survey is a joint project of the University of 
Massachusetts and the Infrared Processing and Analysis Center/California 
Institute of Technology, funded by the National Aeronautics and Space 
Administration and the National Science Foundation.
The NASA/IPAC Extragalactic Database (NED) which is operated 
by the Jet Propulsion Laboratory, California Institute of Technology, under 
contract with the National Aeronautics and Space Administration.

\end{acknowledgements}

\appendix
\onecolumn
\section{The B3-VLA-SDSS sample and notes}
\label{B3_VLA}
\tabcolsep1.2mm
\begin{longtable}{lcrrr|rrrrr|cr|rrr|rcl}
\caption{The B3\_VLA - SDSS sample}\label{B3tab1} \\
\hline
\multicolumn{5}{c|}{Paper I} & \multicolumn{10}{c|}{SDSS} & \multicolumn{3}{c}{adopted} \\
B3 Name   &Id&    m$_{\rm R}$& z\ \ \   & Ref& $u$\ \ \   &  $g$\ \ \   &  $r$\ \ \   &  $i$\ \ \   &   $z$\ \ \  & Id& z$_{\rm{sp}}$& z$_{\rm {ph}}$& z$_{\rm {ph2}}^{\rm {cc2}}$& z$_{\rm {ph2}}^{\rm {d1}}$
& z\ \ \ & Id& m$_{\rm{r}}$\\
\hline\hline
\endfirsthead
\multicolumn{18}{c}%
{{\bf \tablename\ \thetable{}} {\rm The B3\_VLA - SDSS sample (cont.)}} \\
\\
\hline
\multicolumn{5}{c|}{Paper I} & \multicolumn{10}{c|}{SDSS} & \multicolumn{3}{c}{adopted} \\
B3 Name   &Id&    m$_{\rm R}$& z\ \ \   & Ref& $u$\ \ \   &  $g$\ \ \   &  $r$\ \ \   &  $i$\ \ \   &   $z$\ \ \  & Id& z$_{\rm{sp}}$& z$_{\rm {ph}}$& z$_{\rm {ph2}}^{\rm {cc2}}$& z$_{\rm {ph2}}^{\rm {d1}}$
& z\ \ \ & Id& m$_{\rm{r}}$\\
\hline\hline
\endhead
  \object{0701+392}~ & Q&   18.7&  1.283&  1&   20.40& 19.90& 19.10& 18.88& 18.72& Q&      -  &  -  &  -  &  - & 1.283&  Q &19.10\\
          &  &       &       &   &    0.06&  0.01&  0.01&  0.01&  0.01&  &         &  -  &  -  &  -  \\
\vspace {-0.3cm}
           &  &  &  &  &  &  &  &  &  &  &  &  &  & \\
  \object{0744+464} & G&     - &  2.926&  1&   24.44& 22.35& 21.38& 20.98& 20.76& G&      -  & 0.11& 0.12& 0.43& 2.926&  G &21.38\\
           &  &       &       &   &    1.38&  0.10&  0.07&  0.08&  0.08&  &         & 0.04& 0.15& 0.14\\
\vspace {-0.3cm}
             &  &  &  &  &  &  &  &  &  &  &  &  &  & \\
\object{0754+396} & G&     - &  2.119&  1&      - &    - &     -&   -  &   -  & E&     -   &  -  &  -  &  -  & 2.119&  G &  - \\
           &  &       &       &   &      - &    - &   -  &  -   &   -  &  &         &  -  &  -  &  -  \\
\vspace {-0.3cm}
           &  &  &  &  &  &  &  &  &  &  &  &  &  & \\
  \object{0800+472} & E&     - &   -   &  1&  20.57& 20.26& 20.02& 19.08& 19.11& Q&  0.509&  - &  - & - & 0.509&  Q &20.02\\
       &  &       &       &   &    0.08&  0.02&  0.02&  0.02&  0.02&  &       &      - &    - &  -\\
\vspace {-0.3cm}
           &  &  &  &  &  &  &  &  &  &  &  &  &  & \\
  \object{0805+406} & Q&     - &  1.780&  2&   20.89& 20.74& 20.61& 20.26& 20.31& Q&    - &  - &  - & -  & 1.8&  Q &20.61\\
       &  &       &       &   &    0.08&  0.03&  0.04&  0.03&  0.03&  &       &  -&   -&  -\\
\vspace {-0.3cm}
           &  &  &  &  &  &  &  &  &  &  &  &  &  & \\
  \object{0809+404} & G&     - &  0.551&  1&   21.63& 20.69& 19.50& 18.63& 18.53& G&  0.551& 0.47& 0.53& 0.44& 0.551&  G &19.50\\
       &  &       &       &   &    0.14&  0.03&  0.02&  0.01&  0.01&  &       & 0.15& 0.05& 0.06\\
\vspace {-0.3cm}
           &  &  &  &  &  &  &  &  &  &  &  &  &  & \\
  \object{0810+460B} & G&    -&  0.620&  2&   20.69& 20.43& 19.37& 19.01& 18.84& G&   - & 0.31& 0.41& 0.40& 0.620&  G &19.37\\
       &  &       &       &   &    0.13&  0.04&  0.03&  0.03&  0.03&  &       & 0.12& 0.08& 0.06\\
\vspace {-0.3cm}
           &  &  &  &  &  &  &  &  &  &  &  &  &  & \\
  \object{0814+441} & E&     - &   -   &  1&   23.31& 23.08& 22.95& 21.38& 21.49& G&   - & 0.12&  -  &  - &(1.2) &  G &22.95\\
       &  &       &       &   &    0.80&  0.30&  0.40&  0.15&  0.15&  &       & 0.10& -     &  -  \\
\vspace {-0.3cm}
           &  &  &  &  &  &  &  &  &  &  &  &  &  & \\
  \object{0822+394} & G&     - &  1.2$k$& 2&   -    &   -  & -    &  -   &   - & E     &  -  & -   &  -  & - & 1.2$k$&  G & - \\
       &  &       &       &   &       -&  - &  -  &  -   &  -   &     &       &  -  &  -  &  - \\
\vspace {-0.3cm}
           &  &  &  &  &  &  &  &  &  &  &  &  &  & \\
  \object{0840+424A} & E&    - &   -&  1&   25.70& 22.74& 21.35& 21.36& 20.28& G&   -  & 0.22& 0.84& 0.76& 1.0&  G &21.35\\
       &  &       &       &   &    1.62&  0.33&  0.14&  0.19&  0.19&  &       & 0.08& 0.18& 0.17\\
\vspace {-0.3cm}
           &  &  &  &  &  &  &  &  &  &  &  &  &  & \\
  \object{0856+406} & G&   24.3&  2.280&  1&     - &  - &   -  &  -  &   -   & E&    - &  - &   -  &   - & 2.280&  G & 24.3\\
       &  &       &       &   &     -  &     -  &    -  &   -  &   -   &  &       &  - &  -&  -\\
\vspace {-0.3cm}
           &  &  &  &  &  &  &  &  &  &  &  &  &  & \\
  \object{0902+416} & E&    - &   - &  1&   22.53& 22.67& 21.78& 21.08& 20.28& G&  - & 0.53& 0.77& 0.79& 1.0&  G &21.78\\
       &  &       &       &   &    0.30&  0.13&  0.08&  0.07&  0.07&  &       & 0.08& 0.13& 0.11\\
\vspace {-0.3cm}
           &  &  &  &  &  &  &  &  &  &  &  &  &  & \\
  \object{0930+389} & G&     - &  2.395&  1&     -  &   -  &  -  &    - &  -  & E&     - &  -  &  -  &  - & 2.395&  G &  - \\
       &  &       &       &   &         -  &   -  &  -  &    - &    -&  &       &    - &   - &  - \\
\vspace {-0.3cm}
           &  &  &  &  &  &  &  &  &  &  &  &  &  & \\
  \object{0935+428A} & G&   24.0&  1.291&  1&    -  &   -  &  -  &    - &   - & E&   - &  -  &   - &  -  & 1.291&  G &24.00\\
       &  &       &       &   &         -  &   -  &  -  &    - &   - &  &     &  -  &   - &  -\\
\vspace {-0.3cm}
           &  &  &  &  &  &  &  &  &  &  &  &  &  & \\
  \object{0951+422} & Q&   20.5&  1.783&  1&   21.92& 21.22& 20.91& 20.19& 20.20& Q&      -  &   - &  - & - & 1.783&  Q &20.91\\
       &  &       &       &   &    0.19&  0.04&  0.05&  0.04&  0.04&  &       &  -  &   - &  - \\
\vspace {-0.3cm}
           &  &  &  &  &  &  &  &  &  &  &  &  &  & \\
  \object{0955+390} & E&     - &    -  &  1&   22.80& 24.12& 21.45& 20.36& 19.94& G&   - & 0.52& 0.70& 0.72& 0.9&  G &21.45\\
       &  &       &       &   &    0.72&  0.87&  0.15&  0.09&  0.09&  &       & 0.16& 0.11& 0.08\\
\vspace {-0.3cm}
           &  &  &  &  &  &  &  &  &  &  &  &  &  & \\
  \object{1007+422} & E&     - &    -  &  1&   23.25& 21.76& 20.67& 19.77& 19.20& G&   -     & 0.36& 0.59& 0.57& 0.75&  G &20.67\\
       &  &       &       &   &    1.11&  0.10&  0.07&  0.04&  0.04&  &       & 0.14& 0.08& 0.06\\
\vspace {-0.3cm}
           &  &  &  &  &  &  &  &  &  &  &  &  &  & \\
  \object{1014+392} & G&     - &  0.536&  2&   26.08& 22.25& 21.33& 19.69& 18.04& G&   -     & 0.99& 0.84& 0.66& 0.536&  G &21.33\\
       &  &       &       &   &    2.13&  0.44&  0.32&  0.13&  0.13&  &       & 0.13& 0.05& 0.05\\
\vspace {-0.3cm}
           &  &  &  &  &  &  &  &  &  &  &  &  &  & \\
  \object{1016+443} & G&    19.7& 0.33R &  1&   25.17& 22.61& 21.77& 21.65& 20.88& G&   -     & 0.09& 0.55& 0.53& 0.7&  G &21.77\\
           &  &        &       &   &    1.29&  0.20&  0.15&  0.24&  0.24&  &         & 0.05& 0.31& 0.24\\
\vspace {-0.3cm}
           &  &  &  &  &  &  &  &  &  &  &  &  &  & \\
  \object{1025+390B} & G&   18.4&  0.361&  1&   20.24& 19.39& 18.27& 17.75& 17.31& G&   -    & 0.31& 0.28& 0.27& 0.361&  G &18.27\\
       &  &       &       &   &    0.07&  0.02&  0.01&  0.01&  0.01&  &       & 0.08& 0.04& 0.04\\
\vspace {-0.3cm}
           &  &  &  &  &  &  &  &  &  &  &  &  &  & \\
  \object{1027+392} & E&     -  &   -   &  1&   23.64& 22.42& 21.01& 19.95& 19.59& G&   -     & 0.56& 0.67& 0.59& 0.8&  G &21.01\\
       &  &       &       &   &    0.78&  0.12&  0.05&  0.03&  0.03&  &       & 0.05& 0.05& 0.05\\
\vspace {-0.3cm}
           &  &  &  &  &  &  &  &  &  &  &  &  &  & \\
  \object{1044+454} & G&   24.8& 4.1$k$&  1&     -  &   -  &   -  &   -  &   -  & E&   -   &  -  &  -  &  -  & 4.1$k$&  G &24.80\\
           &  &       &       &   &     -  &   -  &   -  &   -  &   -  &  &       &  -  &  -  &  -  \\
\vspace {-0.3cm}
           &  &  &  &  &  &  &  &  &  &  &  &  &  & \\
  \object{1049+384} & G&   20.9&  1.018&  1&   21.61& 21.34& 20.77& 20.27& 19.72& G&   -   & 0.43& 0.69& 0.66& 1.018&  G &20.77\\
           &  &       &       &   &    0.16&  0.04&  0.04&  0.04&  0.04&  &       & 0.05& 0.12& 0.12\\
\vspace {-0.3cm}
           &  &  &  &  &  &  &  &  &  &  &  &  &  & \\
  \object{1055+404A} & E&    - &  -    &  1&   23.12& 23.19& 23.60& 22.62& 21.60& Q?& -    &  -  &  -  &  -  &  -   &  Q? &23.60\\
            &  &       &       &   &    0.57&  0.22&  0.45&  0.23&  0.23&  &      &  -  &  -  &  - \\
\vspace {-0.3cm}
           &  &  &  &  &  &  &  &  &  &  &  &  &  & \\
  \object{1128+455} & G&   18.7&  0.404&  1&   22.02& 20.92& 19.61& 19.06& 18.65& G&   -   & 0.30& 0.31& 0.34& 0.404&  G &19.61\\
           &  &       &       &   &    0.29&  0.04&  0.02&  0.02&  0.02&  &       & 0.07& 0.08& 0.09\\
\vspace {-0.3cm}
           &  &  &  &  &  &  &  &  &  &  &  &  &  & \\
  \object{1136+420}& G&   21.7&  0.829&  1&   23.95& 22.75& 21.55& 20.70& 19.93& G&    -   & 0.43& 0.58& 0.64& 0.829&  G &21.55\\
          &  &       &       &   &    1.33&  0.21&  0.10&  0.07&  0.07&  &        & 0.09& 0.10& 0.11\\
\vspace {-0.3cm}
           &  &  &  &  &  &  &  &  &  &  &  &  &  & \\
  \object{1141+466} & G&   15.5&  0.116&  2&   18.76& 16.48& 15.48& 14.95& 14.58& G&  0.116& 0.11& 0.10& 0.07& 0.116&  G &15.48\\
           &  &       &       &   &    0.08&  0.01&  0.01&  0.01&  0.01&  &       & 0.03& 0.02& 0.01\\
\vspace {-0.3cm}
           &  &  &  &  &  &  &  &  &  &  &  &  &  & \\
  \object{1143+456} & G&   24.0&  0.762&  1&     -  &   -  &   -  &   -  &   -  & E&   -   &  -  &  -  &  -  & 0.762&  G &24.00\\
           &  &       &       &   &     -  &   -  &   -  &   -  &   -  &  &       &  -  &  -  &  -  \\
\vspace {-0.3cm}
           &  &  &  &  &  &  &  &  &  &  &  &  &  & \\
  \object{1157+460} & G&   21.3&  0.743&  1&   22.65& 22.17& 21.49& 20.76& 20.55& G&   -   & 0.40& 0.70& 0.64& 0.743&  G &21.49\\
           &  &       &       &   &    0.48&  0.10&  0.08&  0.07&  0.07&  &       & 0.14& 0.11& 0.12\\
\vspace {-0.3cm}
           &  &  &  &  &  &  &  &  &  &  &  &  &  & \\
  \object{1159+395} & G&   23.4&  2.370&  1&   23.78& 23.85& 23.30& 22.24& 20.03& G&   -   & 0.62& -& -& 2.370&  G &23.30\\
           &  &       &       &   &    1.43&  0.51&  0.55&  0.34&  0.34&  &       & 0.02& -& -\\
\vspace {-0.3cm}
           &  &  &  &  &  &  &  &  &  &  &  &  &  & \\
  \object{1201+394} & G&   19.0&  0.445&  1&   24.05& 21.21& 19.51& 18.70& 18.36& G&  0.445& 0.44& 0.44& 0.44& 0.445&  G &19.51\\
           &  &       &       &   &    2.64&  0.09&  0.03&  0.02&  0.02&  &       & 0.02& 0.03& 0.02\\
\vspace {-0.3cm}
           &  &  &  &  &  &  &  &  &  &  &  &  &  & \\
  \object{1204+401} & G&   23.0&  2.066&  1&     -  &   -  &   -  &   -  &   -  & E&   -   &  -  &  -  &  -  & 2.066&  G &23.00\\
           &  &       &       &   &     -  &   -  &   -  &   -  &   -  &  &       &  -  &  -  &  -  \\
\vspace {-0.3cm}
           &  &  &  &  &  &  &  &  &  &  &  &  &  & \\
  \object{1212+380} & G&   24.0& 1.5$k$&  1&     -  &   -  &   -  &   -  &   -  & E&   -   &  -  &  -  &  -  & 1.5$k$&  G & 24.0\\
           &  &       &       &   &     -  &   -  &   -  &   -  &   -  &  &       &  -  &  -  &  -  \\
\vspace {-0.3cm}
           &  &  &  &  &  &  &  &  &  &  &  &  &  & \\
  \object{1216+402} & G&   22.1&  0.756&  1&   22.56& 22.54& 21.65& 21.35& 20.35& G&   -   & 0.34& 0.64& 0.68& 0.756&  G &21.65\\
           &  &       &       &   &    0.80&  0.25&  0.17&  0.19&  0.19&  &       & 0.04& 0.21& 0.19\\
\vspace {-0.3cm}
           &  &  &  &  &  &  &  &  &  &  &  &  &  & \\
  \object{1220+408} & G&   23.2&   -   &  1&     -  &   -  &   -  &   -  &   -  & E&   -   &  -  &  -  &  -  &  -   &  G &23.20\\
           &  &       &       &   &     -  &   -  &   -  &   -  &   -  &  &       &  -  &  -  &  -  \\
\vspace {-0.3cm}
           &  &  &  &  &  &  &  &  &  &  &  &  &  & \\
  \object{1225+442} & G&   18.2&  0.348&  2&   22.53& 20.31& 18.71& 18.31& 17.80& G&   -   & 0.26& 0.26& 0.27& 0.348&  G &18.71\\
           &  &       &       &   &    0.54&  0.04&  0.01&  0.01&  0.01&  &       & 0.04& 0.02& 0.02\\
\vspace {-0.3cm}
           &  &  &  &  &  &  &  &  &  &  &  &  &  & \\
  \object{1233+418} & G&   18.4& 0.25$R$&  1&   20.99& 19.18& 17.86& 17.35& 16.83& G&  -   & 0.18& 0.20& 0.20& 0.25 &  G &17.86\\
           &  &       &       &   &    0.37&  0.02&  0.01&  0.01&  0.01&  &       & 0.04& 0.02& 0.01\\
\vspace {-0.3cm}
           &  &  &  &  &  &  &  &  &  &  &  &  &  & \\
  \object{1241+411} & G&   17.7&  0.259&  1&   19.94& 18.63& 17.55& 17.11& 16.80& G&  0.259& 0.20& 0.20& 0.20& 0.259&  G &17.55\\
           &  &       &       &   &    0.08&  0.01&  0.01&  0.01&  0.01&  &       & 0.03& 0.02& 0.02\\
\vspace {-0.3cm}
           &  &  &  &  &  &  &  &  &  &  &  &  &  & \\
  \object{1242+410} & Q&   19.7&  0.813&  1&   21.00& 20.39& 20.14& 19.80& 19.30& Q&   -   &  -  &  -  &  -  & 0.813&  Q &20.14\\
           &  &       &       &   &    0.10&  0.02&  0.02&  0.02&  0.02&  &       &  -  &  -  &  -  \\
\vspace {-0.3cm}
           &  &  &  &  &  &  &  &  &  &  &  &  &  & \\
  \object{1314+453A} & G&   21.8&  1.544&  1&   22.82& 22.09& 21.70& 21.50& 21.35& G&  -   & 0.06& 0.23& 0.53& 1.544&  G &21.70\\
            &  &       &       &   &    0.40&  0.07&  0.07&  0.09&  0.09&  &       & 0.06& 0.31& 0.35\\
\vspace {-0.3cm}
           &  &  &  &  &  &  &  &  &  &  &  &  &  & \\
  \object{1340+439} & E&     - &    -  &  1&   24.33& 23.27& 22.91& 22.94& 21.98& Q?&  -   &  -   &  -  &  -   &  - &  Q? &22.91\\
           &  &       &       &   &    0.73&  0.18&  0.17&  0.27&  0.27&  &       &  -   &  -  &  -  \\
\vspace {-0.3cm}
           &  &  &  &  &  &  &  &  &  &  &  &  &  & \\
  \object{1343+386} & Q&   17.5&  1.844&  1&   18.40& 18.24& 18.06& 17.60& 17.53& Q&  1.844&  -  &  -  &  -  & 1.844&  Q &18.06\\
           &  &       &       &   &    0.02&  0.01&  0.01&  0.01&  0.01&  &       &  -  &  -  &  -\\
\vspace {-0.3cm}
           &  &  &  &  &  &  &  &  &  &  &  &  &  & \\
  \object{1441+409} & E&     - &   -   &  1&   23.52& 23.00& 23.77& 22.55& 21.92& G&   -   & 0.15&  -  &  -  & (2.0)   &  G &23.77\\
           &  &       &       &   &    0.80&  0.17&  0.49&  0.29&  0.29&  &       & 0.21&  -  &  -  \\
\vspace {-0.3cm}
           &  &  &  &  &  &  &  &  &  &  &  &  &  & \\
  \object{1445+410} & G&   17.9&  0.180&  1&   20.39& 19.02& 17.87& 17.38& 17.05& G&  0.195& 0.21& 0.20& 0.20& 0.195&  G &17.87\\
           &  &       &       &   &    0.08&  0.01&  0.01&  0.01&  0.01&  &       & 0.03& 0.03& 0.03\\
\vspace {-0.3cm}
           &  &  &  &  &  &  &  &  &  &  &  &  &  & \\
  \object{1458+433} & G&   22.3&  0.927&  1&   22.66& 21.94& 20.97& 19.92& 19.40& G&   - & 0.63& 0.77& 0.67& 0.927&  G &20.97\\
           &  &       &       &   &    1.17&  0.24&  0.16&  0.11&  0.11&  &       & 0.10& 0.07& 0.08\\
\hline
\end{longtable}  

Columns are:

\noindent
Col. 1: B3 source name.

\noindent
Cols. 2-5: earlier data: optical identification (G galaxy, Q quasar, E empty
field), red magnitude and redshift for
identifications prior to this work (the $K$ and $R$ subscripts are for 
redshifts 
derived from the $K$- or $R$ bands respectively) and references: (1 
from Paper I, 2  from NED).

\noindent
Col. 6-10: SDSS magnitudes (first line) and errors (second line).

\noindent
Cols.  11 and 12: SDSS optical identification and spectroscopic redshift 
(z$_{\rm sp}$). 

\noindent
Col. 13-15: SDSS photometric redshifts, z$_{\rm ph}$, z$_{\rm ph2}^{\rm cc2}$ 
and z$_{\rm ph2}^{\rm d1}$ (Sect. \ref{phrs}).

\noindent
Cols. 16-18: final status of the identification: z$_{\rm sp}$  (3 decimal 
digits) or z$_{\rm ph2}$ (1 or 2 decimal digits) corrected {\bz for the offset}
discussed in Sect. \ref{phrs}, in parenthesis two redshifts estimated from 
$r$, $i$, $z$ magnitudes (Sect. \ref{Hdia}); optical classification 
(Q? quasar candidate); SDSS $r$ magnitude or, in its absence, 
m$_{\rm R}$ from the literature. 

\twocolumn

\begin{table}[h]
\caption{Unidentified radio sources in the SDSS B3-VLA CSS sample\hfill}
\medskip
 \begin{tabular}{lll}
\hline
\hline
\smallskip
\object{0703+468} &   \object{1008+423} & \object{1217+427}\\
\smallskip
\object{0722+393A} &  \object{1039+424} & \object{1350+432}\\
\smallskip
\object{0729+437} &   \object{1133+432} & \object{1432+428B} \\
\smallskip
\object{0748+413B} &   \object{1136+383} & \object{1449+421}\\
\hline
\end{tabular}
\label{EF}
\end{table}

\subsection{Notes to Table \ref{B3tab1} and Table \ref{EF}}
\label{notes}

\noindent  {\bf \object{0701+392}}:  Spectroscopic redshift from \citet{Lah91}.
The old red magnitude is a visual estimate from the Palomar Sky Survey 
Prints \citep{vig1}.

\noindent {\bf \object{0703+468}}:   \citet{St93} report a possible 
identification with a 23 mag. object.

\noindent  {\bf \object{0744+464}}:  Redshift by \citet{McCa}
from Ly$_{\alpha}$, [\ion{C}{iv}]1549 and \ion{He}{ii}1640. McCarthy 
classified the optical 
             object as Broad Line Radio Galaxy (see Sect. \ref{Hdia}).   
             
\noindent  {\bf \object{0754+396}}:  Redshift from Vigotti (priv.comm.).

\noindent  {\bf \object{0800+472}}:  Classified as quasar in the SDSS, from 
the optical spectrum and  
point-like appearance; in the Hubble diagram (Fig.~\ref{H-dia}) it is located 
in the galaxy area.    

\noindent  {\bf \object{0805+406}}:  Photometric redshift by \citet{Rich04}. 
A range of values 1.5 -1.95 
is quoted, with 0.935 probability.    

\begin{figure*}[ht]
{\includegraphics[width=16.0cm]{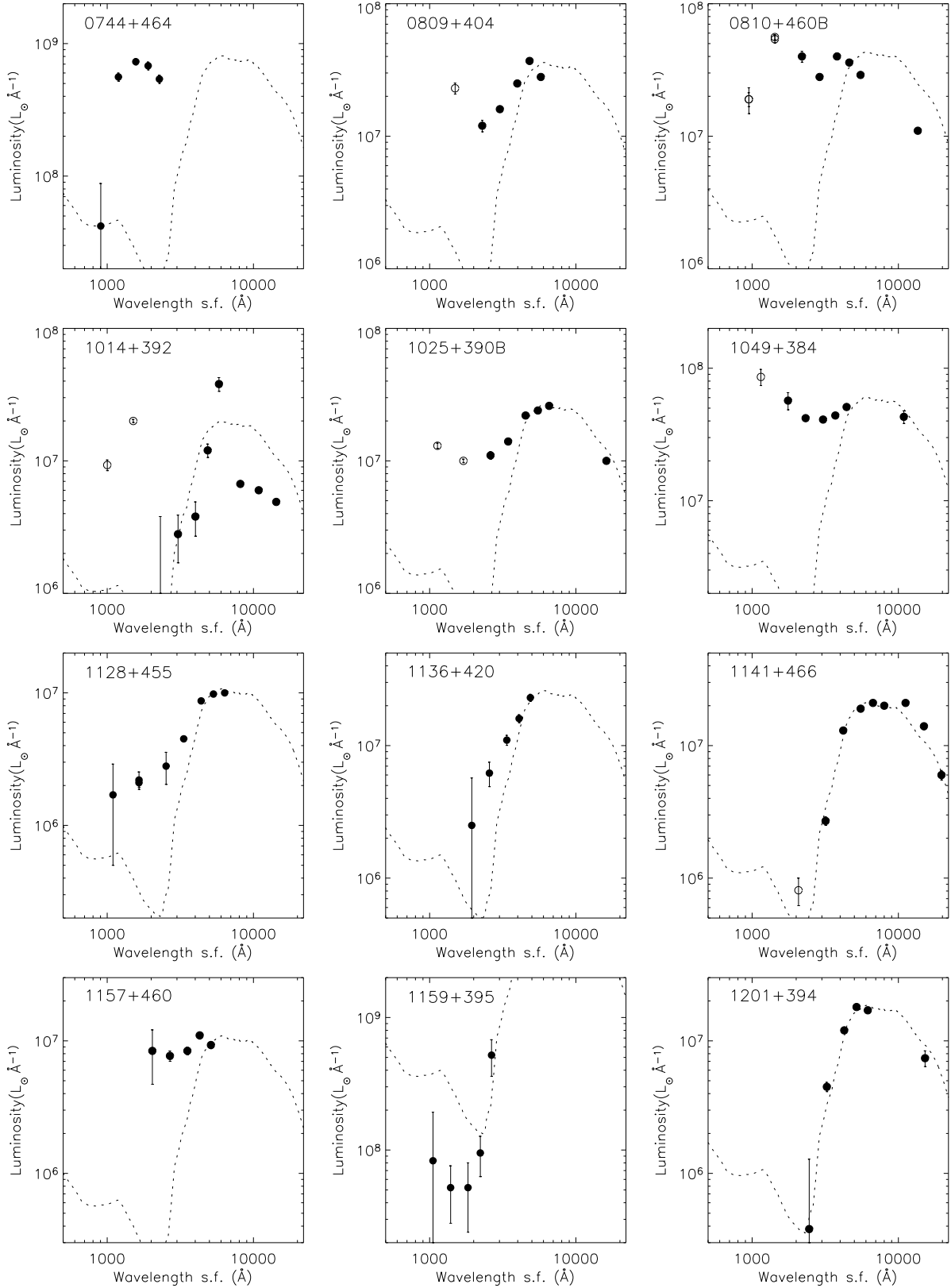}}
\label{plot_sp}
\caption{ Individual SEDs for galaxies with {\it spectroscopic}
redshift. Empty circles are GALEX data. The dotted line represents a B\&C 
model normalized to the long-wavelength SDSS data {\it only} 
(see notes to \object{1014+392}). } 
\end{figure*}

\noindent  {\bf \object{0809+404}}:  The spectra in \citet{vig2}
and in the  SDSS show strong [\ion{O}{iii}]5007 and [\ion{O}{ii}]3727
lines. \citet{vig2} say that the object is point-like, {\ba but} 
in the SDSS it is 
{\ba clearly}  extended. Also the UV-O-SED in Fig. \ref{plot_sp}.1 clearly
supports the galaxy 
classification. In the magnitude-redshift diagram  (Fig.~\ref{H-dia}) it is 
located at the separation 
between quasars and galaxies.    

\noindent  {\bf \object{0810+460B}}: Identification and redshift from 
             \citet{Cr06}. Strong [\ion{O}{ii}]3727, [\ion{Ne}{iii}]3869,
             H$_{\gamma}$, [\ion{O}{iii}] 4959,5007 lines are present. The 
             [\ion{O}{ii}/[\ion{O}{iii}] ratio (2.8) shows a low level of 
             excitation consistent with a shock heating mechanism. 

\noindent  {\bf \object{0822+394}}:  A photometric redshift z=1.18 (1.0-1.4) 
             is reported by \citet{Li89}; (in a note  z $\approx$ 1.2 is 
             given). Later authors  (e.g. \cite{Law95} and  \cite{Ro98})
             report z=1.21, without specifying that it is a photometric 
             redshift.            

\noindent  {\bf \object{0856+406}}:  Redshift by \citet{PD}
             from $H_\alpha$ line.
             It is erroneously reported as a QSO in NED. Old red magnitude 
             from Vigotti (private communication).

\noindent  {\bf \object{0930+389}}: Redshift  by \citet{ER96}
             from $H_\alpha$ line.

\noindent  {\bf \object{0935+428A}}: Redshift and $r$ magnitude by 
             \citet{Th94}. Emission lines of \ion{C}{ii}]2326 
             and \ion{Mg}{ii}2798. 

\noindent  {\bf \object{0951+422}}:  Redshift by \citet{Fal98}, from 
             \ion{Si}{iv}, \ion{C}{iv}, [\ion{C}{iii}] and \ion{Mg}{ii} lines. 

\noindent  {\bf \object{1014+392}}:  The reported optical identification and 
             the redshift are from  \citet{gandhi}, 
             who classify this object as a strongly obscured type 2 quasar 
             with an intrinsic 2-10 kev luminosity 
             L$_{2-10} = 5~10^{44}$~h$^{-2}_{0.7}$ erg~s$^{-1}$. 
             [\ion{O}{ii}]3727  and [\ion{O}{iii}]5007 lines, 
             with ratio $\approx$ 2, are present. Near-IR magnitudes are 
             given in \cite{gandhi04}.

\noindent  {\bf \object{1016+443}}:  The identification reported in Paper 1 
             (m$_{\rm R}$ = 19.7, z = 0.33)
             is rejected because of radio-optical  positional disagreement. 
             The new identification  reported here is from the SDSS.

\noindent {\bf \object{1025+390B}}: $r$ magnitude and redshift from 
             \citet{Alli88}.
             [\ion{Ne}{iii}]3869 and [\ion{O}{iii}]4959,5007 are seen.

\noindent  {\bf \object{1044+454}}:  Photometric redshift from K-mag. from 
             Vigotti (private communication).

\noindent  {\bf \object{1049+384}}:   Redshift and $r$ magnitude from 
             \citet{Alli88}. High excitation narrow lines system (large 
             [\ion{O}{iii}]/[\ion{O}{ii}] ratio) and broad
             \ion{Mg}{ii}2799 emission.   

\noindent  {\bf \object{1055+404A}}:  Point-like in the SDSS and therefore 
             quasar candidate.

\noindent  {\bf \object{1128+455}}:  Redshift from \citet{vig2}. 
             [\ion{O}{ii}]3727 and [\ion{O}{iii}]5007 present.

\noindent  {\bf \object{1136+420}}:  Redshift and red magnitude from Vigotti 
             (private communication).

\noindent  {\bf \object{1141+466}}:  Redshift from SDSS4. Strong 
             [\ion{O}{ii}]3727 and H$_\alpha$, very weak [\ion{O}{iii}]5007.

\noindent  {\bf \object{1143+456}}:  $r$ magnitude and redshift, from 
             [\ion{O}{ii}]3727 line, by \citet{Th94}.

\noindent  {\bf \object{1157+460}}:  $r$ magnitude and redshift, from 
             [\ion{O}{ii}]3727 line,  by \citet{Max}.

\noindent  {\bf \object{1159+395}}:  Redshift by Vigotti (private 
             communication);  $r$ magnitude from \citet{Max}.

\noindent  {\bf \object{1201+394}}:  The SDSS spectrum shows [\ion{O}{ii}]3727
             and [\ion{O}{iii}]5007 lines, with low excitation status. 

\noindent  {\bf \object{1204+401}}:  Redshift and $r$ magnitude by 
             \citet{Th94}; Ly$_\alpha$, \ion{C}{iv}, [\ion{C}{iii}] and 
             \ion{C}{ii}] lines are seen.

\noindent  {\bf \object{1212+380}}:  K-mag. photometric redshift from Vigotti 
             (private communication).

\noindent  {\bf \object{1216+402}}:  $r$ magnitude and Redshift, from 
             [\ion{O}{ii}]3727 and [\ion{Ne}{iii}]3869, by \citet{Th94}.

\noindent  {\bf \object{1217+427}}:  \citet{Max} report a possible 
             identification  with $g, r \ge$ 23.0 and $i \ge$ 22.5.

\noindent  {\bf \object{1225+442}}:  Redshift from \citet{Wegn03}; the optical
             object is  classified as Seyfert~2. Red magnitude from Vigotti 
             (private communication).

\noindent  {\bf \object{1241+411}}:  \citet{vig2} classify the optical object 
             as Seyfert. In  the SDSS it has an extended image  and, according
             to strong broad lines in the spectrum, it is classified as QSO. 
             In the m-z diagram (Fig.~\ref{H-dia}) it is located at the 
             separation between quasars and galaxies. [\ion{O}{ii}]3727 and 
             [\ion{O}{iii}]5007 present.
  
\noindent  {\bf \object{1242+410}}:  Redshift by \citet{Xu94} and  
             \citet{vig2}. [\ion{C}{iii}]1909, \ion{Mg}{ii}2798, H$_{\beta}$, 
             [\ion{O}{ii}]3727, [\ion{O}{iii}]4959,5007 are seen.

\noindent  {\bf \object{1314+453A}}: Redshift, red magnitude and optical 
             classification from Vigotti (private communication).
             The UV-O-SED is consistent with a quasar classification
             (Sect.\ref{SEDG}).
             However the object is clearly  extended in the SDSS image.

\noindent  {\bf \object{1340+439}}:   Point-like in the SDSS and therefore 
             quasar candidate; reported as G in NED, perhaps from \citet{Max}
             who report an object of 23.4, 22.8, and 22.1 in the $g, r$ and 
             $i$ bands. 

\noindent  {\bf \object{1343+386}}:  Redshift and red magnitude from 
             \cite{vig2}. The SDSS spectrum shows strong and broad 
             [\ion{Si}{iv}], \ion{C}{iv}, [\ion{C}{iii}] and \ion{Mg}{ii}.   

\noindent  {\bf \object{1350+432}}:  The identification in Paper I is now 
             rejected because of radio-optical  positional disagreement.

\noindent  {\bf \object{1432+428B}}: \citet{Max} report a possible 
             identification  with $g, r \ge$ 23.0 and $i \ge$ 22.5.
               
\noindent  {\bf \object{1441+409}}: \citet{Max} report a possible 
             identification  with $g, r, i $ = 23.0, 22.6 and 22.1.
             However the positional agreement is obscure.

\noindent  {\bf \object{1445+410}}:  Redshift from SDSS4. H$_\alpha$, 
             [\ion{O}{ii}]3727 and weak [\ion{O}{iii}]5007 lines.
  
\noindent  {\bf \object{1458+433}}:  Redshift and red magnitude from Vigotti 
             (private communication).

\onecolumn

\section{UV-GALEX and near-IR data}
\label{GalexIR}

\begin{table} [h]
\begin{center}
\caption{ GALEX and near-IR data \hfill}
 \begin{tabular}{lcccl|cccc}
\hline
Name              &Id &  m$_{NUV}$     & m$_{FUV}$      &Survey&  m$_J$        & m$_H  $       &  m$_K$        &Ref     \\
\hline
\hline
\vspace {0.05cm}
\object{0701+392} & Q & 21.93$\pm$0.14&               & MIS &               &               &               &\\
\vspace {0.05cm}
\object{0800+472} & Q & 20.38$\pm$0.06& 20.70$\pm$0.24& AIS &               &               &               & \\
\vspace {0.05cm}
\object{0805+406} & Q & 22.49$\pm$0.24&               & MIS &               &               &               & \\
\vspace {0.05cm}
\object{0809+404} & G & 21.99$\pm$0.15&               & MIS &               &               &               & \\
\vspace {0.05cm}
\object{0810+460B}& G & 21.41$\pm$0.07& 23.46$\pm$0.27& NGS &               &               & 16.00$\pm$0.03 & 1\\
\vspace {0.05cm}
\object{1014+392} & G & 23.39$\pm$0.14& 23.49$\pm$0.11& GI  &18.30$\pm$0.004&17.30$\pm$0.02 & 16.47$\pm$0.02 & 1\\
\vspace {0.05cm}
\object{1025+390B}& G & 21.31$\pm$0.04& 21.93$\pm$0.05& GI  &               &               & 14.48$\pm$0.05 & 1\\
\vspace {0.05cm}
\object{1049+384} & G & 22.24$\pm$0.17&               & MIS &               &               & 16.12$\pm$0.12 & 1\\
\vspace {0.05cm}
\object{1128+455} & G & 23.36$\pm$0.26& 24.00$\pm$0.60& NGS &               &               &                &\\
\vspace {0.05cm}
\object{1141+466} & G & 21.20$\pm$0.30& 21.26$\pm$0.42& AIS &12.92$\pm$0.04 &12.24$\pm$0.06 & 12.11$\pm$0.09 & 2\\
\vspace {0.05cm}
\object{1201+394} & G &     und.      &     und.      & AIS &               &               & 15.46$\pm$0.15 & 1\\
\vspace {0.05cm}
\object{1225+442} & G & 23.30$\pm$0.37& 21.77$\pm$0.20& NGS &               &               &                & \\
\vspace {0.05cm}
\object{1241+411} & G & 21.54$\pm$0.15& 21.97$\pm$0.20& NGS &               &               &                & \\
\vspace {0.05cm}
\object{1242+410} & Q & 20.92$\pm$0.13&     und.      & AIS &               &               &               & \\
\vspace {0.05cm}
\object{1343+386} & Q & 20.00$\pm$0.15&     und.      & AIS & 16.42$\pm$0.09& 16.45$\pm$0.20& 15.34$\pm$0.13 & 2\\
\vspace {0.05cm}
\object{1445+410} & G & 21.96$\pm$0.16&               & MIS &               &               &                &\\
\hline
\vspace {0.05cm}
\object{3C186}    & Q & 18.56$\pm$0.01& 20.22$\pm$0.04& MIS & 16.84$\pm$0.06& 16.46$\pm$0.05& 15.68$\pm$0.06 & 1\\  
\vspace {0.05cm}
\object{3C190}    & Q & 21.41$\pm$0.31&   und.        & AIS & 17.47$\pm$0.04& 16.63$\pm$0.03& 15.84$\pm$0.04 & 1\\  
\vspace {0.05cm}
\object{1153+31}  & Q & 19.32$\pm$0.02& 19.63$\pm$0.03& GI  & 16.82$\pm$0.16& 16.01$\pm$0.15& 15.36$\pm$0.16 & 2\\  
\vspace {0.05cm}
\object{3C237   } & G & 21.89$\pm$0.28&   und.        & AIS &               &               &               & \\
\vspace {0.05cm}
\object{3C241}    & G &    und.       &   und.        & AIS & 18.86$\pm$0.20& 18.16$\pm$0.34& 17.29$\pm$0.15 & 1\\
\vspace {0.05cm}
\object{3C268.3}  & G & 23.31$\pm$0.40&   und.        & MIS &               &               &                & \\  
\vspace {0.05cm}
\object{3C277.1}  & Q &               &               &       & 16.48$\pm$0.10& 16.15$\pm$0.18& 14.98$\pm$0.13 & 2\\
\vspace {0.05cm}
\object{3C286}    & Q & 17.68$\pm$0.04& 18.91$\pm$0.10& AIS   & 15.94$\pm$0.06& 15.64$\pm$0.13& 14.63$\pm$0.06 & 2\\
\vspace {0.05cm}
\object{3C287}    & Q &               &               &       & 16.44$\pm$0.14& 16.06$\pm$0.19& 15.10$\pm$0.15 & 2\\
\vspace {0.05cm}
\object{1358+624}  & G & 24.24$\pm$0.39& 23.92$\pm$0.22& GI    &               &               &                &\\  
\vspace {0.05cm}
\object{OQ208}    & G & 18.28$\pm$0.01& 20.15$\pm$0.05& GI    & 12.91$\pm$0.04& 12.20$\pm$0.05& 11.52$\pm$0.05 & 2\\  
\vspace {0.05cm}
\object{1442+101} & Q & 20.22$\pm$0.04& 21.01$\pm$0.09& MIS   & 16.80$\pm$0.20& 15.98$\pm$0.18& 15.77$\pm$0.32 & 2\\  
\vspace {0.05cm}
\object{1509+054} & G & 21.29$\pm$0.16& 22.37$\pm$0.43& MIS   & 14.23$\pm$0.08& 13.18$\pm$0.06& 12.08$\pm$0.05 & 2\\
\vspace {0.05cm}
\object{3C298}    & Q & 18.20$\pm$0.01& 19.17$\pm$0.04& MIS   & 15.13$\pm$0.04& 14.48$\pm$0.05& 15.19$\pm$0.06 & 2\\  
\vspace {0.05cm}
\object{3C318}    & Q & 22.57$\pm$0.39&   und.        & AIS   &               &               & 16.96$\pm$0.15 & 1\\  

\hline
\end{tabular}
\label{GalexIR_Tab}
\end{center}
\smallskip
 G = Galaxy, Q = Quasar; Ref: 1 = UKIRT (references in NED, 1014+392 
\citep{gandhi04}), 2 = 2MASS; Blank = not observed, und. = undetected.
\end{table}

\newpage
\section{The combined CSOs/MSOs Quasar sample}
\label{QSS}
\begin{table} [h]
\begin{center}
\caption{The combined quasar sample \hfill}
 \begin{tabular}{lccc|cr}
\hline
Name        & ~~ z&log P$_{1.4~{\rm GHz}}$ &LS~~ & $\alpha^1_\lambda$~~~~~~~~~$\alpha^2_\lambda$          & L $^*_{0.30}$~~~~ \\
            &     &      W~Hz$^{-1}$~    & kpc & $\ga 0.30\mu$m~~~$\la 0.30\mu$m  & 10$^{7}\rm L_{\sun}$ \AA$^{-1}$  \\
\hline
\hline
\vspace {0.05cm}
\object{0701+392}    & 1.283 &  27.60       & 14.8  &  ~~---~~~~~~~~~~ 0.50      & 39.0 ~~~~\\
\vspace {0.05cm}
\object{0800+472}    & 0.509 &  26.90       &  6.0  &~$\approx$ 0~~~~~~~~~~\, 1.9  &    ~1.7 ~~~~\\
\vspace {0.05cm}
\object{0805+406}    & 1.8   &  27.90       & 20.4  &  ~~---~~~~~~~~~~ 1.56      &   20.0 ~~~~\\
\vspace {0.05cm}
\object{0951+422}    & 1.783 &  27.87       & 16.6  &  ~~---~~~~~~~~~~ 0.24      &   21.0 ~~~~\\
\vspace {0.05cm}
\object{1242+410}    & 0.813 &  27.70       &  0.52 &~$\approx$ 0~~~~~~~~~~\, 1.5  &    4.37 ~~~~\\
\vspace {0.05cm}
\object{1343+386}    & 1.844 &  28.23       &  1.1  & ~1.80~~~~~~~~~ 1.06         &  245.0 ~~~~\\
\vspace {0.05cm}
\object{3C\,186}     & 1.067 &  27.82       & 17.9  &     1.70                   &   64.6 ~~~~\\
\vspace {0.05cm}
\object{3C\,190}     & 1.194 &  28.29       & 32.0  &  \,0.85~~~~~~~~~ $\approx$~0 &   20.4 ~~~~\\
\vspace {0.05cm}
\object{1153+31}     & 0.417 &  27.18       &  6.1  &  ~0.95~~~~~~~~~ 1.60       &    3.37 ~~~~\\
\vspace {0.05cm}
\object{1225+36}     & 1.973 &  28.66       &  0.46 &  ~~---~~~~~~~~~~ 0.77      &   12.0 ~~~~\\
\vspace {0.05cm}
\object{3C\,277.1}   & 0.320 &  26.85       &  8.0  &1.60~~~~~~~~~~\, ---       &    5.0 ~~~~\\
\vspace {0.05cm}
\object{3C\,286}     & 0.849 &  28.63       &  0.5  &       1.61                 &   72.4 ~~~~\\
\vspace {0.05cm}
\object{3C\,287}     & 1.06  &  28.54       &  1.0  &       1.30                 &   53.7 ~~~~\\
\vspace {0.05cm}
\object{3C\,298}     & 1.437 &  28.79       & 12.45 &  ~1.55~~~~~~~~~ 0.81       &  416.9 ~~~~\\
\vspace {0.05cm}
\object{3C\,318}     & 1.570 &  28.54       &  5.99 &  ~1.70~~~~~~~~~ 0.00       &   39.9 ~~~~\\    
\vspace {0.05cm}
\object{1442+10}     & 3.53  &  28.90       &  0.14 &  ~1.65~~~~~~~~~ 0.31       & 1698.0 ~~~~\\
\vspace {0.05cm}
\object{1200+045}    & 1.211 &  27.90       &  1.45 &  \,~~---~~~~~~~~~~ 1.41      &   45.71 ~~~~\\    
\hline
\end{tabular}
\label{QssTab}
\end{center}
\smallskip
The SEDs of 3C186, 3C286 and 3C287 are well fitted by  single power laws, 
whose $\alpha_{\lambda}$ are  reported in the table.
\end{table}

\newpage
\section{The CSO/MSO composite SDSS galaxy sample} 
\label{Galaxies}

\begin{table*}[h]
\tabcolsep 0.6mm
\begin{center}
\caption{The CSO/MSO SDSS galaxy sample. Tabulated values are from fits 
without GALEX data. \hfill}
 \begin{tabular}{llcrccr|lcc|lcr}
\hline

 &\multicolumn{6}{c|}{Obs. data}&\multicolumn{3}{c|}{YSP models}&\multicolumn{3}{c}{P.L. models}\\
\hline
Name     & ~~~ z & P$_{1.4~\rm{GHz}}$ &LS~   &L$_{0.30}$~ & L$_{0.45}$~ &R$^{0.30}_{0.45}$& Age$_{\rm {YSP}}$ & M$_{\rm {YSP}}$ & M$_{\rm {OSP}}$ & $\alpha_\lambda$ &{L$^*$}$_{0.30}$ & M$_{\rm {OSP}}$~~~\\
     &      &($10^{27}$W\,Hz$^{-1}$)& kpc~ &  10$^7$L$_\odot$A$^{-1}$ & 10$^7$L$_\odot$A$^{-1}$ &               & Gyr&10$^{11}$M$_\odot$&10$^{11}$M$_\odot$& & 10$^7$L$_\odot$A$^{-1}$&10$^{11}$M$_\odot$\\
\hline\hline\\
0744+464  &  2.926&29.00& 11.04&   $-$  &  $-$ &   $-$  &       ~~~~~~0.1    &    ~~0.21         &      $-$      &   $-$  &   $-$ &   $-$~~~~\\
\vspace {0.05cm}
{\it 0809+404}  &  0.551& 1.20&  7.68&  1.60& 3.7&  0.43&   0.30 $-$ 0.6$^{b1}$&  0.300 $-$ 1.00&   6.7 $-$  0.7& 1.00& 1.00&  13.7~~~~\\
\vspace {0.05cm}
{\it 0810+460B} &  0.620& 1.70&  4.27&  2.90& 3.6&  0.79&   0.06 $-$  0.1$^a$&  0.095 $-$ 0.17&   7.5 $-$  6.6&  1.20$^e$& 2.60&   5.2~~~~\\
\vspace {0.05cm}
{\it 1014+392}  &  0.536& 1.40& 38.50&  0.28& 0.9&  0.31&   0.03 $-$ 0.1  & 0.008 $-$ 0.05 & 2.9 $-$ 2.6 &    $-$  &  $-$ &     $-$~~~~\\
\vspace {0.05cm}
{\it 1025+390B} &  0.361& 0.27& 16.07&  1.20& 2.2&  0.55&   0.06 $-$ 0.1$^{b2}$&  0.038 $-$ 0.07&   8.9 $-$  8.5&  1.00& 1.10&   8.9~~~~\\
\vspace {0.05cm}
{\it 1049+384}  &  1.018& 3.10&  0.81&  4.10& 5.1&  0.80&   0.06 $-$  0.1$^a$&  0.100 $-$ 0.20&  15.0 $-$ 12.4&  0.80& 3.40&  11.9~~~~\\
\vspace {0.05cm}
{\it 1128+455}  &  0.404& 1.10&  4.85&  0.40& 0.9&  0.44&   0.06 $-$ 0.1$^c$&  0.010 $-$ 0.02&   4.1 $-$  3.8& 0.80& 0.31&   3.8~~~~\\
\vspace {0.05cm}
1136+420  &  0.829& 1.30&  7.61&  0.90& 2.2&  0.41&   0.30 $-$ 0.6&  0.140 $-$ 0.55&   ~~5.8 $-$ (1.3)&  0.70& 0.47&   8.4~~~~\\
\vspace {0.05cm}
1141+466  &  0.116& 0.03& 16.91&  0.25& 1.5&  0.17&        ~~~~~~~$-$       &        $-$       &     11.5     &    $-$ &   $-$ &    $-$~~~~\\
\vspace {0.05cm}
1157+460  &  0.743& 2.70&  5.85&  0.80& 1.0&  0.80&   0.10 $-$ 0.3&  0.050 $-$ 0.18&   ~~1.8 $-$  (0.6)&  0.50& 0.75&   1.8~~~~\\
\vspace {0.05cm}
1159+395  &  2.370&21.00&  0.41&   $-$  &  $-$ &   $-$  &        ~~~~~~$-$     &        $-$       &       $-$      &    $-$ &   $-$ &    $-$~~~~\\
\vspace {0.05cm}
1201+394  &  0.445& 0.31& 11.99&  0.30& 1.4&  0.21&   0.60 $-$ 1.0&  0.120 $-$ 0.27&   6.2 $-$  5.5&  0.75& 0.09&   7.3~~~~\\
\vspace {0.05cm}
1216+402  &  0.756& 0.82& 27.97&  0.65& 0.9&  0.72&   0.06 $-$ 0.1&  0.018 $-$ 0.03&      3.0     &  1.00& 0.50&   2.7~~~~\\
\vspace {0.05cm}
{\it 1225+442}  &  0.348& 0.19&  0.98&  0.32& 1.3&  0.25&   0.60 $-$ 1.0$^a$&  0.200 $-$ 0.50&   3.5 $-$  2.0&  0.75& 0.17&   6.3~~~~\\
\vspace {0.05cm}
{\it 1241+411}  &  0.259& 0.07&  4.00&  0.65& 1.9&  0.34&   0.30 $-$ 0.6$^a$&  0.150 $-$ 0.35&   6.1 $-$  4.6&  0.70& 0.70&   7.4~~~~\\
\vspace {0.05cm}
1314+453A &  1.544&10.00&  1.37&  5.00&  $-$ &   $-$  &   0.10 $-$ 0.3&  0.250 $-$ 1.40&   (7.8)$-$(3.1)&  1.0 & 4.60&   (0.4)~~~~\\
\vspace {0.05cm}
{\it 1445+410}  &  0.180& 0.03& 24.64&  0.24& 0.7&  0.34&   0.06 $-$ 0.1&   ~0.006 $-$ 0.01$^c$&      3.5       &  0.50$^e$& 0.18&   3.3~~~~\\
\vspace {0.05cm}
1458+433  &  0.927& 1.40& 12.61&  2.50& 5.2&  0.48&   0.30 $-$ 0.6&  0.550 $-$ 1.60&   ~~8.2 $-$ (2.2)&  1.00& 1.60&  19.7~~~~\\
\hline
\vspace {0.05cm}
3C 237    &  0.880&22.00& 19.40&  2.20& 3.4&  0.65&   0.10 $-$ 0.3&  0.110 $-$ 0.52&   ~~9.0 $-$ (1.7)&  0.80& 1.75&  10.1~~~~\\
\vspace {0.05cm}
3C 241    &  1.610&24.00& 17.96&  2.20& 7.0&  0.31&   0.03 $-$ 0.3&  0.020 $-$ 0.50&  12.9 $-$ 9.5&  1.20& 1.30&  12.9~~~~\\
{\it 3C 268.3}  &  0.370& 1.60&  7.65&  0.20& 0.5&  0.40&   0.10 $-$ 0.3$^c$&  0.012 $-$ 0.04&   1.9 $-$  1.5&  0.60& 0.16&   1.9~~~~\\
\hline
\vspace {0.05cm}
4C 14.41  &  0.362& 1.00&  0.43&  0.10& 0.4&  0.25&   0.30 $-$ 1.0&  0.018 $-$ 0.10&   1.5 $-$  1.0&  0.80& 0.07&   1.8~~~~\\
\vspace {0.05cm}
1323+321  &  0.370& 2.00&  0.31&  0.35& 1.0&  0.34&   0.30 $-$ 1.0&  0.060 $-$ 0.37&   3.9 $-$  1.9&  0.70& 0.22&   4.9~~~~\\
\vspace {0.05cm}
{\it 1358+624}  &  0.430& 2.70&  0.39&  0.47& 1.4&  0.35&   0.03 $-$ 0.1$^a$&  0.004 $-$ 0.01&   5.6 $-$  5.5&  1.00& 0.20&   5.6~~~~\\
\vspace {0.05cm}
{\it OQ208}     &  0.077& 0.07*&  0.01&  0.66& 1.5&  0.44&    ~~~~~~0.1$^c$&  ~~0.05&     5.9      &  1.00$^e$& 0.65&   6.1~~~~\\
\vspace {0.05cm}
1607+268  &  0.470& 2.00&  0.29&  0.30& 1.3&  0.24&   0.03 $-$ 0.1&  0.003 $-$ 0.01&     5.1      &  1.00& 0.15&   5.3~~~~\\
\hline
\vspace {0.05cm}
1108+201  &  0.299& 0.32&  0.08&  0.37& 1.2&  0.30&   0.30 $-$ 0.6&  0.044 $-$ 0.12&   4.9 $-$  4.2&  0.50& 0.15&   5.4~~~~\\
\vspace {0.05cm}
{\it 1509+054}  &  0.084& 0.04*&  0.01&  0.15& 0.4&  0.38&  0.30 $-$ 0.6$^d$&  0.020 $-$ 0.05&   2.3 $-$  2.1&  0.50& 0.11&   2.3~~~~\\
\vspace {0.05cm}
1543+005  &  0.556& 2.00&  0.07&  0.58& 2.2&  0.26&   0.00 $-$ 0.6&  0.000 $-$ 0.26&     5.8      &  0.60& 0.04&   8.6~~~~\\
\hline
\vspace {0.05cm}
1503+4528 &  0.521& 0.58&  4.66&  0.65& 2.3&  0.28&   0.60 $-$ 1.0&  ~0.350 $-$ 0.900&   5.6 $-$  2.8&  0.75& 0.30&  10.1~~~~\\
\hline
\end{tabular}
\end{center}
Name in italics: GALEX data available.

Results from GALEX: $^a$: model unchanged; $^{b1}$: 0.6 Gyr excluded;  
$^{b2}$: 0.06 Gyr excluded; $^c$: 0.3-0.6 Gyr required; $^d$: 0.6-1.0 Gyr 
required; 1014+392: no possible model with SDSS data alone, model from GALEX 
and near-IR only; $^e$: power law excluded.
\label{SDSS_sample_2}
\end{table*}  

\noindent
The content of the table is:  

\noindent
Col. 1: Source name (in italics for sources with GALEX data).

\noindent
Col. 2: redshift.

\noindent
Cols. 3 and 4:  Radio luminosity (W~Hz$^{-1}$) at 1.4 GHz and linear 
size (kpc). 
{\bx For OQ208 and 1509+054, self-absorbed  at $\nu >$ 1.4 GHz, the luminosity
was extrapolated from the transparent  part of the spectrum. }

\noindent
Cols. 5-7: Luminosities at 0.30$\mu$m (L$_{0.30}$) and 0.45 $\mu$m 
(L$_{0.45}$), interpolated from the data.

\noindent
Cols. 8-10: range of parameters for a {\it two 
stellar population galaxy model} (Sect. \ref{YSP}) from SDSS and, when 
available, near-IR data. Values of M$_{\rm OSP}$ in parenthesis represent 
upper limits. Superscripts refer to table notes for model changes required by 
the addition of GALEX data. 

\noindent
Cols. 11-13: parameters for an OSP - power law model (see Sect. 
\ref{AGN-contr.}). A superscript refers to table note for model changes 
required by the addition of GALEX data. 

\smallskip
\noindent
The samples are separated by horizontal lines, in order from top to bottom: 
B3-VLA, 3C, PW, Labiano et al. (2007) and 9C.

\newpage
\section{UV-O-SEDs of B3\_CSS VLA and other compact sources }

\addtocounter{figure}{-1}
\begin{figure*}[ht]
{\includegraphics[width=16.0cm]{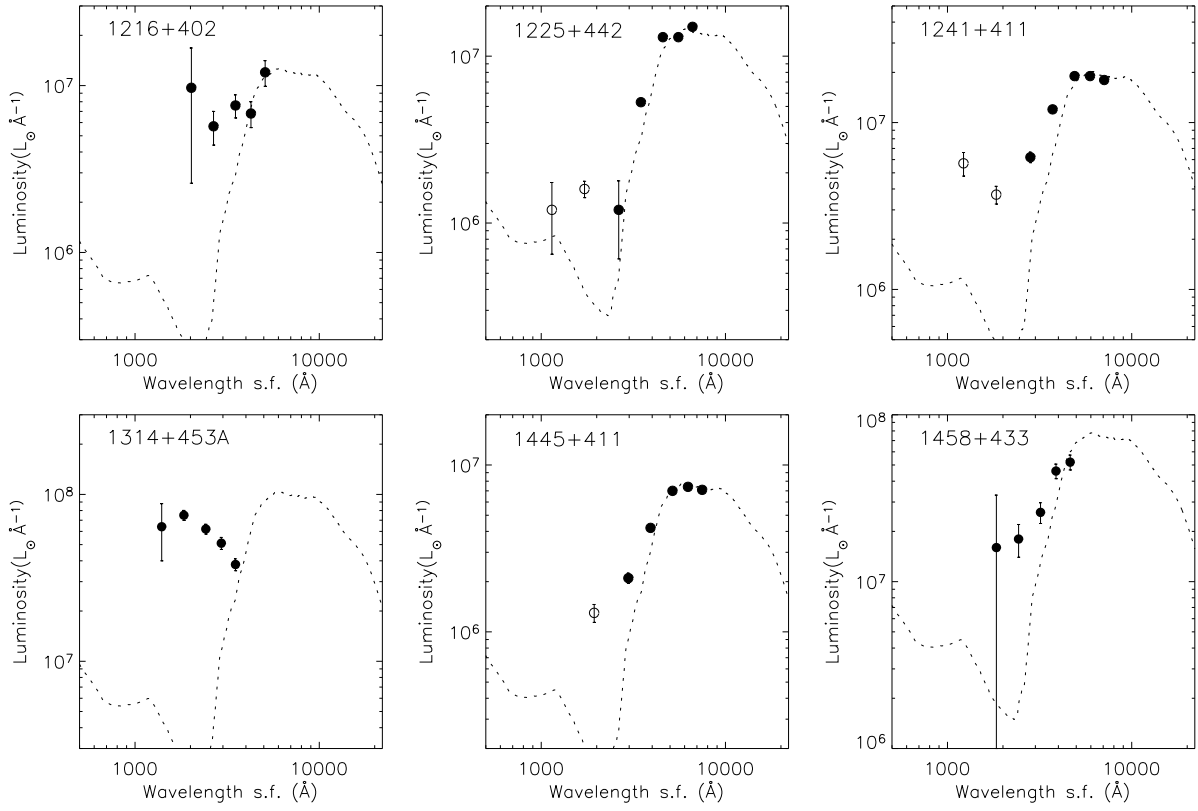}}
\caption{ Individual SEDs for galaxies with {\it spectroscopic}
 redshift (cont.)} 
\end{figure*}

\begin{figure*}[ht]
{\includegraphics[width=16.0cm]{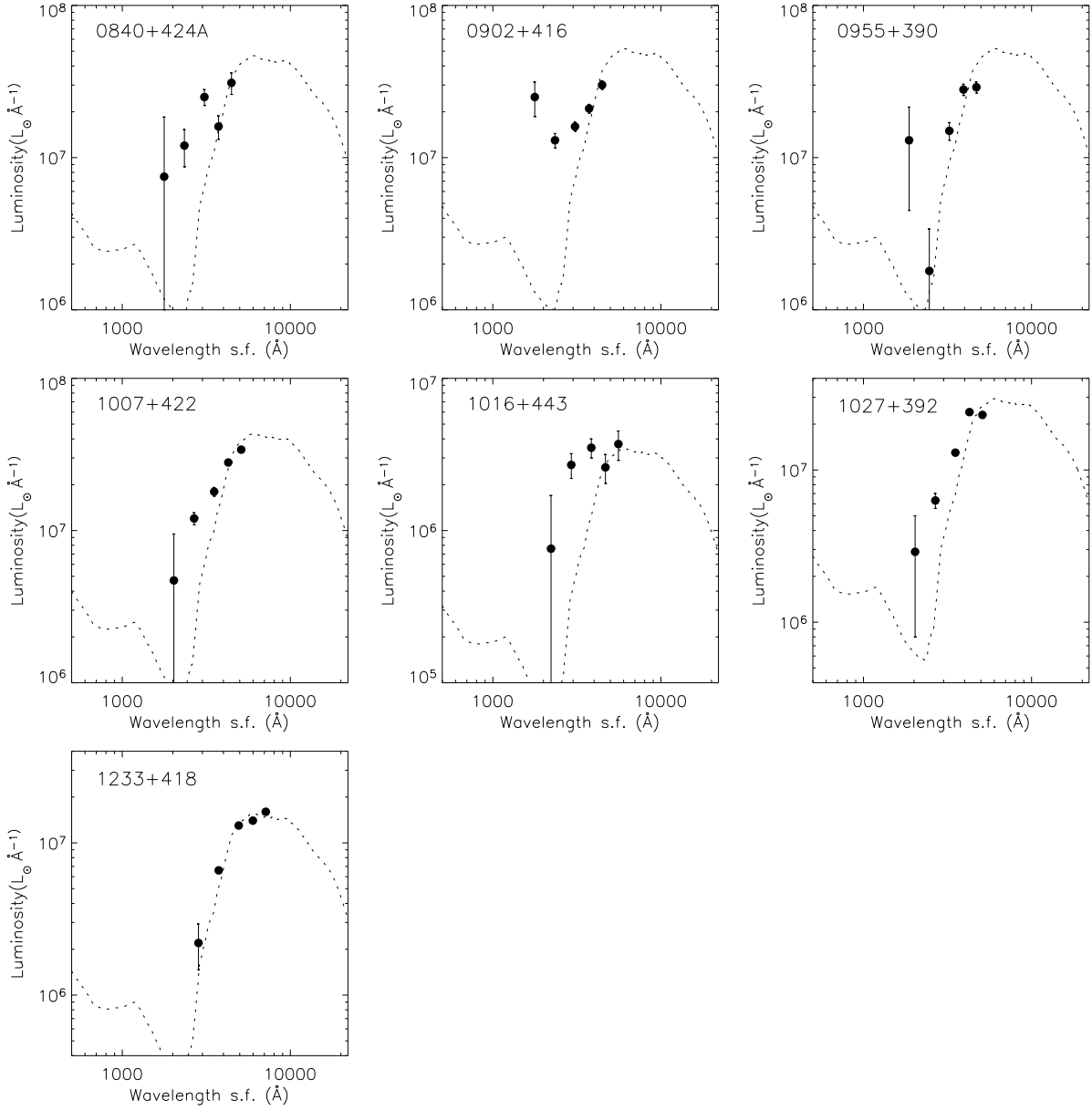}}
\label {plot_ph}
\caption{ Individual SEDs for galaxies with {\it photometric}
 redshift. The dotted line represents a B\&C model normalized
  to the long-wavelength SDSS data {\it only}.  } 
\end{figure*}

\begin{figure*}[ht]
{\includegraphics[width=16.0cm]{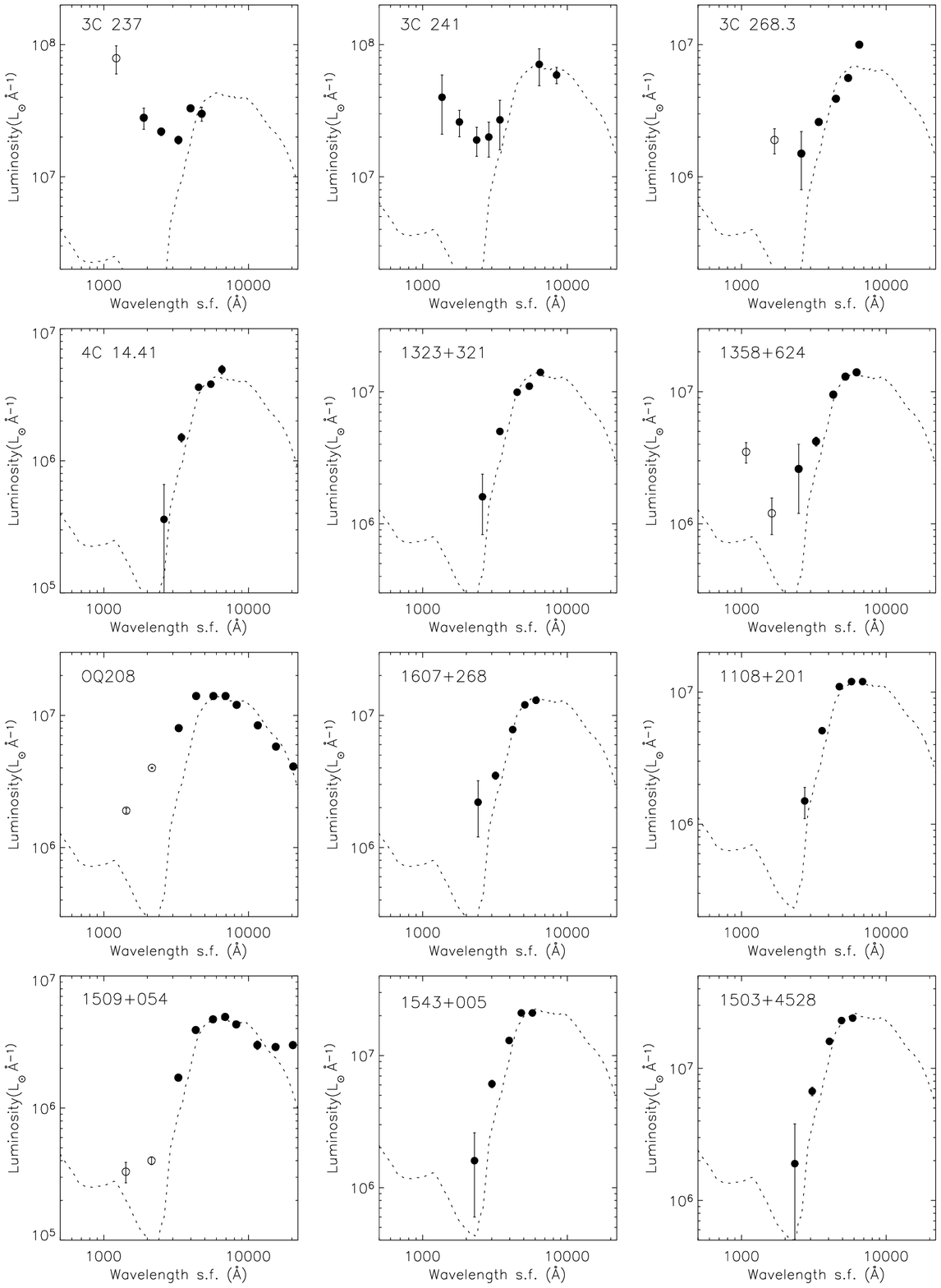}}
\label{RG_Seds}
\caption{ Individual SEDs for the {\it NON B3} galaxies with {\it 
spectroscopic} redshift. Empty circles are the GALEX data. The dotted line 
represents a B\&C model normalized to the long-wavelength SDSS data 
{\it only}. }
\end{figure*}

\onecolumn
\newpage

\section{The 3C/6C LSO sample}
\label{LSO}
\begin{table*}[h]
\begin{center}
\caption{Extended radio sources}
\begin{tabular}{llrr|llrr}
\hline
\hline
 Name &             z&   P$_{1.4~{\rm GHz}}$ &      LS& Name &             z&   P$_{1.4~{\rm GHz}}$ &      LS\\
      &              &    (10$^{27}$W~Hz$^{-1}$&           (kpc)&      &              &    (10$^{27}$W~Hz$^{-1}$&           (kpc)\\
\hline
\vspace {0.05cm}
 \object{3C192}&   0.060&  0.04&  210 & \object{3C280}&   0.996& 23.00&  104   \\
\vspace {0.05cm}
 \object{3C200}&   0.458&  1.50&  139 & \object{3C284}&   0.240&  0.31&  667   \\
\vspace {0.05cm}
 \object{3C217}&   0.898&  7.40&  101 & \object{3C285}&   0.079&  0.03&  251   \\
\vspace {0.05cm}
 \object{3C219}&   0.170&  0.72&  514 & \object{3C289}&   0.967& 10.00&  104   \\
\vspace {0.05cm}
 \object{3C223}&   0.140&  0.19&  585 & \object{3C292}&   0.71 &  4.30&  713\\
\vspace {0.05cm}
 \object{3C225B}&  0.580&  4.70&   38 & \object{3C295}&   0.464& 17.00&   27  \\  
\vspace {0.05cm}
 \object{3C226}&   0.818&  6.50&  538 & \object{3C299}&   0.367&  1.30&   61   \\
\vspace {0.05cm}
 \object{3C228}&   0.550&  4.40&  280 & \object{3C300}&   0.270&  0.77&  391   \\
\vspace {0.05cm}
 \object{3C234}&   0.185&  0.46&  330 & \object{3C303}&   0.14 &  0.15&   41\\
\vspace {0.05cm}
 \object{3C236}&   0.101&  0.10& 4285 & \object{3C319}&   0.192&  0.35&  197.0\\
\vspace {0.05cm}
 \object{3C239}&   1.780& 26.00&   96 & \object{3C321}&   0.096&  0.08&  522\\   
\vspace {0.05cm}
 \object{3C244.1}& 0.43 &  2.80&  287 & \object{3C322}&   1.68 & 30.00&  275\\  
\vspace {0.05cm}
 \object{3C247}&   0.742&  6.20&  102 &  \object{3C324}&   1.206& 17.00&   91\\   
 \vspace {0.05cm}
\object{3C252}&   1.105&  7.00&  460 &  \object{3C325}&   1.14 & 23.00&  129\\
\vspace {0.05cm}
 \object{3C263.1}& 0.82 &  9.00&   43 & \object{3C326}&   0.09 &  0.07& 1878\\ 
\vspace {0.05cm}
 \object{3C265}&   0.811&  7.70&  514 & \object{6C1011+36}&  1.042&  1.90&  398\\   
\vspace {0.05cm}
 \object{3C266}&   1.272& 12.00&   35 & \object{6C1017+37}&  1.053&  1.90&   57\\
\vspace {0.05cm}
 \object{3C267}&   1.144& 12.00&  302 & \object{6C1019+39}&  0.922&  1.90&   63\\   
\vspace {0.05cm}
 \object{3C272}&   0.94 &  6.90&  456 & \object{6C1129+37}&  1.060&  2.20&  122\\   
\vspace {0.05cm}
 \object{3C274.1}& 0.422&  1.70&  841 & \object{6C1217+36}&  1.088&  2.60&   25\\   
\vspace {0.05cm}
 \object{3C277.2}& 0.766&  4.30&  385 & \object{6C1257+36}&  1.004&  0.92&  306\\   
\vspace {0.05cm}
 \object{3C277.3}& 0.08 &  0.05&   34 \\
\hline
\end{tabular}
\end{center}
\label{Tab_LSO}
\end{table*}

\newpage
\section{UV-O-SEDs of 3CR \& 6C LSOs }

\begin{figure*}[h]
{\includegraphics[width=16.0cm]{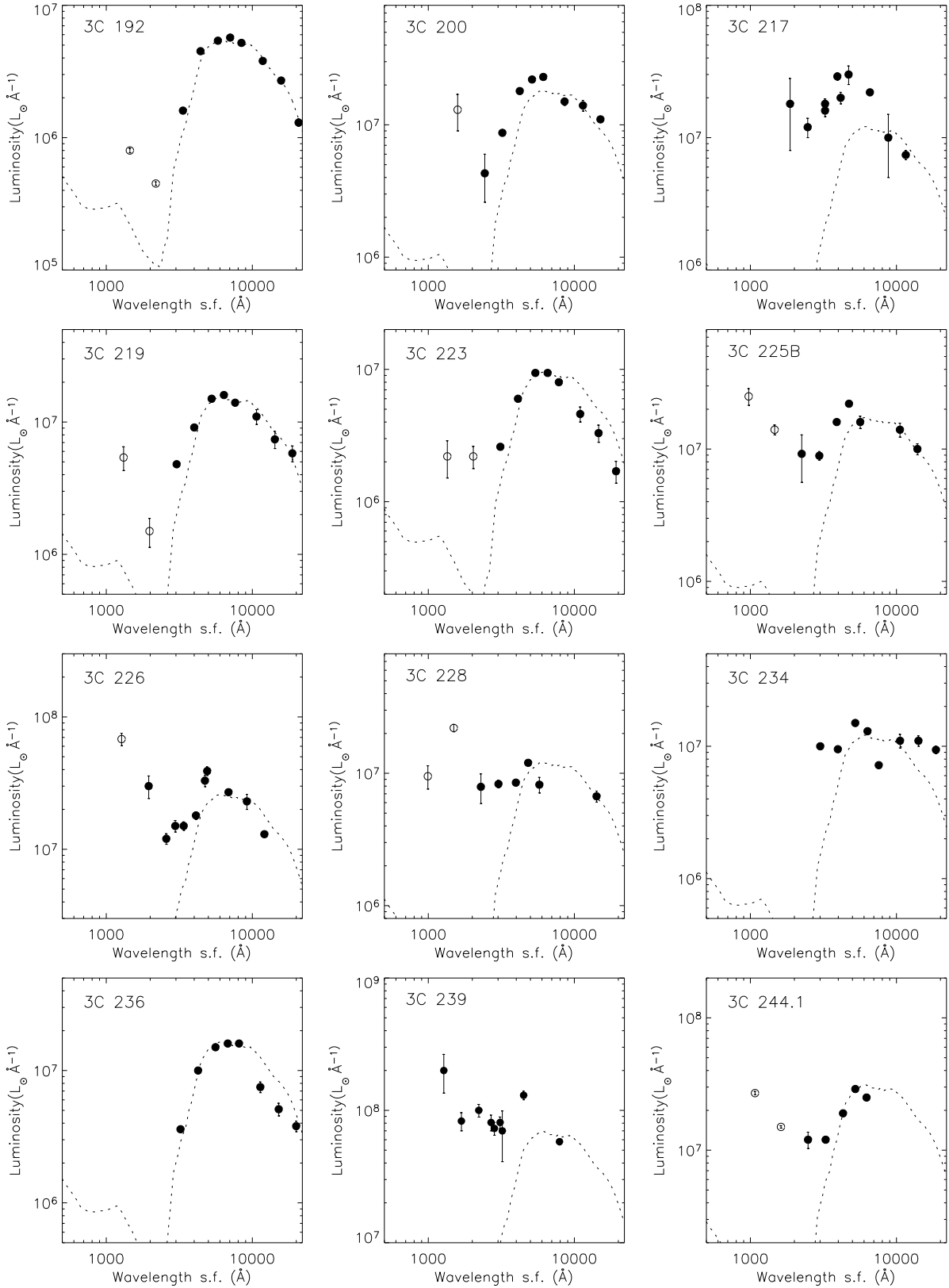}}
\label{3C_ext_SEDs}
\caption{ Individual 3CR and 6C LSO from Table  \ref{Tab_LSO}.1. The empty 
circles are GALEX data. The dotted line represents a B\&C model normalized to 
the long-wavelength SDSS data {\it only}.} 
\end{figure*}
\addtocounter{figure}{-1}
\begin{figure*}[p]
{\includegraphics[width=16.0cm]{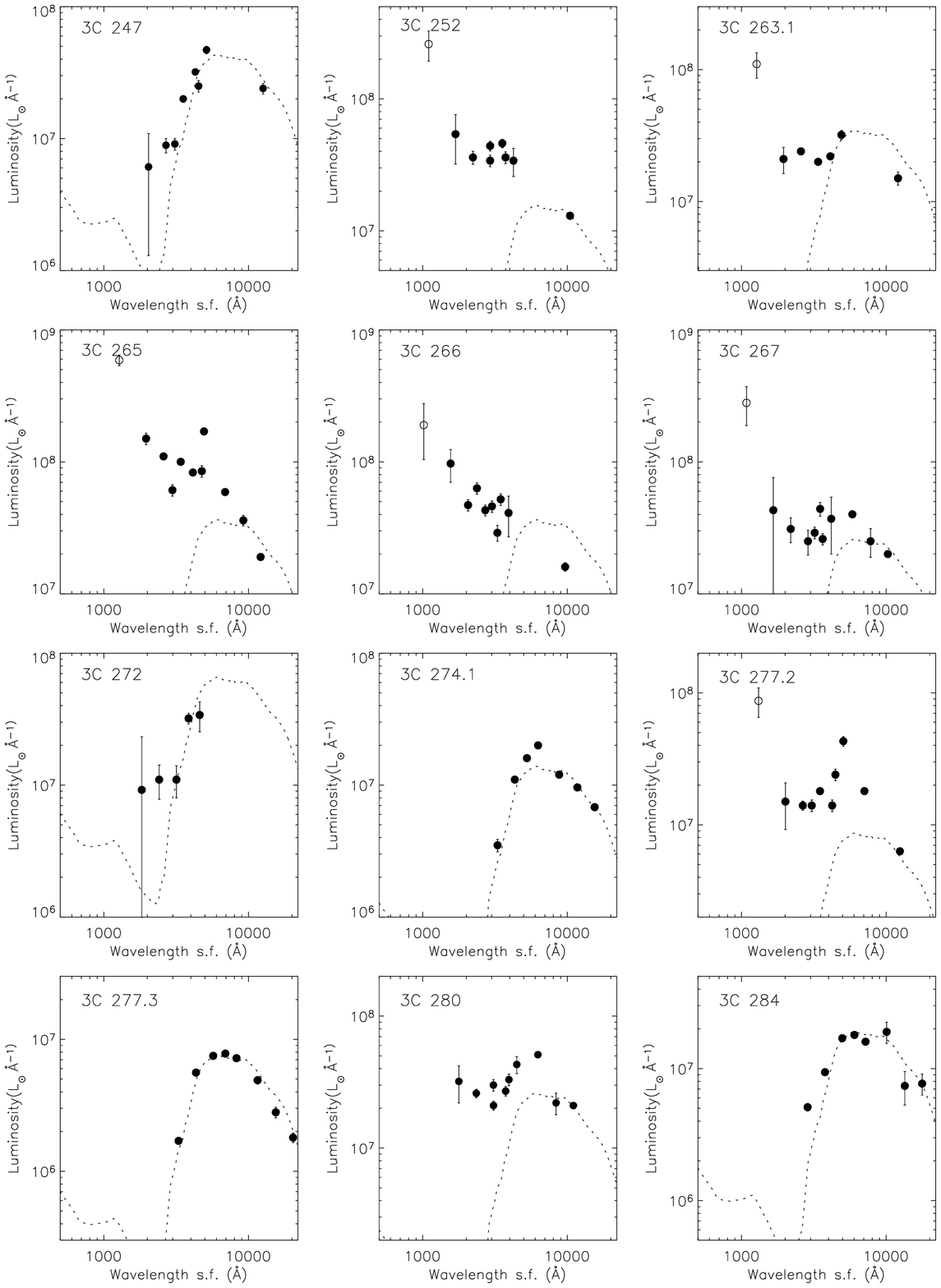}}
\caption{ Individual 3CR and 6C LSO from Table \ref{Tab_LSO}.1 (cont.)} 
\end{figure*}
\addtocounter{figure}{-1}
\begin{figure*}[p]
{\includegraphics[width=16.0cm]{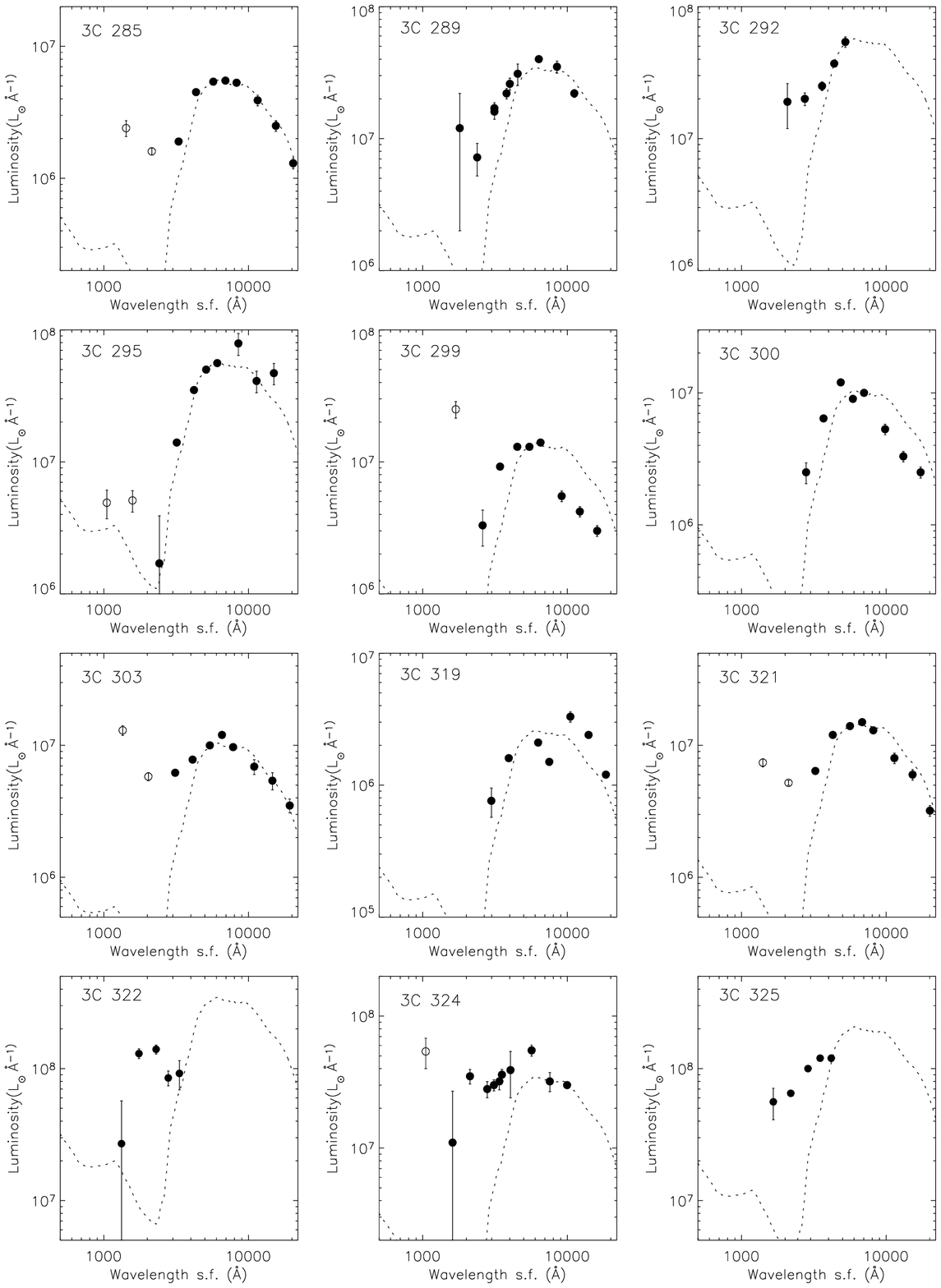}}
\caption{ Individual 3CR and 6C LSO from Table \ref{Tab_LSO}.1 (cont.)} 
\end{figure*}
\addtocounter{figure}{-1}
\begin{figure*}[p]
{\includegraphics[width=16.0cm]{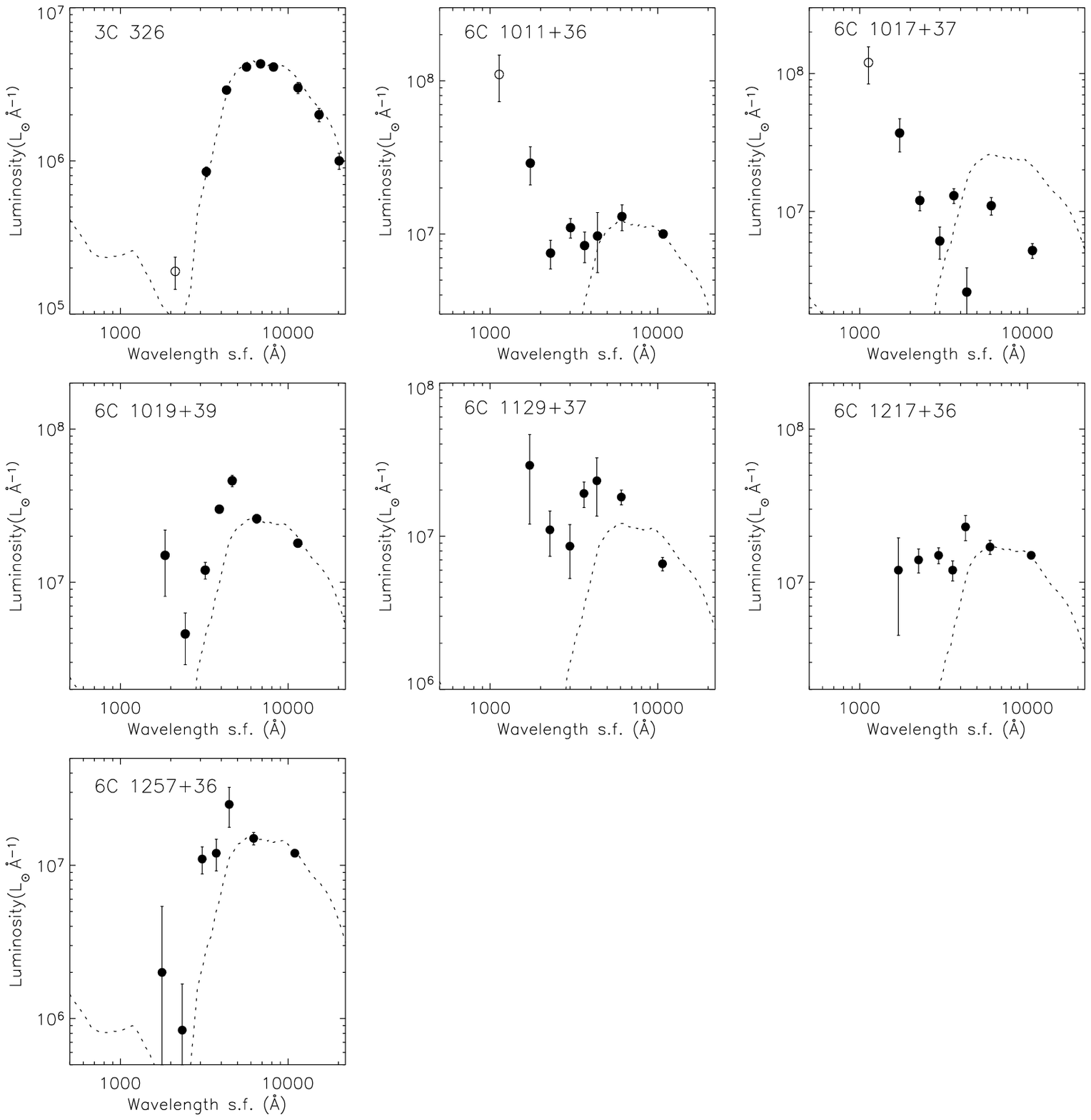}}
\caption{ Individual 3CR and 6C LSO from Table \ref{Tab_LSO}.1 (cont.)} 
\end{figure*}

\twocolumn
\listofobjects

\end{document}